\documentclass[onecolumn,aps,prx,showpacs,amsmath,superscriptaddress,longbibliography,notitlepage]{revtex4-1}
\usepackage{amssymb}
\usepackage{mathrsfs}
\usepackage{graphicx}
\usepackage{float}
\usepackage{verbatim}
\usepackage{epstopdf}
\usepackage[normalem]{ulem}
\usepackage{xcolor}
\usepackage{bm}
\usepackage{soul}
\usepackage{ulem}

\usepackage{graphicx,epsfig,xcolor}
\usepackage[T1]{fontenc}
\usepackage[latin1]{inputenc}
\usepackage[english]{babel}
\usepackage{amsmath}
\usepackage{amssymb}
\usepackage{graphicx}
\usepackage{babel}
\usepackage{upgreek}
\usepackage{mathrsfs}
\usepackage{setspace}
\usepackage{threeparttable}
\usepackage{amsmath,mathtools,amsthm}
\usepackage{array}
\usepackage{extarrows}
\usepackage{amsfonts}
\usepackage{epstopdf}
\usepackage{caption}
\usepackage{multirow}
\usepackage{xcolor}
\usepackage{enumitem}
\usepackage{amsmath}
\usepackage{bm}
\usepackage{graphicx}
\usepackage{subfigure}
\usepackage{ mathrsfs }
\usepackage{epstopdf}
\usepackage{ marvosym }
\usepackage{ tipa }
\usepackage{geometry}
\usepackage{ wasysym }
\usepackage{ulem}
\usepackage{indentfirst}
\usepackage{amsfonts,amssymb,dsfont}
\usepackage{accents}
\begin{document}
\title{Two-particle self-consistency in a system without condensation }

\author{Chen-Huan Wu$^{}$} \email{chenhuanwu1@gmail.com}
\affiliation{ College of Physics and Electronic Engineering, Northwest Normal University, Lanzhou 730070, China}

\begin{abstract}
		In a generic random system, the coexistence of extended and localized states can be evidenced by the subextensive width of energy distribution of a physical initial state in, for example, the quantum quenches which involving the local Hamiltonian.
The robust thermalization is also evidenced in terms of the microscopic canonical ensemble average in the thermalization limit\cite{Shiraishi},
which satisfies the weak eigenstate thermalization hypothesis (ETH).
In this article, we study the method of two-particle self-consistency 
for a system without condensation,
i.e., without the inaccessible localizations that violating the ETH.
We provide another pespective that considering the local conservation of an nonintegrable system the stubborn correlations between the three kinds of decompositions for the four-point functions,
which can be regarded as the elements in a product of the self-energy and Green function matrices (i.e., the two-particle correlations in Kadanoff and Baym notation\cite{Vilk Y M}).

\end{abstract}

\date{\today}
\maketitle

\section{Introduction}

In this article, we consider the many-body fermion system with conservation restrictions and using the 
functional-derivative-based nonperturbative approach,
which also called two-particle self-consistent approach (TPSC)\cite{Allen S}.
The system is divided into multi-subsystems with the non-Abelian-type symmetries (like the SU(2) symmetry in irreducible representation)
acting on these correlated subsystems.
In other word,
for an arbitary operator picking from a pair,
its second order derivative equivalents to a certain combination of its first and zeroth order time derivatives, and only depends on the another operator within this pair.
The second order derivatives of the two operators thus form the subspace
whose state expectation only related to the 
dynamical property about the time difference,
i.e., the overall character described by the single-particle density of this pair.
Similarly,
for a four-point function which owning two distinct sets of coefficients
(e.g., labelled by $\alpha$ and $\beta$;
and this is what the above decompositions are accoding to),
the correlation between these two sets of coefficients
guarantees the symmetry sector formed by $\pmb{\Psi}^{(1)}$
and $\pmb{\Psi}^{(2)}$.
And such correlation between $\pmb{\Psi}^{(1)}$
and $\pmb{\Psi}^{(2)}$ can only be observed in terms of the operators labelled by the imaginary time in string form.
This guarantees the U(1) conservation between two spin-components
(otherwise a constant ratio between the two spin components will be fixed).
Interestingly, for the Green functions labelled by imaginary time in string form,
the diagonal Green functions form such conservation through its two representations in different spin basis,
while the off-diagonal Green functions form such conservation through 
itself with its time-exchanged version,
i.e.,
$G_{21}(\tau_{1},\tau_{2})$ and $G_{12}^{*}(\tau_{1},\tau_{2})$,
or $G_{12}(\tau_{1},\tau_{2})$ and $G_{21}^{*}(\tau_{1},\tau_{2})$.
Thus a selectrive (restricted) sum is required in the fluctuation part of Eq.(\ref{6153}),
to reflect a fully thermalized equilibrium state where 
$\langle \hat{\sum}_{\alpha}c^{*}_{\alpha}\psi^{*}_{\alpha}(t_{1})\rangle=0$.
As there is finite correlation between the operators at $t_{1}$ and $t_{2}$,
thus an average over the selected $\alpha$'s which builts a nonlocal conservation at a single certain time reflecting a subsystem without the lowest energy state.

On the contrary,
a subsystem has a lowest energy state,
where the system Hamiltonian has zero expectation on each one of the mutually-independent eigenstates,
is extensive and nonergodic,.i.e.,
without a maximal number of the principal components. 
Here, in noninteracting case, each eigenstate follows the same pattern,
i.e.,
the product of creation operators (with the power determined by the corresponding noninteracting particle number) of the corresponding generator (lowest energy state).
In interacting case,
the corresponding integrable eigenstates are the superpositions of the products in the noninteracting case,
with the expansion coefficients on each product representing the particle-number fluctuations and two-particle correlations\cite{Kita,Bi}.
Note that an nonzero averaged value of nonfluctuation part (the condensate wave function) requires such superpositions as well as the finite two-particle correlation.

\section{Model}

Firstly we consider the many-body fermion system with conservation restrictions and using the 
functional-derivative-based nonperturbative approach,
which also called the two-particle self-consistent approach (TPSC)\cite{Allen S}.
The Green's function matrix reads
\begin{equation} 
\begin{aligned}
\label{1301}
G(\tau,\tau^{+})=
\begin{pmatrix}
G_{11}(\tau,\tau^{\pm}) & G_{12}(\tau,\tau^{\pm})\\
G_{21}(\tau,\tau^{\pm}) & G_{22}(\tau,\tau^{\pm})
\end{pmatrix},
\end{aligned}
\end{equation}
with
\begin{equation} 
\begin{aligned}
&G_{11}(\tau,\tau^{\pm})=-\mathcal{T}c_{\uparrow}(\tau)c_{\uparrow}^{\dag}(\tau^{\pm}),\\
&G_{22}(\tau,\tau^{\pm})=G_{11}(\tau,\tau^{\mp})=-G^{*}_{11}(\tau,\tau^{\pm})
=\mathcal{T}c_{\uparrow}(\tau^{\pm})c^{\dag}_{\uparrow}(\tau)
=-\mathcal{T}c^{\dag}_{\downarrow}(\tau)c_{\downarrow}(\tau^{\pm}),\\
&G_{12}(\tau,\tau^{\pm})=-\mathcal{T}c_{\uparrow}(\tau)c_{\downarrow}(\tau^{\pm}),\\
&G_{21}(\tau,\tau^{\pm})=-\mathcal{T}c^{\dag}_{\downarrow}(\tau)c_{\uparrow}^{\dag}(\tau^{\pm}),
\end{aligned}
\end{equation}
The property of the fermion product here is different from the fermion bilinear,
or the commutation-case (with $\tau\neq \tau'$ far away from the IR limit) unless
in the special case with $n_{\uparrow}=n_{\downarrow}=\frac{1}{2}$.
We note the following properties for the exchanging between two fermion operators (satisfying the 
TPSC) with difference of time components in IR limit,
\begin{equation} 
\begin{aligned}
\label{1191}
&\mathcal{T}c_{a}(\tau)c_{a}^{\dag}(\tau^{\pm})=-\mathcal{T}c^{\dag}_{a}(\tau)c_{a}(\tau^{\pm}),\\
&\mathcal{T}c_{a}(\tau)c_{a}^{\dag}(\tau^{\pm})=\mathcal{T}c_{b}(\tau^{\pm})c^{\dag}_{b}(\tau),\\
&\mathcal{T}c_{a}(\tau)c_{a}^{\dag}(\tau^{+})=\mathcal{T}c_{a}^{\dag}(\tau^{-})c_{a}(\tau)
=-\mathcal{T}c_{a}(\tau^{-})c^{\dag}_{a}(\tau),\\
&\mathcal{T}c_{a}(\tau^{+})c_{a}^{\dag}(\tau)=\mathcal{T}c_{a}^{\dag}(\tau)c_{a}(\tau^{-})
=-\mathcal{T}c_{a}(\tau)c^{\dag}_{a}(\tau^{-}),
\end{aligned}
\end{equation}
with spin-index $a,b=\uparrow,\downarrow\ (a\neq b)$.
Thus replacing the imaginary time componets between two fermion operators
equivalents to spin flip for both the two operators,
and there is a non-local effect
for the imaginary time evolution (antisymmetry).

Considering the Hartree-Fock-type conservation approximation, we
introduce the fermion number operators $n_{\uparrow}$ and $n_{\downarrow}$,
which has $n_{\uparrow}+n_{\downarrow}=1$ and
$G_{11}(\tau,\tau^{+})=n_{\uparrow}$,
$G_{22}(\tau,\tau^{+})=G_{11}(\tau,\tau^{-})=-n_{\downarrow}=n_{\uparrow}-1$.
For $n_{\uparrow}$ and $n_{\downarrow}$,
we have the following equivalent expressions for the number operators
of up-spin and down-spin (according to Eq.(\ref{1191})),
\begin{equation} 
\begin{aligned}
n_{\uparrow}
&=-\mathcal{T}c_{\uparrow}(\tau)c^{\dag}_{\uparrow}(\tau^{+})
=-\mathcal{T}c_{\downarrow}(\tau^{+})c^{\dag}_{\downarrow}(\tau)\\
&=\mathcal{T}c^{\dag}_{\uparrow}(\tau)c_{\uparrow}(\tau^{+})
=\mathcal{T}c^{\dag}_{\downarrow}(\tau^{+})c_{\downarrow}(\tau)\\
&=\mathcal{T}c_{\uparrow}(\tau^{-})c^{\dag}_{\uparrow}(\tau)
=\mathcal{T}c_{\downarrow}(\tau)c^{\dag}_{\downarrow}(\tau^{-})\\
&=-\mathcal{T}c^{\dag}_{\uparrow}(\tau^{-})c_{\uparrow}(\tau)
=-\mathcal{T}c^{\dag}_{\downarrow}(\tau)c_{\downarrow}(\tau^{-}),
\end{aligned}
\end{equation}
While the expressions for $n_{\downarrow}$ can be obtained by directly replacing the $\uparrow$
($\downarrow$) subscript in above expression by $\downarrow$ ($\uparrow$),
or equivalently, they can also be obtained by $n_{\downarrow}=n_{\uparrow}^{*}$
or $n_{\downarrow}=-G_{11}(\tau,\tau^{-})=1-n_{\uparrow}$.
It is easy to know that the operations of complex conjugation on $n_{\uparrow}$ is the same with 
replacing the $\uparrow$ ($\downarrow$) subscript by $\downarrow$ ($\uparrow$),
where we then arrives at
\begin{equation} 
\begin{aligned}
\label{1221}
n_{\downarrow}
&=-\mathcal{T}c_{\downarrow}(\tau)c^{\dag}_{\downarrow}(\tau^{+})
=-\mathcal{T}c_{\uparrow}(\tau^{+})c^{\dag}_{\uparrow}(\tau)\\
&=\mathcal{T}c^{\dag}_{\downarrow}(\tau)c_{\downarrow}(\tau^{+})
=\mathcal{T}c^{\dag}_{\uparrow}(\tau^{+})c_{\uparrow}(\tau)\\
&=\mathcal{T}c_{\downarrow}(\tau^{-})c^{\dag}_{\downarrow}(\tau)
=\mathcal{T}c_{\uparrow}(\tau)c^{\dag}_{\uparrow}(\tau^{-})\\
&=-\mathcal{T}c^{\dag}_{\downarrow}(\tau^{-})c_{\downarrow}(\tau)
=-\mathcal{T}c^{\dag}_{\uparrow}(\tau)c_{\uparrow}(\tau^{-}).
\end{aligned}
\end{equation}
In terms of $n_{\downarrow}=-G_{11}(\tau,\tau^{-})$,
we have
\begin{equation} 
\begin{aligned}
n_{\downarrow}
&=\mathcal{T}c_{\uparrow}(\tau)c^{\dag}_{\uparrow}(\tau^{-})
=\mathcal{T}c_{\downarrow}(\tau^{-})c^{\dag}_{\downarrow}(\tau)\\
&=-\mathcal{T}c^{\dag}_{\uparrow}(\tau)c_{\uparrow}(\tau^{-})
=-\mathcal{T}c^{\dag}_{\downarrow}(\tau^{-})c_{\downarrow}(\tau)\\
&=-\mathcal{T}c_{\uparrow}(\tau^{+})c^{\dag}_{\uparrow}(\tau)
=-\mathcal{T}c_{\downarrow}(\tau)c^{\dag}_{\downarrow}(\tau^{+})\\
&=\mathcal{T}c^{\dag}_{\uparrow}(\tau^{+})c_{\uparrow}(\tau)
=\mathcal{T}c^{\dag}_{\downarrow}(\tau)c_{\downarrow}(\tau^{+}),
\end{aligned}
\end{equation}
which can be verified that equivalents to Eq.(\ref{1221}).
As $n_{\downarrow}+n_{\uparrow}=1$,
we have the following commutation relations between fermion fields
(not the anticommutation due to a small slice in imaginary time)
\begin{equation} 
\begin{aligned}
&[c_{\uparrow}^{\dag}(\tau),c_{\uparrow}(\tau^{+})]=1,
[c_{\downarrow}^{\dag}(\tau^{+}),c_{\downarrow}(\tau)]=1,\\
&[c_{\uparrow}^{\dag}(\tau^{+}),c_{\uparrow}(\tau)]=1,
[c_{\downarrow}^{\dag}(\tau),c_{\downarrow}(\tau^{+})]=1,\\
&[c_{\uparrow}(\tau^{-}),c_{\uparrow}^{\dag}(\tau)]=1,
[c_{\downarrow}(\tau),c_{\downarrow}^{\dag}(\tau^{-})]=1,\\
&[c_{\uparrow}(\tau),c_{\uparrow}^{\dag}(\tau^{-})]=1,
[c_{\downarrow}(\tau^{-}),c_{\downarrow}^{\dag}(\tau)]=1.
\end{aligned}
\end{equation}

We can rewrite the Green's function in Eq.(\ref{1301}) as
\begin{equation} 
\begin{aligned}
\label{241}
{\bf G}(\tau,\tau^{+};\theta)
&=
\begin{pmatrix}
\mathcal{T}c^{\dag}_{\downarrow}(\tau^{+})c_{\downarrow}(\tau)
 & \mathcal{T}c_{\downarrow}(\tau^{+})c_{\uparrow}(\tau)\\
 \mathcal{T}c^{\dag}_{\uparrow}(\tau^{+})c_{\downarrow}^{\dag}(\tau) & -\mathcal{T}c^{\dag}_{\uparrow}(\tau^{+})c_{\uparrow}(\tau)
\end{pmatrix}
=\begin{pmatrix}
\psi^{\dag}_{1s}\psi_{2s}
 & \psi_{1s}\psi_{2s'}\\
\psi^{\dag}_{1s'}\psi^{\dag}_{2s} & -\psi^{\dag}_{1s'}\psi_{2s'}
\end{pmatrix}
=\begin{pmatrix}
G_{11}(\tau,\tau^{+};\theta) & G_{12}(\tau,\tau^{+};\theta)\\
G_{21}(\tau,\tau^{+};\theta) & G_{22}(\tau,\tau^{+};\theta) 
\end{pmatrix},
\end{aligned}
\end{equation}
where the off-diagonal terms originate from the pairing field
$\theta(\tau,\tau)$.
While the inversed Green's function in perturbative theory 
reads\cite{Vilk Y M,Allen S} 
\begin{equation} 
\begin{aligned}
\label{321}
{\bf G}^{-1}(\tau,\tau^{+};\theta)=
\begin{pmatrix}
G_{11}^{-1}(\tau,\tau^{+})-\Sigma_{11}(\tau,\tau^{+};\theta) & 
- \Sigma_{12}(\tau,\tau^{+};\theta)-\theta(\tau,\tau^{+})\\
-\Sigma_{21}(\tau,\tau^{+};\theta)-\theta^{*}(\tau,\tau^{+})  & G_{22}^{-1}(\tau,\tau^{+})-\Sigma_{22}(\tau,\tau^{+};\theta)
\end{pmatrix},
\end{aligned}
\end{equation}
and the susceptibility reads
\begin{equation} 
\begin{aligned}
\sum_{\tau'}\Sigma(\tau,\tau';\theta)G(\tau',\tau^{+};\theta)
&=U\begin{pmatrix}
\psi^{\dag}_{1s}\psi_{2s}\psi^{\dag}_{1s}\psi_{2s}
 & \psi^{\dag}_{1s}\psi_{2s}\psi_{1s}\psi_{2s'}\\
-\psi^{\dag}_{1s'}\psi_{2s'}\psi^{\dag}_{1s'}\psi^{\dag}_{2s} & \psi^{\dag}_{1s'}\psi_{2s'}\psi^{\dag}_{1s'}\psi_{2s'}
\end{pmatrix}.
\end{aligned}
\end{equation}

\section{The self-energy matrix with pairing field at $\tau$}

By introducing an auxiliary pairing field,
the self-energy can be solved from above self-consistent equation,
where  we firstly consider the self-energy matrix containing two kinds of time trajectories,
\begin{equation} 
\begin{aligned}
\begin{pmatrix}
\label{5571}
\Sigma_{11}(\tau,\tau_{f};\theta) & \Sigma_{12}(\tau,\tau_{f};\theta)\\
\Sigma_{21}(\tau^{+},\tau_{f};\theta) & \Sigma_{22}(\tau^{+},\tau_{f};\theta)
\end{pmatrix}
&=U\sum_{\tau'}\begin{pmatrix}
-\mathcal{T}\dot{\psi}^{\dag}_{1s}\ddot{\psi}_{2s}\ddot{\psi}_{2s'}\dot{\psi}^{\dag}_{\tau's'}
 & -\mathcal{T}\dot{\psi}^{\dag}_{1s}\ddot{\psi}_{2s}\ddot{\psi}_{2s'}\dot{\psi}_{\tau' s}\\
-\mathcal{T}\dot{\psi}_{1s'}\ddot{\psi}^{\dag}_{2s'}\ddot{\psi}^{\dag}_{2s}\dot{\psi}^{\dag}_{\tau' s'} & -\mathcal{T}\dot{\psi}_{1s'}\ddot{\psi}^{\dag}_{2s'}\ddot{\psi}^{\dag}_{2s}\dot{\psi}_{\tau' s}
\end{pmatrix}
{\bf G}^{-1}(\tau',\tau_{f};\theta)\\
&=-U\sum_{\tau'}\begin{pmatrix}
p_{21}(\tau^{+},\tau',\theta^{*})
 & p_{22}(\tau^{+},\tau',\theta^{*})\\
p_{11}(\tau^{+},\tau',\theta)
 & p_{12}(\tau^{+},\tau',\theta)
\end{pmatrix}
{\bf G}^{-1}(\tau',\tau_{f};\theta),
\end{aligned}
\end{equation}
where the pairing field 
$\theta^{*}(\tau,\tau)$ is related to the generation
of pairing $\psi_{2s}\psi_{2s'}$ in the instantaneous approximation of propagator $ \lim_{\tau^{+}\rightarrow\tau}G_{12}(\tau,\tau^{+};\theta)$,
while 
$\theta(\tau,\tau)$ is related to 
$\psi^{\dag}_{2s'}\psi^{\dag}_{2s}=\lim_{\tau^{+}\rightarrow\tau}G_{21}(\tau,\tau^{+};\theta)$.
And the functional derivative operators
$\frac{\delta}{\delta \theta^{*}(\tau,\tau)}$ and $\frac{\delta}{\delta \theta(\tau,\tau)}$ correspond to the 
initial point and middle point in the time trajectories $\tau \rightarrow \tau^{+}\rightarrow\tau'$
and $\tau^{+}\rightarrow\tau\rightarrow\tau'$.
In the above expression,
there is a replacement of time component $\tau^{+}\rightarrow
\tau'$ in the first line,
and $\tau\rightarrow\tau^{+},\tau^{+}\rightarrow\tau'$ in the third line,
which can be seen by comparing with the original expression Eq.(\ref{241}).
Note that in any case $\tau'\neq \tau$.

For convinience, we define the matrix within above expression as $\Psi(\tau^{+},\tau';\tau)$
which represent the functioanl derivative with respect to the static source field imaginary time $\tau$,
and using the following property of functional derivative,
\begin{equation} 
\begin{aligned}
	\label{3291}
\frac{\delta F[\theta]}{\delta \theta^{*}(\tau^{+})}=[\theta(\tau^{+}),F[\theta]],
\ 
\frac{\delta F[\theta^{*}]}{\delta \theta(\tau^{+})}=-[\theta^{*}(\tau^{+}),F[\theta^{*}]],
\end{aligned}
\end{equation}
where $F[\theta]$ is a functional of $\theta$ (Green's function here).
Each element of the matrix ${\bf \Psi}(\tau,\tau^{+},\tau')$ 
can be represented by functional derivative
or the commutation relations between pairing field and components of Green's function,
\begin{equation} 
\begin{aligned}
&	\label{2171}
{\bf \Psi}(\tau^{+},\tau';\tau)=
\begin{pmatrix}
-\mathcal{T}\dot{\psi}^{\dag}_{1s}\ddot{\psi}_{2s}\ddot{\psi}_{2s'}\dot{\psi}^{\dag}_{\tau's'}
 & -\mathcal{T}\dot{\psi}^{\dag}_{1s}\ddot{\psi}_{2s}\ddot{\psi}_{2s'}\dot{\psi}_{\tau' s}\\
-\mathcal{T}\dot{\psi}_{1s'}\ddot{\psi}^{\dag}_{2s'}\ddot{\psi}^{\dag}_{2s}\dot{\psi}^{\dag}_{\tau' s'} & -\mathcal{T}\dot{\psi}_{1s'}\ddot{\psi}^{\dag}_{2s'}\ddot{\psi}^{\dag}_{2s}\dot{\psi}_{\tau' s}
\end{pmatrix}
=-\begin{pmatrix}
p_{21}(\tau^{+},\tau',\theta^{*})
 & p_{22}(\tau^{+},\tau',\theta^{*})\\
p_{11}(\tau^{+},\tau',\theta)
 & p_{12}(\tau^{+},\tau',\theta)
\end{pmatrix}\\
=&\begin{pmatrix}
{}[G_{21}(\tau^{+},\tau'),\theta(\tau,\tau)] & [G_{22}(\tau^{+},\tau'),\theta(\tau,\tau)]\\
[\theta^{*}(\tau,\tau),G_{11}(\tau^{+},\tau')] & 
[\theta^{*}(\tau,\tau),G_{12}(\tau^{+},\tau')]
\end{pmatrix}
=\begin{pmatrix}
	{}[-\psi^{\dag}_{1s}\psi^{\dag}_{\tau's'},-\psi_{2s}\psi_{2s'}] & [-\psi^{\dag}_{1s}\psi_{\tau's},-\psi_{2s}\psi_{2s'}]\\
	[-\psi^{\dag}_{2s'}\psi^{\dag}_{2s},-\psi_{1s'}\psi^{\dag}_{\tau's'}] & 
	[-\psi^{\dag}_{2s'}\psi^{\dag}_{2s},-\psi_{1s'}\psi_{\tau's}]
\end{pmatrix},
\end{aligned}
\end{equation}
where we define the source fields $\theta(\tau,\tau)
=-\psi_{2s}\psi_{2s'}=-G_{12}(\tau,\tau)$,
$\theta^{*}(\tau,\tau)
=-\psi^{\dag}_{2s'}\psi^{\dag}_{2s}=-G_{21}(\tau,\tau)$. 
%While $\theta(\tau^{+},\tau')=-G_{12}(\tau^{+},\tau')=\psi_{1s'}\psi_{\tau's}$,
%$\theta^{*}(\tau',\tau^{+})=-G_{21}(\tau^{+},\tau')=\psi^{\dag}_{1s}\psi^{\dag}_{\tau's'}$.

In terms of the correlation between the source field and the normal component of Green's function,
it can be described by other decompositions where
the source field is a product of creation and annihilation operators 
(instead of a pair of creation operator or annihilation operator).
Then combined with the deduction in Appendix.A,
we obtain that the relation between different decompositions for each element
(we only consider the normal case here where
$\psi_{\tau's}\psi_{2s}^{\dag}$
and $\psi^{\dag}_{2s}\psi_{\tau's}$
be the annihilational termfor $\Psi_{11}$
and $\Psi_{22}$, respectively)
\begin{equation} 
	\begin{aligned}
		\label{641}
		&\Psi_{11}(\tau^{+},\tau';\tau)=
		-\mathcal{T}\dot{\psi}^{\dag}_{1s}\ddot{\psi}_{2s}\ddot{\psi}_{2s'}\dot{\psi}^{\dag}_{\tau's'}
		=
		\left[-\mathcal{T}
		\dot{\psi}^{\dag}_{1s}\dot{\psi}_{2s}\ddot{\psi}_{2s'}\ddot{\psi}^{\dag}_{\tau's'}\right]-
		\left[-\mathcal{T}\dot{\psi}^{\dag}_{1s}\ddot{\psi}_{2s}\dot{\psi}_{2s'}\ddot{\psi}^{\dag}_{\tau' s'}
		\right],\\
		&\Psi_{12}(\tau^{+},\tau';\tau)
		=-\mathcal{T}\dot{\psi}_{1s'}\ddot{\psi}^{\dag}_{2s'}\ddot{\psi}^{\dag}_{2s}\dot{\psi}^{\dag}_{\tau' s'}
		=\left[-\mathcal{T}\dot{\psi}_{1s'}\dot{\psi}^{\dag}_{2s'}\ddot{\psi}^{\dag}_{2s}\ddot{\psi}^{\dag}_{\tau' s'}\right]
		-
		\left[-\mathcal{T}\dot{\psi}_{1s'}\ddot{\psi}^{\dag}_{2s'}\dot{\psi}^{\dag}_{2s}\ddot{\psi}^{\dag}_{\tau' s'}\right],\\
		&\Psi_{21}(\tau^{+},\tau';\tau)
		=-\mathcal{T}\dot{\psi}_{1s'}\ddot{\psi}^{\dag}_{2s'}\ddot{\psi}^{\dag}_{2s}\dot{\psi}^{\dag}_{\tau' s'}
		=
		\left[-\mathcal{T}\dot{\psi}_{1s'}\dot{\psi}^{\dag}_{2s'}\ddot{\psi}^{\dag}_{2s}\ddot{\psi}^{\dag}_{\tau' s'}\right]
		+
		\left[-\mathcal{T}\dot{\psi}_{1s'}\ddot{\psi}^{\dag}_{2s'}\dot{\psi}^{\dag}_{2s}\ddot{\psi}^{\dag}_{\tau' s'}\right],\\
		&\Psi_{22}(\tau^{+},\tau';\tau)
		=-\mathcal{T}\dot{\psi}_{1s'}\ddot{\psi}^{\dag}_{2s'}\ddot{\psi}^{\dag}_{2s}\dot{\psi}_{\tau's}
		=\left[-\mathcal{T}\dot{\psi}_{1s'}\dot{\psi}^{\dag}_{2s'}\ddot{\psi}^{\dag}_{2s}\ddot{\psi}_{\tau's}\right]
		+\left[-\mathcal{T}\dot{\psi}_{1s'}\ddot{\psi}^{\dag}_{2s'}\dot{\psi}^{\dag}_{2s}\ddot{\psi}_{\tau's}\right].
	\end{aligned}
\end{equation}
The above two equations show the same result of the limit of zero pairing field,
i.e.,
\begin{equation} 
\begin{aligned}
\Sigma(\tau^{+},\tau;\theta)
&=U\sum_{\tau'}
\begin{pmatrix}
n_{\downarrow}-n_{\uparrow}
 & -\frac{1}{3}\psi_{2s'}\psi_{\tau's}\\
\psi_{2s}^{\dag}\psi^{\dag}_{\tau's'} & n_{\downarrow}-n_{\uparrow}
\end{pmatrix}
{\bf G}^{-1}(\tau',\tau^{+};\theta).
\end{aligned}
\end{equation}

\section{Functional property in terms of the propagators}

Using the Luttinger-Ward identity base on the Green's function in Eq.(\ref{321}),
\begin{equation} 
	\begin{aligned}
1=\frac{\delta (-G_{12}^{-1})}{\delta\theta}+\frac{\delta (-\Sigma_{12})}{\delta\theta},\ 
1=\frac{\delta (-G_{21}^{-1})}{\delta\theta^{*}}+\frac{\delta (-\Sigma_{21})}{\delta\theta^{*}},\\
\frac{\delta G_{11}^{-1}}{\delta\theta}=\frac{\delta (-\Sigma_{11})}{\delta\theta},\ 
\frac{\delta G_{22}^{-1}}{\delta\theta^{*}}=\frac{\delta (-\Sigma_{22})}{\delta\theta^{*}},\\
	\end{aligned}
\end{equation}
where $\frac{\delta {\bf G}_{0}^{-1}}{\delta\theta}=0$,
and note that ${\bf G}\frac{\partial}{\partial \theta^{*}}{\bf G}^{-1}
=\frac{\partial}{\partial \theta^{*}}{\rm ln}{\bf G}^{-1}=-{\bf G}^{-1}
\frac{\partial}{\partial \theta^{*}}{\bf G}$,
we can reproduce the above results in terms of the functional derivatives 
of ${\bf G}$ (the susceptibility) and ${\bf G}^{-1}$ (the self-energy).
\begin{equation} 
\begin{aligned}
\label{331}
\pmb{\Sigma}=
\frac{-1}{U}
&\begin{pmatrix}
	\Sigma_{11}(\tau,\tau_{f};\theta) & \Sigma_{12}(\tau,\tau_{f};\theta)\\
	\Sigma_{21}(\tau^{+},\tau_{f};\theta) & \Sigma_{22}(\tau^{+},\tau_{f};\theta) 
	\end{pmatrix}=
\sum_{\tau'}\begin{pmatrix}
p_{21}(\tau^{+},\tau',\theta^{*}) 
& p_{22}(\tau^{+},\tau',\theta^{*}) \\
p_{11}(\tau^{+},\tau',\theta) 
& p_{12}(\tau^{+},\tau',\theta) 
\end{pmatrix}{\bf G}^{-1}(\tau',\tau_{f};\theta)\\
&=\sum_{\tau'}\begin{pmatrix}
p_{21}(\tau^{+},\tau',\theta^{*}) G^{-1}_{11}(\tau',\tau_{f};\theta)
& p_{21}(\tau^{+},\tau',\theta^{*})G_{12}^{-1}(\tau',\tau_{f};\theta) \\
p_{12}(\tau^{+},\tau',\theta) G_{21}^{-1}(\tau',\tau_{f};\theta)
& p_{12}(\tau^{+},\tau',\theta) G_{22}^{-1}(\tau',\tau_{f};\theta)
\end{pmatrix}\\
&+
\sum_{\tau'}\begin{pmatrix}
	p_{22}(\tau^{+},\tau',\theta^{*}) G^{-1}_{21}(\tau',\tau_{f};\theta)
	& p_{22}(\tau^{+},\tau',\theta^{*})G_{22}^{-1}(\tau',\tau_{f};\theta) \\
	p_{11}(\tau^{+},\tau',\theta) G_{11}^{-1}(\tau',\tau_{f};\theta)
	& p_{11}(\tau^{+},\tau',\theta) G_{12}^{-1}(\tau',\tau_{f};\theta)
\end{pmatrix}
\\
&=\sum_{\tau'}\begin{pmatrix}
-G_{22}(\tau^{+},\tau';\theta)
q_{21}(\tau',\tau_{f},\theta^{*}) & 
-G_{22}(\tau^{+},\tau';\theta)
q_{22}(\tau',\tau_{f},\theta^{*}) \\
-G_{11}(\tau^{+},\tau';\theta)
q_{11}(\tau',\tau_{f},\theta) & 
-G_{11}(\tau^{+},\tau';\theta)
q_{12}(\tau',\tau_{f},\theta) 
\end{pmatrix}\\
&
+\sum_{\tau'}\begin{pmatrix}
	-G_{21}(\tau^{+},\tau';\theta)
	q_{11}(\tau',\tau_{f},\theta^{*}) & 
	-G_{21}(\tau^{+},\tau';\theta)
	q_{12}(\tau',\tau_{f},\theta^{*}) \\
	-G_{12}(\tau^{+},\tau';\theta)
	q_{21}(\tau',\tau_{f},\theta) & 
	-G_{12}(\tau^{+},\tau';\theta)
	q_{22}(\tau',\tau_{f},\theta) 
\end{pmatrix},
\end{aligned}
\end{equation}
where the functional derivatives with respect to inverse Green function read
\begin{equation} 
\begin{aligned}
	\label{6281}
&\frac{\delta {\bf G}^{-1}(\tau',\tau_{f};\theta)}
{\delta \theta^{*}(\tau,\tau)}
=-\delta_{\tau',\tau}\delta_{\tau_{f},\tau}+\begin{pmatrix}
-\frac{\delta \Sigma_{11}(\tau',\tau_{f};\theta)}{\delta G_{21}(\tau_{3},\tau_{4};\theta)}
\frac{\delta G_{21}(\tau_{3},\tau_{4};\theta)}{\delta \theta^{*}(\tau,\tau)} 
& -\frac{\delta \Sigma_{12}(\tau',\tau_{f};\theta)}{\delta G_{21}(\tau_{3},\tau_{4};\theta)}
\frac{\delta G_{21}(\tau_{3},\tau_{4};\theta)}{\delta \theta^{*}(\tau,\tau)}  \\
-\frac{\delta \Sigma_{21}(\tau',\tau_{f};\theta)}{\delta G_{21}(\tau_{3},\tau_{4};\theta)}
\frac{\delta G_{21}(\tau_{3},\tau_{4};\theta)}{\delta \theta^{*}(\tau,\tau)} 
&-\frac{\delta \Sigma_{22}(\tau',\tau_{f};\theta)}{\delta G_{21}(\tau_{3},\tau_{4};\theta)}
\frac{\delta G_{21}(\tau_{3},\tau_{4};\theta)}{\delta \theta^{*}(\tau,\tau)} 
\end{pmatrix},\\
&\frac{\delta {\bf G}^{-1}(\tau',\tau_{f};\theta)}
{\delta \theta(\tau,\tau)}
=-\delta_{\tau',\tau}\delta_{\tau_{f},\tau}+\begin{pmatrix}
-\frac{\delta \Sigma_{11}(\tau',\tau_{f};\theta)}{\delta G_{12}(\tau_{3},\tau_{4};\theta)}
\frac{\delta G_{12}(\tau_{3},\tau_{4};\theta)}{\delta \theta(\tau,\tau)} 
&-\frac{\delta \Sigma_{12}(\tau',\tau_{f};\theta)}{\delta G_{12}(\tau_{3},\tau_{4};\theta)}
\frac{\delta G_{12}(\tau_{3},\tau_{4};\theta)}{\delta \theta(\tau,\tau)}  \\
-\frac{\delta \Sigma_{21}(\tau',\tau_{f};\theta)}{\delta G_{12}(\tau_{3},\tau_{4};\theta)}
\frac{\delta G_{12}(\tau_{3},\tau_{4};\theta)}{\delta \theta(\tau,\tau)} 
&-\frac{\delta \Sigma_{22}(\tau',\tau_{f};\theta)}{\delta G_{12}(\tau_{3},\tau_{4};\theta)}
\frac{\delta G_{12}(\tau_{3},\tau_{4};\theta)}{\delta \theta(\tau,\tau)} 
\end{pmatrix}.
\end{aligned}
\end{equation}

Using Eq.(\ref{331}), we can further obtain
\begin{equation} 
	\begin{aligned}
		\label{6246}
		&p_{21}(\tau^{+},\tau_{f},\theta^{*})
		=\sum_{\tau',\tau''}
	-G_{22}(\tau^{+},\tau';\theta)
q_{21}(\tau',\tau'';\theta^{*})
	G_{11}(\tau'',\tau_{f};\theta)\\
	&=\sum_{\tau',\tau''}
	-G_{22}(\tau^{+},\tau';\theta)
q_{22}^{-1}(\tau',\tau'';\theta^{*})
	G_{12}(\tau'',\tau_{f};\theta),\\
			&p_{12}(\tau^{+},\tau_{f},\theta)
	=\sum_{\tau',\tau''}
	-G_{11}(\tau^{+},\tau';\theta)
q_{12}(\tau',\tau'';\theta)
	G_{22}(\tau'',\tau_{f};\theta)\\
	&	=\sum_{\tau',\tau''}
	-G_{11}(\tau^{+},\tau';\theta)
q_{11}(\tau',\tau'';\theta)
	G_{21}(\tau'',\tau_{f};\theta),
		\end{aligned}
\end{equation}

such that
\begin{equation} 
	\begin{aligned}
		\label{6247}
q_{21}(\tau^{+},\tau_{f},\theta^{*})
%&=		\sum_{\tau',\tau''}-G^{-1}_{22}(\tau^{+},\tau')
%		G_{21}(\tau',\tau'';\theta)G_{11}^{-1}(\tau'',\tau_{f};\theta)\\
&=\sum_{\tau',\tau''}\frac{\delta G_{22}^{-1}(\tau',\tau'';\theta)}
		{\delta \theta^{*}(\tau,\tau)} 
		G_{12}(\tau'',\tau_{f};\theta)	G^{-1}_{11}(\tau'',\tau_{f};\theta)
=\sum_{\tau',\tau''}-G^{-1}_{22}(\tau'',\tau_{f};\theta)
\frac{\delta G_{21}(\tau',\tau'';\theta)}
{\delta \theta^{*}(\tau,\tau)} 
G_{11}^{-1}(\tau'',\tau_{f};\theta),\\
q_{12}(\tau^{+},\tau_{f},\theta)
%&=\sum_{\tau',\tau''}-G_{11}^{-1}(\tau^{+},\tau')
%G_{12}(\tau',\tau'';\theta)G_{22}^{-1}(\tau'',\tau_{f};\theta)\\
&=\sum_{\tau',\tau''}\frac{\delta G_{11}^{-1}(\tau',\tau'';\theta)}
{\delta \theta(\tau,\tau)} 
G_{21}(\tau'',\tau_{f};\theta)	G^{-1}_{22}(\tau'',\tau_{f};\theta)
=\sum_{\tau',\tau''}-G^{-1}_{11}(\tau'',\tau_{f};\theta)
\frac{\delta G_{12}(\tau',\tau'';\theta)}
{\delta \theta(\tau,\tau)} 
G_{22}^{-1}(\tau'',\tau_{f};\theta),\\
q_{11}(\tau^{+},\tau_{f},\theta)
&=\sum_{\tau',\tau''}
-G^{-1}_{11}(\tau^{+},\tau';\theta)
\frac{\delta G_{12}(\tau',\tau'';\theta)}
{\delta \theta(\tau,\tau)} 
G^{-1}_{21}(\tau'',\tau_{f};\theta)
=\sum_{\tau',\tau''}\frac{\delta G_{12}^{-1}(\tau',\tau'';\theta)}
{\delta \theta(\tau,\tau)} 
G_{22}(\tau'',\tau_{f};\theta)G^{-1}_{21}(\tau'',\tau_{f};\theta),\\
q_{22}(\tau^{+},\tau_{f},\theta^{*})
&=\sum_{\tau',\tau''}
-G^{-1}_{22}(\tau^{+},\tau';\theta)
\frac{\delta G_{21}(\tau',\tau'';\theta)}
{\delta \theta^{*}(\tau,\tau)} 
G^{-1}_{12}(\tau'',\tau_{f};\theta)
=\sum_{\tau',\tau''}\frac{\delta G_{21}^{-1}(\tau',\tau'';\theta)}
{\delta \theta^{*}(\tau,\tau)} 
G_{11}(\tau'',\tau_{f};\theta)G^{-1}_{12}(\tau'',\tau_{f};\theta).
	\end{aligned}
\end{equation}

Combining Eqs.(\ref{331}) and apply the integration by parts in last step, we obtain
\begin{equation} 
	\begin{aligned}
		\label{6245}
\pmb{\Sigma}_{11}&=\sum_{\tau'}\left(
p_{21}(\tau^{+},\tau',\theta^{*}) G^{-1}_{11}(\tau',\tau_{f};\theta)
+p_{22}(\tau^{+},\tau',\theta^{*}) G_{21}^{-1}(\tau',\tau_{f};\theta)\right)\\
&
=\sum_{\tau'}\left(
p_{21}(\tau^{+},\tau',\theta^{*}) G^{-1}_{11}(\tau',\tau_{f};\theta)
-G_{21}(\tau^{+},\tau';\theta)
q_{11}(\tau',\tau_{f},\theta^{*})\right)\\
&=\sum_{\tau'}
G_{21}(\tau^{+},\tau';\theta)G^{-1}_{11}(\tau',\tau_{f};\theta),\\
\pmb{\Sigma}_{22}&=
\sum_{\tau'}\left(p_{12}(\tau^{+},\tau',\theta) G_{22}^{-1}(\tau',\tau_{f};\theta)
+ p_{11}(\tau^{+},\tau',\theta) G_{12}^{-1}(\tau',\tau_{f};\theta)\right)\\
&
=\sum_{\tau'}\left(
p_{12}(\tau^{+},\tau',\theta) G_{22}^{-1}(\tau',\tau_{f};\theta)
-G_{12}(\tau^{+},\tau';\theta)
q_{22}(\tau',\tau_{f},\theta) \right)\\
&
=\sum_{\tau'}
G_{12}(\tau^{+},\tau';\theta)G_{22}^{-1}(\tau',\tau_{f};\theta).
\end{aligned}
\end{equation}

For the vanishing part of Eq.(\ref{331}),
we obtain
\begin{equation} 
	\begin{aligned}
p_{11}(\tau^{+},\tau',\theta) 
=&	-G_{12}(\tau^{+},\tau';\theta)
q_{21}(\tau',\tau'',\theta)G_{11}(\tau'',\tau_{f};\theta)\\
=&G_{12}(\tau^{+},\tau';\theta)
G_{11}(\tau',\tau_{f};\theta)
+G_{12}(\tau^{+},\tau';\theta)
\frac{\delta \Sigma_{21}(\tau',\tau_{f};\theta)}
{\delta G_{21}(\tau_{3},\tau_{4};\theta)}
\frac{\delta G_{21}(\tau_{3},\tau_{4};\theta)}
{\delta \theta(\tau,\tau)}
G_{11}(\tau',\tau_{f};\theta)\\
=&	-G_{12}(\tau^{+},\tau';\theta)
q_{22}(\tau',\tau_{f},\theta)G_{12}(\tau',\tau_{f};\theta),\\
p_{22}(\tau^{+},\tau',\theta^{*})
=&	-G_{21}(\tau^{+},\tau';\theta)
q_{12}(\tau',\tau'',\theta^{*})G_{22}(\tau'',\tau_{f};\theta)\\
=&	G_{21}(\tau^{+},\tau';\theta)
G_{22}(\tau',\tau_{f};\theta)
+G_{21}(\tau^{+},\tau';\theta)
\frac{\delta \Sigma_{12}(\tau',\tau_{f};\theta)}
{\delta G_{12}(\tau_{3},\tau_{4};\theta)}
\frac{\delta G_{12}(\tau_{3},\tau_{4};\theta)}
{\delta \theta^{*}(\tau,\tau)}
G_{22}(\tau',\tau_{f};\theta)\\
=&	-G_{21}(\tau^{+},\tau';\theta)
q_{11}(\tau',\tau_{f},\theta^{*})G_{21}(\tau',\tau_{f};\theta),\\
	\end{aligned}
\end{equation}
and then in comparasion with Eq.(\ref{6247}),
we have the following derivatives of inverse Green function 
(with respect to the conjugated source field)
\begin{equation} 
	\begin{aligned}
q_{11}(\tau',\tau_{f};\theta^{*})
%		{\delta \theta^{*}(\tau,\tau)} 	
&=
		\sum_{\tau',\tau''}-G^{-1}_{11}(\tau^{+},\tau';\theta)
		G_{12}(\tau',\tau'';\theta)
		\frac{\delta G_{21}^{-1}(\tau'',\tau_{f};\theta)}
		{\delta\theta(\tau,\tau)},\\
q_{22}(\tau',\tau_{f};\theta)
%\frac{\delta G_{22}^{-1}(\tau',\tau_{f};\theta)}
%		{\delta \theta(\tau,\tau)} 
&=\sum_{\tau',\tau''}-G_{22}^{-1}(\tau^{+},\tau';\theta)
		G_{21}(\tau',\tau'';\theta)
		\frac{\delta G_{12}^{-1}(\tau'',\tau_{f};\theta)}
		{\delta\theta^{*}(\tau,\tau)},\\
q_{21}(\tau',\tau_{f};\theta)
%		\frac{\delta G_{21}^{-1}(\tau^{+},\tau_{f};\theta)}
%		{\delta \theta(\tau,\tau)}
		&=
		\sum_{\tau',\tau''}
		\frac{\delta G_{22}^{-1}(\tau^{+},\tau';\theta)}
		{\delta \theta(\tau,\tau)}G_{12}(\tau',\tau'';\theta)
		G_{11}^{-1}(\tau'',\tau_{f};\theta)
=		-G^{-1}_{12}(\tau^{+},\tau';\theta)
		G_{11}(\tau^{+},\tau';\theta)
		\frac{\delta G_{11}^{-1}(\tau',\tau_{f};\theta)}
		{\delta \theta(\tau,\tau)}\\		
&=-G^{-1}_{12}(\tau^{+},\tau';\theta)
		\frac{\delta G_{11}(\tau',\tau_{f};\theta)}
		{\delta \theta(\tau,\tau)}G^{-1}_{11}(\tau',\tau_{f};\theta),\\
q_{12}(\tau',\tau_{f};\theta^{*})
%		\frac{\delta G_{12}^{-1}(\tau^{+},\tau_{f};\theta)}
%		{\delta \theta^{*}(\tau,\tau)}
		&=		\sum_{\tau',\tau''}
		\frac{\delta G_{11}^{-1}(\tau^{+},\tau';\theta)}
		{\delta \theta^{*}(\tau,\tau)}G_{21}(\tau',\tau'';\theta)
		G^{-1}_{22}(\tau'',\tau_{f};\theta)
=\sum_{\tau',\tau''}-G^{-1}_{21}(\tau^{+},\tau';\theta)
		G_{22}(\tau^{+},\tau';\theta)
		\frac{\delta G_{22}^{-1}(\tau',\tau_{f};\theta)}
		{\delta \theta^{*}(\tau,\tau)}\\
&=	\sum_{\tau',\tau''}-G^{-1}_{21}(\tau^{+},\tau';\theta)
		\frac{\delta G_{22}(\tau',\tau'';\theta)}
		{\delta \theta^{*}(\tau,\tau)}G^{-1}_{22}(\tau'',\tau_{f};\theta).
	\end{aligned}
\end{equation}

\section{Functional property of the propagator labelled by times in string-form: saddle-point fluctuation in terms of an orthonormal basis }

In terms of an orthonormal basis,
we can expand an arbitary operator labelled by imaginary time in string form
as (we use the notaion $\hat{\sum}$ to represent the restricted sum)
\begin{equation} 
	\begin{aligned}
		\label{6153}
		&
		\psi^{\dag}(t_{1})
		=\jupiter^{\dag}(t_{1}-t_{1;0})+\hat{\sum}_{\alpha}c^{*}_{\alpha}\psi^{*}_{\alpha}(t_{1}),\\
		&\psi(t_{2} )
		=\jupiter(t_{2}-t_{2;0})+\hat{\sum}_{\alpha}c_{\alpha}\psi_{\alpha}(t_{2} ),
			\end{aligned}
	\end{equation} 
where we call the first term $\jupiter^{\dag}(t_{1}-t_{1;0})$
($\jupiter(t_{2}-t_{2;0})$) as the non-fluctuation part
(i.e., the condensate wave function\cite{Kita,Zinn-Justin}),
and the rest terms are fluctuation part.
$t_{0}$ represents a family of degenerate saddle points.
$\psi_{\alpha}(t)$ form an orthonormal basis,
and thus the variation of $\jupiter^{\dag}(t-t_{0})$ with respect to $t_{0}$ is
orthogonal to the variation of $\psi^{\dag}(t)$ with respect to $c_{\alpha}^{*}$, for all $\alpha$,
i.e., $\int \dot{\jupiter}^{\dag}(t-t_{0})\psi^{*}_{\alpha}(t)dt=0$
($\forall \alpha$).
Thus a well-defined pairing (whose expectation reflects a single-particle density) $\psi^{\dag}(t_{1})\psi(t_{2})$ requires $\hat{\sum}_{\alpha}|c_{\alpha}^{*}|
=\hat{\sum}_{\alpha}|c_{\alpha}|$.
Next we simply use the symbols $\alpha,\beta$ to distinguish between different pairs, where the operators inside each pair share the same coefficient.
For a well-defined pair $\psi^{\dag}(t_{1})\psi(t_{2})$,
the correlation between them is revealed by the common pattern of the
coefficient set labelled by the $\alpha$'s.
By setting a cutoff at the second order derivative for these two operators,
which equivalents to the first order approximation on the functional expansion,
we obtain
\begin{equation} 
			\begin{aligned}
				\label{6151}
	&
	\ddot{\psi}^{\dag}(t_{1})
	=\ddot{\jupiter}^{\dag}(t_{1}-t_{1;0})+\hat{\sum}_{\alpha}c^{*}_{\alpha}\ddot{\psi}^{*}_{\alpha}(t_{1})=0,\\
	&\ddot{\psi}(t_{2} )
	=\ddot{\jupiter}(t_{2}-t_{2;0})+\hat{\sum}_{\alpha}c_{\alpha}\ddot{\psi}_{\alpha}(t_{2} )=0,
\end{aligned}
\end{equation} 
and the orthonormality enforce
\begin{equation} 
			\begin{aligned}
				\label{6293}
	&
\sum_{t}
\dot{\jupiter}^{\dag}(t-t_{0})\psi^{*}_{\alpha}(t)=0,\\
&
\sum_{t}
\dot{\jupiter}(t-t_{0})\psi_{\alpha}(t)=0,\\
\end{aligned}
\end{equation} 
thus we obtain
\begin{equation} 
	\begin{aligned}
		&
\dot{\jupiter}(t_{1})=-\frac{1}{\psi_{\alpha}(t_{1})}\sum_{t\neq t_{1}}\dot{\jupiter}(t)\psi_{\alpha}(t),\\
&\dot{\jupiter}(t_{2})=-\frac{1}{\psi_{\alpha}(t_{2})}\sum_{t\neq t_{2}}\dot{\jupiter}(t)\psi_{\alpha}(t).
	\end{aligned}
\end{equation} 
Substituting this back to Eq.(\ref{6151}) we arrives at
\begin{equation} 
	\begin{aligned}
		&
		\ddot{\psi}^{\dag}(t_{1})
		=
		\frac{\dot{\psi}_{\alpha}(t_{1})}{\psi^2_{\alpha}(t_{1})}
		\sum_{t\neq t_{1}}\dot{\jupiter}(t)\psi_{\alpha}(t)
		+\hat{\sum}_{\alpha}c^{*}_{\alpha}\ddot{\psi}^{*}_{\alpha}(t_{1})=0,\\
		&\ddot{\psi}(t_{2} )
		=\frac{\dot{\psi}_{\alpha}(t_{2})}{\psi^2_{\alpha}(t_{2})}
		\sum_{t\neq t_{2}}\dot{\jupiter}(t)\psi_{\alpha}(t)
		+\hat{\sum}_{\alpha}c_{\alpha}\ddot{\psi}_{\alpha}(t_{2} )=0,
	\end{aligned}
\end{equation} 
which directly leads to
\begin{equation} 
	\begin{aligned}
		\label{6152}
		&
		\frac{\dot{\psi}_{\alpha}(t_{1})}{\psi^2_{\alpha}(t_{1})}
		\sum_{t\neq t_{1}}\dot{\jupiter}(t)\psi_{\alpha}(t)
		-\frac{\dot{\psi}_{\alpha}(t_{2})}{\psi^2_{\alpha}(t_{2})}
		\sum_{t\neq t_{2}}\dot{\jupiter}(t)\psi_{\alpha}(t)
		=\hat{\sum}_{\alpha}c_{\alpha}\ddot{\psi}_{\alpha}(t_{2} )
		-\hat{\sum}_{\alpha}c^{*}_{\alpha}\ddot{\psi}^{*}_{\alpha}(t_{1})=0.
	\end{aligned}
\end{equation} 
According to this expression,
we can see that the coefficient set does not related to the second order derivatives of neither the operator at $t_{1}$ or that at $t_{2}$,
but related to their difference.
Thus we have
 \begin{equation} 
 	\begin{aligned}
&
\hat{\sum}_{\alpha}c_{\alpha}\ddot{\psi}_{\alpha}(t_{2} )
=\hat{\sum}_{\alpha}\overline{c_{\alpha}}\overline{\ddot{\psi}_{\alpha}(t_{2} )},\\
& 	
	\hat{\sum}_{\alpha}c^{*}_{\alpha}\ddot{\psi}^{*}_{\alpha}(t_{1})
	=	\hat{\sum}_{\alpha}\overline{c^{*}_{\alpha}}\overline{\ddot{\psi}^{*}_{\alpha}(t_{1})}.
 	\end{aligned}
 \end{equation} 
Thus we can rewrite the Eq.(\ref{6152}) as
\begin{equation} 
	\begin{aligned}
		&
		\frac{\dot{\psi}_{\alpha}(t_{1})}{\psi^2_{\alpha}(t_{1})}
		\frac{\sum_{t\neq t_{1}}\dot{\jupiter}(t)\psi_{\alpha}(t)}
			{\hat{\sum}_{\alpha}\overline{c_{\alpha}}
					\hat{\sum}_{\alpha}\overline{c^{*}_{\alpha}}			}
		-\frac{\dot{\psi}_{\alpha}(t_{2})}{\psi^2_{\alpha}(t_{2})}
		\frac{\sum_{t\neq t_{2}}\dot{\jupiter}(t)\psi_{\alpha}(t)}
		{\hat{\sum}_{\alpha}\overline{c_{\alpha}}
		\hat{\sum}_{\alpha}\overline{c^{*}_{\alpha}}}\\
	=&		\frac{\dot{\psi}_{\alpha}(t_{1})}{\psi^2_{\alpha}(t_{1})}
	\frac{\sum_{t\neq t_{1}}\dot{\jupiter}(t)\psi_{\alpha}(t)}
	{\hat{\sum}_{\alpha}n_{\alpha}			}
	-\frac{\dot{\psi}_{\alpha}(t_{2})}{\psi^2_{\alpha}(t_{2})}
	\frac{\sum_{t\neq t_{2}}\dot{\jupiter}(t)\psi_{\alpha}(t)}
	{\hat{\sum}_{\alpha}n_{\alpha}}\\
		=&\frac{\overline{\ddot{\psi}_{\alpha}(t_{2} )}}{	\hat{\sum}_{\alpha}\overline{c^{*}_{\alpha}}	}
		-\frac{\overline{\ddot{\psi}^{*}_{\alpha}(t_{1})}}
	{	\hat{\sum}_{\alpha}\overline{c_{\alpha}}	}=0,
	\end{aligned}
\end{equation} 
where the coefficients set must be related to, individually,
the operators at $t_{1}$ or that at $t_{2}$,
and such dependence on an arbitarily individual time component can be offset
only in terms of a certain combination between the first order and zeroth order derivatives of a operator at a certain time.
In other word,
for an arbitary operator picking from a pair,
its second order derivative equivalents to a certain combination of its first and zeroth order time derivatives, and only depends on the another operator within this pair.
The second order derivatives of the two operators thus form the subspace
whose state expectation only related to the 
dynamical property about the time difference,
i.e., the overall character described by the single-particle density of this pair.
Similarly,
for a four-point function which owning two distinct sets of coefficients
(e.g., labelled by $\alpha$ and $\beta$;
and this is what the above decompositions are accoding to),
the correlation between these two sets of coefficients
guarantees the symmetry sector formed by $\pmb{\Psi}^{(1)}$
and $\pmb{\Psi}^{(2)}$.
And such correlation between $\pmb{\Psi}^{(1)}$
and $\pmb{\Psi}^{(2)}$ can only be observed in terms of the operators labelled by the imaginary time in string form.
This guarantees the U(1) conservation between two spin-components
(otherwise a constant ratio between the two spin components will be fixed).
Interestingly, for the Green functions labelled by imaginary time in string form,
the diagonal Green functions form such conservation through its two representations in different spin basis,
while the off-diagonal Green functions form such conservation through 
itself with its time-exchanged version,
i.e.,
$G_{21}(\tau_{1},\tau_{2})$ and $G_{12}^{*}(\tau_{1},\tau_{2})$,
or $G_{12}(\tau_{1},\tau_{2})$ and $G_{21}^{*}(\tau_{1},\tau_{2})$.
Thus a selectrive (restricted) sum is required in the fluctuation part of Eq.(\ref{6153}),
to reflect a fully thermalized equilibrium state where 
$\langle \hat{\sum}_{\alpha}c^{*}_{\alpha}\psi^{*}_{\alpha}(t_{1})\rangle=0$.
As there is finite correlation between the operators at $t_{1}$ and $t_{2}$,
thus an average over the selected $\alpha$'s which builts a nonlocal conservation at a single certain time reflecting a subsystem without the lowest energy state.

On the contrary,
a subsystem has a lowest energy state,
where the system Hamiltonian has zero expectation on each one of the mutually-independent eigenstates,
is extensive and nonergodic,.i.e.,
without a maximal number of the principal components. 
Here, in noninteracting case, each eigenstate follows the same pattern,
i.e.,
the product of creation operators (with the power determined by the corresponding noninteracting particle number) of the corresponding generator (lowest energy state).
In interacting case,
the corresponding integrable eigenstates are the superpositions of the products in the noninteracting case,
with the expansion coefficients on each product representing the particle-number fluctuations and two-particle correlations\cite{Kita,Bi}.
Note that an nonzero averaged value of nonfluctuation part (the condensate wave function) requires such superpositions as well as the finite two-particle correlation.

To further understand the reason for the absence of lowest energy state
in the vanishing average term 
$\langle \hat{\sum}_{\alpha}c^{*}_{\alpha}\psi^{*}_{\alpha}(t_{1})\rangle=0$,
which can be described by a localized Hamiltonian,
we can treat the time-dependent operator $\psi^{\dag}(t_{1})$ as a Hermitian operator in the eigenkets of a real symmetry random matrix,
and since eigenstates of real symmetric random matrices always form a orthonormal basis, in terms of the Berry's conjecture in semiclassical limit,
we further obtain
\begin{equation} 
	\begin{aligned}
		\psi^{\dag}(t_{1})
&=\jupiter^{\dag}(t_{1}-t_{1;0})+\hat{\sum}_{\alpha}c^{*}_{\alpha}\psi^{*}_{\alpha}(t_{1})\\
&=\digamma(\hat{\sum}_{\alpha}\overline{c_{\alpha}})\delta_{t_{1},t_{1;0}}+\hat{\sum}_{\alpha}c^{*}_{\alpha}\psi^{*}_{\alpha}(t_{1}).
	\end{aligned}
\end{equation} 
Here the $\psi_{\alpha}^{*}(t_{1})$, which form a orthogonal basis,
can be regarded as a random variable selected from a Gaussian random matrix (real for GOE and complex for GUE) with zero mean and unit variance,
by mapping the time-fluctuation of $t_{1}$ into the orthogonal eigenvectors of a matrix, where tiny correlations between the random vectors due to the orthogonality requirement can be ignored due to the large matrix dimension.
While the restricted sum of the coefficients $\hat{\sum}_{\alpha}c^{*}_{\alpha}$ is the term that indicating the classical analog in terms of the average shape of the eigenfunctions in semiclassical limit (AES),
which reveals the character of the corresponding Gaussian distribution ($\beta$-ensemble) by the effecive Hilbert dimension (which making this term be the leading order term).
In first term, there is a arbitary function
$\digamma(\hat{\sum}_{\alpha}\overline{c_{\alpha}})$
which depends on only the averaged coefficient sum,
where the informations about the Gaussian distribution as well as the Hilbert dimension are loss,
and thus indicates nothing about the classical analog characters
(i.e., the common feature shared by the operators of a pair).
But due to the another orthogonality which is enforced to make sure the linear independence between the integral over time in string form
(or summation for the discrete times),
the nonfluctuation term must be as a function of the 
$\hat{\sum}_{\alpha}\overline{c_{\alpha}})$,
to make sure the "center-of-mass" matches the latter fluctuation term,
i.e., the fluctuations around the saddle-point solution can be correctly expanded in the corresponding orthogonal basis,
and in the mean time, be smoothed out by a certain average which depends on the information about the classical analog of AES.
 In fact, beforce comparing with the another operator of the same pair,
 we can savely rewrite the above operator as
 \begin{equation} 
 	\begin{aligned}
 		\psi^{\dag}(t_{1})
 		&=\digamma(\hat{\sum}_{\alpha}\overline{c_{\alpha}})\delta_{t_{1},t_{1;0}}+\hat{\sum}_{\alpha}\overline{c^{*}_{\alpha}}\overline{\psi}^{*}_{\alpha}(t_{1}),
 	\end{aligned}
 \end{equation} 
where the averages are over the $\alpha$'s.
This is because
once the classical analog of AES is identified,
the fluctuation of $\psi_{\alpha}^{*}(t_{1})$ is independent of the $\alpha$ index, as the set of $\alpha$ is already being determined by the classical analog of AES.

For $G_{11}(\tau_{1},\tau_{2})$ and $G_{22}(\tau_{1},\tau_{2})$,
there are originally two independent channels,
i.e., the up-spin and down-spin.
While that for $G_{12}(\tau_{1},\tau_{2})$ and $G_{21}(\tau_{1},\tau_{2})$,
these two independent channels are described by the anomalous propagators with inversed time components.
For $G_{11}(\tau_{1},\tau_{2})$ and $G_{22}(\tau_{1},\tau_{2})$,
the local conservation in each spin channel originates from the 
simple and most basic restriction $n_{\uparrow}+n_{\downarrow}=1$.
The restriction $n_{\uparrow}+n_{\downarrow}=1$
can simply be expressed as, e.g.,
$\psi^{\dag}_{1s'}\psi_{2s'}+\psi_{2s}^{\dag}\psi_{\tau's}=1$,
where one of the single-particle density need to containing $\tau'$ to successfully map to this conservation between two spin channels to 
that within single spin channel.
Within single spin channel,
one of the conservation resulted from the above one reads $-\Theta(t_{1}-t_{2})
\psi^{\dag}_{s'}(t_{1})\psi_{s'}(t_{2})-\Theta(t_{2}-t_{1})
\psi^{\dag}_{s}(t_{2})\psi_{s}(t_{1})=1$.
As the conservation within a single-spin channel process the same physical effects
with their derivatives with respect to the time in string form,
we can consider an open boundary and an periodic boundary 
for the time evolutions within $-\Theta(t_{2}-t_{1})
\psi^{\dag}_{s}(t_{2})\psi_{s}(t_{1})$
and $\psi^{\dag}_{2s}\psi_{\tau's}$, respectively,
as shown in the second terms of the above two conservation expressions.
Thus different to the periodic boundary, which is represented by
$\psi^{\dag}_{2s}\psi_{\tau's}=\psi^{\dag}_{\tau's}\psi_{1s}
=\psi^{\dag}_{1s}\psi_{2s}$ and exhibits a stubborn self-consistency 
(thus smoothing out all the degrees-of-freedom other than the spin components),
the strong conformal boundary potential which guarantees the open boundary
(with reflections) as well as a reversal in time evolution,
and this is similar to the strong scatters in a harmonic waveguide
which increase the number of mutually independent integrable eigenstates\cite{Yurovsky},
by the emergence of local observables.
Thus for the conservation built in a single-spin channel,
the weak correlation between two spin components is overwhelmed by that between the propagators with opposite time evolutions
(with $\Theta(\tau_{1}-\tau_{2})=1$ and $\Theta(\tau_{2}-\tau_{1})=1$,
respectively).
Also, 
the conformal boundary with scaling invariance is good in 
describing a well-defined local observable (as a generalized Gibbs ensemble (GGE) for an integrable system).
In this case,
an initial state whose correlation function become stationary
after a "failed" thermalization which corresponds to
the temperature corresponding to a conserved energy density,
can be well aproximated by a conformal boundary state\cite{Cardy}.

\section{Appendix.A: Proof of Eq.(\ref{641})}

We retain the time ordering notation $\mathcal{T}$ in each element here  
\begin{equation} 
\begin{aligned}
{\bf \Psi}(\tau^{+},\tau';\tau)=&
\begin{pmatrix}
-\mathcal{T}\dot{\psi}^{\dag}_{1s}\ddot{\psi}_{2s}\ddot{\psi}_{2s'}\dot{\psi}^{\dag}_{\tau's'}
 & -\mathcal{T}\dot{\psi}^{\dag}_{1s}\ddot{\psi}_{2s}\ddot{\psi}_{2s'}\dot{\psi}_{\tau' s}\\
-\mathcal{T}\dot{\psi}_{1s'}\ddot{\psi}^{\dag}_{2s'}\ddot{\psi}^{\dag}_{2s}\dot{\psi}^{\dag}_{\tau' s'} & -\mathcal{T}\dot{\psi}_{1s'}\ddot{\psi}^{\dag}_{2s'}\ddot{\psi}^{\dag}_{2s}\dot{\psi}_{\tau' s}
\end{pmatrix}\\
=&-\begin{pmatrix}
p_{21}(\tau^{+},\tau',\theta^{*})
 & p_{22}(\tau^{+},\tau',\theta^{*})\\
p_{11}(\tau^{+},\tau',\theta)
 & p_{12}(\tau^{+},\tau',\theta)
\end{pmatrix}\\
=&\begin{pmatrix}
{}[G_{21}(\tau^{+},\tau'),\theta(\tau,\tau)] & [G_{22}(\tau^{+},\tau'),\theta(\tau,\tau)]\\
[\theta^{*}(\tau,\tau),G_{11}(\tau^{+},\tau')] & 
[\theta^{*}(\tau,\tau),G_{12}(\tau^{+},\tau')]
\end{pmatrix},
\end{aligned}
\end{equation}
There is an important ansatz for the following discussion and the validity of Eq.(\ref{641}):
$\psi^{\dag}_{\tau,s'}\psi_{\tau,s'}=\psi_{\tau,s}\psi_{\tau,s}^{\dag}=0$,
$\psi_{\tau,s'}\psi^{\dag}_{\tau,s'}=\psi^{\dag}_{\tau,s}\psi_{\tau,s}=1$.

Shifting the source field to the left or right of the fermion propagator,
we obtain 
\begin{equation} 
\begin{aligned}
\label{2101}
{\bf \Psi}^{(1)}(\tau^{+},\tau';\tau)
=&\begin{pmatrix}
\mathcal{T}\ddot{\psi}_{2s}\ddot{\psi}_{2s'}\dot{\psi}^{\dag}_{1s}\dot{\psi}^{\dag}_{\tau's'}-\psi_{2s'}\psi^{\dag}_{\tau's'}
 & \mathcal{T}\ddot{\psi}_{2s}\ddot{\psi}_{2s'}\dot{\psi}^{\dag}_{1s}\dot{\psi}_{\tau' s}-\psi_{2s'}\psi_{\tau's}\\
\mathcal{T}\ddot{\psi}^{\dag}_{2s'}\ddot{\psi}^{\dag}_{2s}\dot{\psi}_{1s'}\dot{\psi}^{\dag}_{\tau' s'}
+\psi^{\dag}_{2s}\psi^{\dag}_{\tau's'} & 
\mathcal{T}\ddot{\psi}^{\dag}_{2s'}\ddot{\psi}^{\dag}_{2s}\dot{\psi}_{1s'}\dot{\psi}_{\tau' s}
+\psi^{\dag}_{2s}\psi_{\tau's}
\end{pmatrix}\\
=&
\begin{pmatrix}
0 & \psi_{2s}\psi_{2s'}\\ 
\psi^{\dag}_{2s'}\psi^{\dag}_{2s} & 0
\end{pmatrix}
\begin{pmatrix}
\psi_{1s'}\psi^{\dag}_{\tau' s'}
 & 
\psi_{1s'}\psi_{\tau' s}\\
\psi^{\dag}_{1s}\psi^{\dag}_{\tau's'} & \psi^{\dag}_{1s}\psi_{\tau' s}
\end{pmatrix}
+
\begin{pmatrix}
-\psi_{2s'}\psi^{\dag}_{\tau's'}
 & -\psi_{2s'}\psi_{\tau's}\\
\psi^{\dag}_{2s}\psi^{\dag}_{\tau's'} & 
\psi^{\dag}_{2s}\psi_{\tau's}
\end{pmatrix}\\
=&
\begin{pmatrix}
\psi_{2s}\psi_{2s'} & 0\\ 
0 & \psi^{\dag}_{2s'}\psi^{\dag}_{2s} 
\end{pmatrix}[(-\bm{\sigma_{x}}){\bf G}(\tau^{+},\tau';\theta)]
+
\bm{\sigma_{z}}{\bf G}(\tau,\tau';\theta)
,
\end{aligned}
\end{equation}
and
\begin{equation} 
\begin{aligned}
\label{2102}
&{\bf \Psi}^{(2)}(\tau^{+},\tau';\tau)
=\begin{pmatrix}
\mathcal{T}\dot{\psi}^{\dag}_{1s}\dot{\psi}^{\dag}_{\tau's'}\ddot{\psi}_{2s}\ddot{\psi}_{2s'}+\psi^{\dag}_{1s}\psi_{2s}
 & -\mathcal{T}\dot{\psi}^{\dag}_{1s}\dot{\psi}_{\tau' s}\ddot{\psi}_{2s}\ddot{\psi}_{2s'}\\
-\mathcal{T}\ddot{\psi}_{1s'}\ddot{\psi}^{\dag}_{\tau' s'}\dot{\psi}^{\dag}_{2s'}\dot{\psi}^{\dag}_{2s}  & 
\mathcal{T}\dot{\psi}_{1s'}\dot{\psi}_{\tau' s}\ddot{\psi}^{\dag}_{2s'}\ddot{\psi}^{\dag}_{2s}
-\psi_{1s'}\psi^{\dag}_{2s'}
\end{pmatrix}\\
=&
\begin{pmatrix}
-G_{21}(\tau^{+},\tau';\theta) & G_{22}(\tau^{+},\tau';\theta)\\
0 & 0\end{pmatrix}
\psi_{2s}\psi_{2s'}\bm{\sigma_{0}}
+\begin{pmatrix}
0 & 0\\
G_{11}(\tau^{+},\tau';\theta) & -G_{12}(\tau^{+},\tau';\theta)\end{pmatrix}
\psi^{\dag}_{2s'}\psi^{\dag}_{2s}\bm{\sigma_{0}}
+\begin{pmatrix}
\psi^{\dag}_{1s}\psi_{2s}
 & 0\\
0 & -\psi_{1s'}\psi^{\dag}_{2s'}
\end{pmatrix}\\
=&
\begin{pmatrix}
0 & -1\\
0 & 0\end{pmatrix}{\bf G}(\tau^{+},\tau';\theta)
\begin{pmatrix}
1 & 0\\
0 & -1\end{pmatrix}
\psi_{2s}\psi_{2s'}\bm{\sigma_{0}}
+\begin{pmatrix}
0 & 0\\
-1 & 0\end{pmatrix}{\bf G}(\tau^{+},\tau';\theta)
\begin{pmatrix}
-1 & 0\\
0 & 1\end{pmatrix}\psi^{\dag}_{2s'}\psi^{\dag}_{2s}\bm{\sigma_{0}}
+\begin{pmatrix}
n_{\uparrow} & 0\\
0 & n_{\downarrow}
\end{pmatrix},
\end{aligned}
\end{equation}
respectively,
where $\bm{\sigma_{0}}$ is the zeroth Pauli matrix.
Combining Eqs.(\ref{2171},\ref{2101},\ref{2102}) we obtain
\begin{equation} 
	\begin{aligned}
		\label{2173}
		{\rm For\ \Psi_{11}}&:\ 
		\theta(\tau,\tau)G_{21}(\tau^{+},\tau')=-\psi^{\dag}_{1s}\psi_{2s}=-n_{\uparrow},\ 
		G_{21}(\tau^{+},\tau')\theta(\tau,\tau)=-2\psi^{\dag}_{1s}\psi_{2s}-\psi_{2s'}\psi^{\dag}_{\tau's'}
		=-2n_{\uparrow}+n_{\downarrow},\\
				&	G_{21}(\tau^{+},\tau')\theta(\tau,\tau)
		-2\theta(\tau,\tau)G_{21}(\tau^{+},\tau')=-\psi_{2s'}\psi_{\tau's'}^{\dag}
		=\psi^{\dag}_{2s}\psi_{1s}=n_{\downarrow},\\
	{\rm For\ \Psi_{12}}&:\ 
		G_{22}(\tau^{+},\tau)\theta(\tau,\tau)=\frac{1}{3}\psi_{2s'}\psi_{\tau's},\ 
		\theta(\tau,\tau)G_{22}(\tau^{+},\tau')=\frac{2}{3}\psi_{2s'}\psi_{\tau's},\\
		&-\psi_{2s'}\psi_{\tau's}=G_{22}(\tau^{+},\tau)\theta(\tau,\tau)-2	\theta(\tau,\tau)G_{22}(\tau^{+},\tau')
		=-3G_{22}(\tau^{+},\tau)\theta(\tau,\tau),\\		
		{\rm For\ \Psi_{21}}&:\ 
		\theta^{*}(\tau,\tau)G_{11}(\tau^{+},\tau')=\mathcal{T}\dot{\psi}^{\dag}_{2s'}\dot{\psi}^{\dag}_{2s}\ddot{\psi}_{1s'}\ddot{\psi}^{\dag}_{\tau' s'}=0,\ 
		G_{11}(\tau^{+},\tau')\theta^{*}(\tau,\tau)
		=\psi_{1s'}\psi^{\dag}_{\tau' s'}\psi^{\dag}_{2s'}\psi^{\dag}_{2s} 
		=-\psi_{2s}^{\dag}\psi^{\dag}_{\tau's'},\\
		{\rm For\ \Psi_{22}}&:\ 
		G_{12}(\tau^{+},\tau')\theta^{*}(\tau,\tau)=-\psi_{2s}^{\dag}\psi_{\tau's}=-n_{\uparrow},\ 
		\theta^{*}(\tau,\tau)G_{12}(\tau^{+},\tau')=-2\psi^{\dag}_{2s}\psi_{\tau's}
		-\psi_{1s'}\psi^{\dag}_{2s'}=-2n_{\uparrow}+n_{\downarrow},\\
		&-\psi_{1s'}\psi^{\dag}_{2s'}
		=\theta^{*}(\tau,\tau)G_{12}(\tau^{+},\tau')-2	G_{12}(\tau^{+},\tau')\theta^{*}(\tau,\tau)=n_{\downarrow}.
	\end{aligned}
\end{equation}
The diagonal elements 
satisfy the symmetry property under the exchange of source field and the corresponding (anomalous) propagator
\begin{equation} 
	\begin{aligned}
		\label{2172}
\Psi_{11}(\tau^{+},\tau';\tau)=&-\mathcal{T}\dot{\psi}^{\dag}_{1s}\ddot{\psi}_{2s}\ddot{\psi}_{2s'}\dot{\psi}^{\dag}_{\tau's'}
			=-\mathcal{T}\ddot{\psi}_{2s}\dot{\psi}^{\dag}_{1s}\dot{\psi}^{\dag}_{\tau's'}\ddot{\psi}_{2s'}-\psi_{2s'}\psi^{\dag}_{\tau's'}
			+\psi_{2s}\psi^{\dag}_{1s}		\\
&	=-\mathcal{T}\ddot{\psi}_{2s}\dot{\psi}^{\dag}_{1s}\dot{\psi}^{\dag}_{\tau's'}\ddot{\psi}_{2s'}+\psi_{1s}^{\dag}\psi_{2s}
			-\psi^{\dag}_{\tau's'}\psi_{2s'}
=-\mathcal{T}\dot{\psi}_{2s}\ddot{\psi}^{\dag}_{1s}\ddot{\psi}^{\dag}_{\tau's'}\dot{\psi}_{2s'},\\
	\Psi_{22}(\tau^{+},\tau';\tau)=	&
-\mathcal{T}\psi_{1s'}\psi^{\dag}_{2s'}\psi^{\dag}_{2s}\psi_{\tau' s}
=-\mathcal{T}\psi^{\dag}_{2s'}\psi_{1s'}\psi_{\tau' s}\psi^{\dag}_{2s}+\psi_{2s}^{\dag}\psi_{\tau's}
-\psi^{\dag}_{2s'}\psi_{1s'}\\
&=-\mathcal{T}\psi^{\dag}_{2s'}\psi_{1s'}\psi_{\tau' s}\psi^{\dag}_{2s}-\psi_{1s'}\psi_{2s'}^{\dag}
+\psi_{\tau's}\psi^{\dag}_{2s}
=-\mathcal{T}\psi^{\dag}_{2s'}\psi_{1s'}\psi_{\tau' s}\psi^{\dag}_{2s},
	\end{aligned}
\end{equation}
While for the off-diagonal elements,
such symmetry property of the exchange of source field and (anomalous) propagator is absent,
\begin{equation} 
	\begin{aligned}
		\Psi_{12}(\tau^{+},\tau';\tau)=&-\mathcal{T}\dot{\psi}^{\dag}_{1s}\ddot{\psi}_{2s}\ddot{\psi}_{2s'}\dot{\psi}_{\tau' s}
		=\mathcal{T}\ddot{\psi}_{2s}\dot{\psi}^{\dag}_{1s}\dot{\psi}_{\tau' s}\ddot{\psi}_{2s'}+\psi_{\tau's}\psi_{2s'},\\
		\Psi_{21}(\tau^{+},\tau';\tau)=&-\mathcal{T}\dot{\psi}_{1s'}\ddot{\psi}^{\dag}_{2s'}\ddot{\psi}^{\dag}_{2s}\dot{\psi}^{\dag}_{\tau' s'}
		=\mathcal{T}\ddot{\psi}^{\dag}_{2s'}\dot{\psi}_{1s'}\dot{\psi}^{\dag}_{\tau' s'}\ddot{\psi}^{\dag}_{2s}
		-\psi^{\dag}_{\tau's'}\psi^{\dag}_{2s}.
	\end{aligned}
\end{equation}

Note that for $\tau'$, we have $\psi^{\dag}_{1s}\psi_{\tau's}
=\psi^{\dag}_{\tau's}\psi_{2s}=\psi^{\dag}_{2s}\psi_{1s}$,
then combined with the deductions shown in Appendix.A,
we further obtain
\begin{equation} 
	\begin{aligned}
&\Psi_{11}(\tau^{+},\tau';\tau)=-\mathcal{T}\dot{\psi}^{\dag}_{1s}\ddot{\psi}_{2s}\ddot{\psi}_{2s'}\dot{\psi}^{\dag}_{\tau's'}
=-\mathcal{T}\dot{\psi}_{2s}\ddot{\psi}^{\dag}_{1s}\ddot{\psi}^{\dag}_{\tau's'}\dot{\psi}_{2s'}
=[G_{21}(\tau^{+},\tau'),\theta(\tau,\tau)]\\
&=-\psi_{2s'}\psi_{\tau's'}^{\dag}-\psi^{\dag}_{1s}\psi_{2s}
=1-2n_{\uparrow}=n_{\downarrow}-n_{\uparrow},\\
&\Psi_{12}(\tau^{+},\tau';\tau)=-\mathcal{T}\dot{\psi}_{1s'}\ddot{\psi}^{\dag}_{2s'}\ddot{\psi}^{\dag}_{2s}\dot{\psi}^{\dag}_{\tau' s'}
=[G_{22}(\tau^{+},\tau),\theta(\tau,\tau)]=-\frac{1}{3}\psi_{2s'}\psi_{\tau's},\\
&\Psi_{21}(\tau^{+},\tau';\tau)=-\mathcal{T}\dot{\psi}_{1s'}\ddot{\psi}^{\dag}_{2s'}\ddot{\psi}^{\dag}_{2s}\dot{\psi}^{\dag}_{\tau' s'}
=-G_{11}(\tau^{+},\tau')\theta^{*}(\tau,\tau)=\psi_{2s}^{\dag}\psi^{\dag}_{\tau's'},\\
&\Psi_{22}(\tau^{+},\tau';\tau)=	
-\mathcal{T}\dot{\psi}_{1s'}\ddot{\psi}^{\dag}_{2s'}\ddot{\psi}^{\dag}_{2s}\dot{\psi}_{\tau' s}
=-\mathcal{T}\dot{\psi}^{\dag}_{2s'}\ddot{\psi}_{1s'}\ddot{\psi}_{\tau' s}\dot{\psi}^{\dag}_{2s}
=[\theta^{*}(\tau,\tau),G_{12}(\tau^{+},\tau')]\\
&=-\psi^{\dag}_{2s'}\psi_{1s'}-\psi_{\tau's}\psi^{\dag}_{2s}
=1-2n_{\uparrow}=n_{\downarrow}-n_{\uparrow},\\
	\end{aligned}
\end{equation}

\subsection{ $\dot{\psi}_{s}^{\dag}\dot{\psi}_{s'}^{\dag}\ddot{\psi}_{s}\ddot{\psi}_{s'}$}

For $\Psi_{11}$:
\begin{equation} 
\begin{aligned}
	\label{643}
\Psi_{11}(\tau^{+},\tau';\tau)
&=-\mathcal{T}\dot{\psi}^{\dag}_{1s}\ddot{\psi}_{2s}\ddot{\psi}_{2s'}\dot{\psi}^{\dag}_{\tau's'}
=[G_{21}(\tau^{+},\tau'),\theta(\tau,\tau)]
=[\psi^{\dag}_{1s}\psi^{\dag}_{\tau's'},\psi_{2s}\psi_{2s'}]\\
&=\psi^{\dag}_{1s}\left[
\{\psi^{\dag}_{\tau's'},\psi_{2s}\}\psi_{2s'}-\psi_{2s}\{\psi_{2s'},\psi^{\dag}_{\tau's'}\}
\right]
+\left[
\{\psi^{\dag}_{1s},\psi_{2s}\}\psi_{2s'}-\psi_{2s}\{\psi_{2s'},\psi^{\dag}_{1s}\}\right]\psi^{\dag}_{\tau's'}\\
&=
-\mathcal{T}\dot{\psi}^{\dag}_{1s}\ddot{\psi}_{2s}\dot{\psi}^{\dag}_{\tau's'}\ddot{\psi}_{2s'}
+\mathcal{T}\dot{\psi}_{2s}\ddot{\psi}^{\dag}_{1s}\dot{\psi}_{2s'}\ddot{\psi}^{\dag}_{\tau's'}
=
\mathcal{T}\dot{\psi}^{\dag}_{1s}\dot{\psi}^{\dag}_{\tau's'}
\ddot{\psi}_{2s}\ddot{\psi}_{2s'}
-\mathcal{T}\dot{\psi}_{2s}\dot{\psi}_{2s'}
\ddot{\psi}^{\dag}_{1s}\ddot{\psi}^{\dag}_{\tau's'}\\
&=(-2\psi_{1s}^{\dag}\psi_{2s}-\psi_{2s'}\psi_{\tau's'}^{\dag})
+\psi_{1s}^{\dag}\psi_{2s}
=-\psi_{1s}^{\dag}\psi_{2s}-\psi_{2s'}\psi_{\tau's'}^{\dag}=-n_{\uparrow}+n_{\downarrow},\\
\end{aligned}
\end{equation}

For $\Psi_{12}$:
\begin{equation} 
	\begin{aligned}
\Psi_{12}(\tau^{+},\tau';\tau)
&=-\mathcal{T}\dot{\psi}^{\dag}_{1s}\ddot{\psi}_{2s}\ddot{\psi}_{2s'}\dot{\psi}_{\tau' s}
=[G_{22}(\tau^{+},\tau'),\theta(\tau,\tau)]
=[\psi^{\dag}_{1s}\psi_{\tau's},\psi_{2s}\psi_{2s'}]\\
&=[\{\psi^{\dag}_{1s},\psi_{2s}\}\psi_{2s'}
-\psi_{2s}\{\psi_{2s'},\psi^{\dag}_{1s}\}]\psi_{\tau's}
=\{\psi^{\dag}_{1s},\psi_{2s}\}\psi_{2s'}\psi_{\tau's}\\
&=-2\mathcal{T}\ddot{\psi}_{2s}\ddot{\psi}_{2s'}\dot{\psi}^{\dag}_{1s}\dot{\psi}_{\tau's}+\psi_{2s'}\psi_{\tau's}\\
&
=2\mathcal{T}\dot{\psi}_{1s}^{\dag}\ddot{\psi}_{2s}\ddot{\psi}_{2s'}\dot{\psi}_{\tau's}-\psi_{2s'}\psi_{\tau's}
=-\frac{1}{3}\psi_{2s'}\psi_{\tau's},\\
\end{aligned}
\end{equation}
where
\begin{equation} 
	\begin{aligned}
		\mathcal{T}\dot{\psi}_{1s}^{\dag}\ddot{\psi}_{2s}\ddot{\psi}_{2s'}\dot{\psi}_{\tau's}=\frac{1}{3}\psi_{2s'}\psi_{\tau's}.
	\end{aligned}
\end{equation}
For the another spin configuration,
\begin{equation} 
	\begin{aligned}
		\Psi_{12}(\tau^{+},\tau';\tau)
		&=-\mathcal{T}\dot{\psi}^{\dag}_{\tau's'}\ddot{\psi}_{2s}\ddot{\psi}_{2s'}\dot{\psi}_{1s'}
		=-\frac{\delta(\psi_{\tau's'}^{\dag}\psi_{1s'})}{\delta(\psi_{2s'}^{\dag}\psi_{2s}^{\dag})}
		=[\psi^{\dag}_{\tau's'}\psi_{1s'},\psi_{2s}\psi_{2s'}]\\
		&=[\{\psi^{\dag}_{\tau's'},\psi_{2s}\}\psi_{2s'}
		-\psi_{2s}\{\psi_{2s'},\psi^{\dag}_{\tau's'}\}]\psi_{1s'}\\
&=
		-\psi_{2s}\{\psi_{2s'},\psi^{\dag}_{\tau's'}\}\psi_{1s'}\\
		&=
		-\mathcal{T}\dot{\psi}_{2s}\dot{\psi}_{2s'}\ddot{\psi}^{\dag}_{\tau's'}\ddot{\psi}_{1s'}
		-\mathcal{T}\dot{\psi}_{2s}\ddot{\psi}^{\dag}_{\tau's'}\dot{\psi}_{2s'}\ddot{\psi}_{1s'}\\
				&=
	2	\mathcal{T}\ddot{\psi}^{\dag}_{\tau's'}\dot{\psi}_{2s}\dot{\psi}_{2s'}\ddot{\psi}_{1s'}+\psi_{2s}\psi_{1s'}\\
		&
		=-\frac{1}{3}\psi_{2s'}\psi_{\tau's},
	\end{aligned}
\end{equation}
where
\begin{equation} 
	\begin{aligned}
\psi_{2s}\psi_{1s'}=-\psi_{2s'}\psi_{\tau's}.
	\end{aligned}
\end{equation}

For $\Psi_{21}$:
\begin{equation} 
	\begin{aligned}
\Psi_{21}(\tau^{+},\tau';\tau)
&=-\mathcal{T}\dot{\psi}_{1s'}\ddot{\psi}^{\dag}_{2s'}\ddot{\psi}^{\dag}_{2s}\dot{\psi}^{\dag}_{\tau' s'}\\
&=[\theta^{*}(\tau,\tau),G_{11}(\tau^{+},\tau')]
=[\psi^{\dag}_{2s'}\psi^{\dag}_{2s},\psi_{1s'}\psi^{\dag}_{\tau's'}]
=\{\psi^{\dag}_{2s'},\psi_{1s'}\}\psi^{\dag}_{\tau's'}\psi^{\dag}_{2s}\\
&=\mathcal{T}\dot{\psi}^{\dag}_{2s'}\ddot{\psi}_{1s'}\ddot{\psi}^{\dag}_{\tau's'}\dot{\psi}^{\dag}_{2s}
+\mathcal{T}\dot{\psi}_{1s'}\ddot{\psi}^{\dag}_{2s'}\dot{\psi}^{\dag}_{\tau's'}\ddot{\psi}^{\dag}_{2s}\\
&=-2\mathcal{T}\ddot{\psi}_{1s'}\dot{\psi}^{\dag}_{2s'}\dot{\psi}^{\dag}_{2s}\ddot{\psi}^{\dag}_{\tau's'}-\psi^{\dag}_{2s}\psi^{\dag}_{\tau's'}\\
&=-2\mathcal{T}
\ddot{\psi}_{1s'}\ddot{\psi}_{\tau's'}^{\dag}\dot{\psi}^{\dag}_{2s'}\dot{\psi}^{\dag}_{2s}+\psi^{\dag}_{\tau's'}\psi_{2s}^{\dag}
=\psi^{\dag}_{2s}\psi^{\dag}_{\tau's'},
\end{aligned}
\end{equation}
where
\begin{equation} 
	\begin{aligned}
-2\mathcal{T}
\ddot{\psi}_{1s'}\ddot{\psi}_{\tau's'}^{\dag}\dot{\psi}^{\dag}_{2s'}\dot{\psi}^{\dag}_{2s}=2\psi^{\dag}_{2s}\psi^{\dag}_{\tau's'}.
	\end{aligned}
\end{equation}
For the another spin configuration
\begin{equation} 
	\begin{aligned}
		\Psi_{21}(\tau^{+},\tau';\tau)
		&=-\mathcal{T}\dot{\psi}_{\tau's}\ddot{\psi}^{\dag}_{2s'}\ddot{\psi}^{\dag}_{2s}\dot{\psi}^{\dag}_{1s}\\
		&=-\frac{\delta(\psi_{\tau's}\psi^{\dag}_{1s})}{\delta(\psi_{2s}\psi_{2s'})}
		=[\psi^{\dag}_{2s'}\psi^{\dag}_{2s},\psi_{\tau's}\psi^{\dag}_{1s}]
		=\psi^{\dag}_{2s'}\{\psi^{\dag}_{2s},\psi_{\tau's}\}\psi^{\dag}_{1s}\\
		&
=\mathcal{T}\dot{\psi}^{\dag}_{2s'}\dot{\psi}^{\dag}_{2s}\ddot{\psi}_{\tau's}\ddot{\psi}^{\dag}_{1s}
+\mathcal{T}\dot{\psi}^{\dag}_{2s'}\ddot{\psi}_{\tau's}\dot{\psi}^{\dag}_{2s}\ddot{\psi}^{\dag}_{1s}\\
		&
=-2\mathcal{T}\ddot{\psi}_{\tau's}\dot{\psi}^{\dag}_{2s'}\dot{\psi}^{\dag}_{2s}\ddot{\psi}^{\dag}_{1s}+\psi_{2s'}^{\dag}\psi_{1s}^{\dag}\\
&		=\psi^{\dag}_{2s}\psi^{\dag}_{\tau's'},
	\end{aligned}
\end{equation}
where
\begin{equation} 
	\begin{aligned}
\psi_{2s'}^{\dag}\psi_{1s}^{\dag}=-\psi^{\dag}_{2s}\psi^{\dag}_{\tau's'}=\psi_{\tau's'}^{\dag}\psi_{2s}^{\dag}.
	\end{aligned}
\end{equation}

For $\Psi_{22}$:
\begin{equation} 
	\begin{aligned}
		\label{645}
\Psi_{22}(\tau^{+},\tau';\tau)
&=-\mathcal{T}\dot{\psi}_{1s'}\ddot{\psi}^{\dag}_{2s'}\ddot{\psi}^{\dag}_{2s}\dot{\psi}_{\tau's}
=[\theta^{*}(\tau,\tau),G_{12}(\tau^{+},\tau')]
=[\psi^{\dag}_{2s'}\psi^{\dag}_{2s},\psi_{1s'}\psi_{\tau' s}]\\
&=\psi^{\dag}_{2s'}
\left[\{\psi_{2s}^{\dag},\psi_{1s'}\}\psi_{\tau's}
-\psi_{1s'}\{\psi_{\tau's},\psi_{2s}^{\dag}\}
\right]
+\left[\{\psi^{\dag}_{2s'},\psi_{1s'}\}\psi_{\tau's}
-\psi_{1s'}\{\psi_{\tau's},\psi^{\dag}_{2s'}\}
\right]\psi^{\dag}_{2s}\\
&=-\psi^{\dag}_{2s'}
\psi_{1s'}\{\psi_{\tau's},\psi_{2s}^{\dag}\}
+\{\psi^{\dag}_{2s'},\psi_{1s'}\}\psi_{\tau's}\psi^{\dag}_{2s}\\
&=-\mathcal{T}\dot{\psi}^{\dag}_{2s'}
\ddot{\psi}_{1s'}\dot{\psi}_{2s}^{\dag}\ddot{\psi}_{\tau's}
+\mathcal{T}\dot{\psi}_{1s'}\ddot{\psi}^{\dag}_{2s'}\dot{\psi}_{\tau's}\ddot{\psi}^{\dag}_{2s}\\
&=-\psi^{\dag}_{2s'}\psi_{1s'}-\psi_{\tau's}\psi^{\dag}_{2s}
=-n_{\uparrow}+n_{\downarrow},
\end{aligned}
\end{equation}

\subsection{
	 $\dot{\psi}^{\dag}_{s}\ddot{\psi}_{s'}^{\dag}\ddot{\psi}_{s'}\dot{\psi}_{s}$}

For $\Psi_{11}$, there are two cases:

$Case\ (i)$: $\psi_{2s'}\psi^{\dag}_{\tau's'}$ be the annihilational term,
\begin{equation} 
	\begin{aligned}
&
-\mathcal{T}
\dot{\psi}^{\dag}_{2s'}\dot{\psi}_{1s'}\ddot{\psi}_{2s'}\ddot{\psi}^{\dag}_{\tau's'}
=\mathcal{T}
\dot{\psi}^{\dag}_{2s'}\ddot{\psi}_{2s'}\ddot{\psi}^{\dag}_{\tau's'}\dot{\psi}_{1s'}
\\
&=\frac{\delta(\psi^{\dag}_{2s'}\psi_{1s'})}{\delta(\psi_{\tau's'}\psi^{\dag}_{2s'})}
=[\psi_{2s'}\psi^{\dag}_{\tau's'},\psi_{2s'}^{\dag}\psi_{1s'}]\\
&=\{\psi_{2s'},\psi_{2s'}^{\dag}\}\psi_{1s'}\psi^{\dag}_{\tau's'}
-\psi_{2s'}\psi_{2s'}^{\dag}\{\psi_{1s'},\psi^{\dag}_{\tau's'}\}\\
&=\mathcal{T}\dot{\psi}_{2s'}^{\dag}\ddot{\psi}_{2s'}\dot{\psi}_{1s'}\ddot{\psi}^{\dag}_{\tau's'}
-\mathcal{T}\dot{\psi}_{2s'}\ddot{\psi}_{2s'}^{\dag}\dot{\psi}^{\dag}_{\tau's'}\ddot{\psi}_{1s'}\\
&=\mathcal{T}\dot{\psi}_{2s'}^{\dag}\ddot{\psi}_{2s'}\ddot{\psi}^{\dag}_{\tau's'}\dot{\psi}_{1s'}-\psi_{2s'}^{\dag}\psi_{2s'}
-\mathcal{T}\dot{\psi}_{2s'}\ddot{\psi}_{2s'}^{\dag}\dot{\psi}^{\dag}_{\tau's'}\ddot{\psi}_{1s'}\\
&=\psi_{\tau's'}^{\dag}\psi_{1s'}=n_{\downarrow},
	\end{aligned}
\end{equation}
where
\begin{equation} 
	\begin{aligned}
&
\mathcal{T}\dot{\psi}_{2s'}^{\dag}\ddot{\psi}_{2s'}\dot{\psi}_{1s'}\ddot{\psi}^{\dag}_{\tau's'}
=\mathcal{T}\dot{\psi}_{2s'}^{\dag}\ddot{\psi}_{2s'}\ddot{\psi}^{\dag}_{\tau's'}\dot{\psi}_{1s'}=\psi_{\tau's'}^{\dag}\psi_{1s'}
=n_{\downarrow},\\
&
\mathcal{T}\dot{\psi}_{2s'}\ddot{\psi}_{2s'}^{\dag}\dot{\psi}^{\dag}_{\tau's'}\ddot{\psi}_{1s'}
=
\mathcal{T}\dot{\psi}_{2s'}\ddot{\psi}_{2s'}^{\dag}\ddot{\psi}_{1s'}\dot{\psi}^{\dag}_{\tau's'}+\psi_{2s'}\psi_{2s'}^{\dag}
=\mathcal{T}\dot{\psi}_{2s'}\dot{\psi}^{\dag}_{\tau's'}\ddot{\psi}_{2s'}^{\dag}\ddot{\psi}_{1s'}=0,\\
&
\mathcal{T}\dot{\psi}_{2s'}\ddot{\psi}_{2s'}^{\dag}\ddot{\psi}_{1s'}\dot{\psi}^{\dag}_{\tau's'}=-1.
	\end{aligned}
\end{equation}
For this case, combined with the result in Eqs.(\ref{643},\ref{642}),
the self-consistency relation is satisfied,
\begin{equation} 
	\begin{aligned}
&\Psi_{11}(\tau^{+},\tau';\tau)=
-\mathcal{T}\dot{\psi}^{\dag}_{1s}\ddot{\psi}_{2s}\ddot{\psi}_{2s'}\dot{\psi}^{\dag}_{\tau's'}
=
\left[-\mathcal{T}
\dot{\psi}^{\dag}_{1s}\dot{\psi}_{2s}\ddot{\psi}_{2s'}\ddot{\psi}^{\dag}_{\tau's'}\right]-
\left[-\mathcal{T}\dot{\psi}^{\dag}_{1s}\ddot{\psi}_{2s}\dot{\psi}_{2s'}\ddot{\psi}^{\dag}_{\tau' s'}
\right]
=n_{\downarrow}-n_{\uparrow}.
	\end{aligned}
\end{equation}

$Case\ (ii)$: $\psi_{2s'}\psi^{\dag}_{\tau's'}$ be the creational term,
\begin{equation} 
	\begin{aligned}
		&
		-\mathcal{T}
		\dot{\psi}^{\dag}_{2s'}\dot{\psi}_{1s'}\ddot{\psi}_{2s'}\ddot{\psi}^{\dag}_{\tau's'}
		=\mathcal{T}
		\dot{\psi}^{\dag}_{2s'}\ddot{\psi}_{2s'}\ddot{\psi}^{\dag}_{\tau's'}\dot{\psi}_{1s'}
		=\frac{\delta(\psi^{\dag}_{2s'}\psi_{1s'})}{\delta(\psi_{\tau's'}\psi^{\dag}_{2s'})}\\
&=[\psi_{2s'}^{\dag}\psi_{1s'},\psi_{2s'}\psi^{\dag}_{\tau's'}]\\
&=[\{\psi_{2s'}^{\dag},\psi_{2s'}\}\psi^{\dag}_{\tau's'}
-\psi_{2s'}\{\psi^{\dag}_{\tau's'},\psi_{2s'}^{\dag}\}]\psi_{1s'}
+\psi_{2s'}^{\dag}[\{\psi_{1s'},\psi_{2s'}\}\psi^{\dag}_{\tau's'}
-\psi_{2s'}\{\psi^{\dag}_{\tau's'},\psi_{1s'}\}]\\
&=\{\psi_{2s'}^{\dag},\psi_{2s'}\}\psi^{\dag}_{\tau's'}\psi_{1s'}
-\psi^{\dag}_{2s'}\psi_{2s'}\{\psi^{\dag}_{\tau's'},\psi_{1s'}\}\\
&=\mathcal{T}\ddot{\psi_{2s'}}\dot{\psi}_{2s'}^{\dag}\ddot{\psi}^{\dag}_{\tau's'}\dot{\psi}_{1s'}
-\mathcal{T}\ddot{\psi}^{\dag}_{2s'}\dot{\psi}_{2s'}\ddot{\psi}_{1s'}\dot{\psi}^{\dag}_{\tau's'}\\
&=\psi^{\dag}_{\tau's'}\psi_{1s'}-2\mathcal{T}\dot{\psi}_{2s'}^{\dag}\ddot{\psi_{2s'}}\ddot{\psi}^{\dag}_{\tau's'}\dot{\psi}_{1s'}\\
&=\frac{1}{3}\psi_{\tau's'}^{\dag}\psi_{1s'}=\frac{1}{3}n_{\downarrow},
	\end{aligned}
\end{equation}
where
\begin{equation} 
	\begin{aligned}
		&
\mathcal{T}\ddot{\psi_{2s'}}\dot{\psi}_{2s'}^{\dag}\ddot{\psi}^{\dag}_{\tau's'}\dot{\psi}_{1s'}=2\mathcal{T}\ddot{\psi}^{\dag}_{2s'}\dot{\psi}_{2s'}\dot{\psi}^{\dag}_{\tau's'}\ddot{\psi}_{1s'}=\frac{2}{3}\psi_{\tau's'}^{\dag}\psi_{1s'}.
	\end{aligned}
\end{equation}
While for this case, 
the self-consistency relation is only possible to be satisfied in the following way
(similar to that of $\Psi_{21}$ and $\Psi_{22}$),
\begin{equation} 
	\begin{aligned}
		&\Psi_{11}(\tau^{+},\tau';\tau)=
		-\mathcal{T}\dot{\psi}^{\dag}_{1s}\ddot{\psi}_{2s}\ddot{\psi}_{2s'}\dot{\psi}^{\dag}_{\tau's'}
		=
		\left[-\mathcal{T}
		\dot{\psi}^{\dag}_{1s}\dot{\psi}_{2s}\ddot{\psi}_{2s'}\ddot{\psi}^{\dag}_{\tau's'}\right]+
		\left[-\mathcal{T}\dot{\psi}^{\dag}_{1s}\ddot{\psi}_{2s}\dot{\psi}_{2s'}\ddot{\psi}^{\dag}_{\tau' s'}
		\right]\\
		&=\frac{1}{3}n_{\downarrow}+n_{\uparrow}
		=n_{\downarrow}-n_{\uparrow},
	\end{aligned}
\end{equation}
which enforce a fixed ratio between the up and down spin components such that $\frac{1}{3}n_{\downarrow}=n_{\uparrow}=\frac{1}{4}$.

In $Case\ (i)$, the other spin configurations,
\begin{equation} 
	\begin{aligned}
	(ss):\ 	& -\mathcal{T}\dot{\psi}_{1s}^{\dag}\dot{\psi}_{2s}\ddot{\psi}_{\tau's}\ddot{\psi}_{2s}^{\dag}\\
		&
=-\mathcal{T}\dot{\psi}_{1s}^{\dag}\ddot{\psi}_{\tau's}\ddot{\psi}_{2s}^{\dag}\dot{\psi}_{2s}+\psi^{\dag}_{1s}\psi_{\tau's}
	=\frac{\delta(\psi_{\tau's}\psi_{2s}^{\dag})}{\delta(\psi_{2s}^{\dag}\psi_{1s})}\\
	&
=\mathcal{T}\ddot{\psi}_{\tau's}\dot{\psi}_{1s}^{\dag}\dot{\psi}_{2s}\ddot{\psi}_{2s}^{\dag}+\psi_{2s}\psi_{2s}^{\dag}
=\frac{\delta(\psi_{\tau's}\psi_{2s}^{\dag})}{\delta(\psi_{2s}^{\dag}\psi_{1s})}
=n_{\downarrow},\\
(ss'):\ &		
-\mathcal{T}
\dot{\psi}^{\dag}_{1s}\dot{\psi}_{2s}\ddot{\psi}_{2s'}\ddot{\psi}^{\dag}_{\tau's'}\\
&=
-\mathcal{T}
\dot{\psi}^{\dag}_{1s}\ddot{\psi}_{2s'}\ddot{\psi}^{\dag}_{\tau's'}\dot{\psi}_{2s}
=
-\mathcal{T}
\ddot{\psi}_{2s'}\dot{\psi}^{\dag}_{1s}\dot{\psi}_{2s}\ddot{\psi}^{\dag}_{\tau's'}\\
&
=-\frac{\delta(\psi^{\dag}_{1s}\psi_{2s})}{\delta(\psi_{\tau' s'}\psi^{\dag}_{2s'})}
=-\frac{\delta(\psi_{2s'}\psi^{\dag}_{\tau's'})}{\delta(\psi^{\dag}_{2 s}\psi_{1s})}
=n_{\downarrow},\\
(s's):\ &		
-\mathcal{T}
\dot{\psi}^{\dag}_{2s'}\dot{\psi}_{1s'}\ddot{\psi}_{\tau's}\ddot{\psi}^{\dag}_{2s}\\
&=-\mathcal{T}
\dot{\psi}^{\dag}_{2s'}\ddot{\psi}_{\tau's}\ddot{\psi}^{\dag}_{2s}\dot{\psi}_{1s'}
=-\mathcal{T}
\ddot{\psi}_{\tau's}\dot{\psi}^{\dag}_{2s'}\dot{\psi}_{1s'}\ddot{\psi}^{\dag}_{2s}\\
&
=-\frac{\delta(\psi^{\dag}_{2s'}\psi_{1s'})}{\delta(\psi_{2 s}\psi^{\dag}_{\tau's})}
=-\frac{\delta(\psi_{\tau's}\psi^{\dag}_{2s})}{\delta(\psi^{\dag}_{1 s'}\psi_{2s'})}
=n_{\downarrow},
\end{aligned}
\end{equation}
where
\begin{equation} 
	\begin{aligned}
		&
\mathcal{T}\dot{\psi}_{1s}^{\dag}\ddot{\psi}_{\tau's}\ddot{\psi}_{2s}^{\dag}\dot{\psi}_{2s}=0,\\
		&
\mathcal{T}\ddot{\psi}_{\tau's}\dot{\psi}_{1s}^{\dag}\dot{\psi}_{2s}\ddot{\psi}_{2s}^{\dag}=\frac{\delta(\psi_{\tau's}\psi_{2s}^{\dag})}{\delta(\psi_{2s}^{\dag}\psi_{1s})}=n_{\downarrow},
\end{aligned}
\end{equation}

For $\Psi_{12}$:
\begin{equation} 
	\begin{aligned}
(1):\ 
&-\mathcal{T}
\dot{\psi}^{\dag}_{1s}\dot{\psi}_{2s}\ddot{\psi}_{2s'}\ddot{\psi}_{\tau's}
=-\mathcal{T}
\dot{\psi}^{\dag}_{1s}\ddot{\psi}_{2s'}\ddot{\psi}_{\tau's}\dot{\psi}_{2s}
=-\frac{\delta(\psi^{\dag}_{1s}\psi_{2s})}{\delta(\psi_{\tau's}^{\dag}\psi_{2s'}^{\dag})}
=[\psi^{\dag}_{1s}\psi_{2s},\psi_{2s'}\psi_{\tau' s}]\\
&=-\psi_{2s'}\{\psi_{\tau's},\psi_{1s}^{\dag}\}\psi_{2s}
=-2\mathcal{T}\ddot{\psi}_{2s'}\ddot{\psi}_{\tau's}\dot{\psi}^{\dag}_{1s}\dot{\psi}_{2s}
-\psi_{2s'}\psi_{2s}\\
&=2\mathcal{T}\dot{\psi}^{\dag}_{1s}\ddot{\psi}_{2s'}\ddot{\psi}_{\tau's}\dot{\psi}_{2s}
+\psi_{2s'}\psi_{2s}\\
(2):\ 
&-\mathcal{T}
\dot{\psi}^{\dag}_{2s'}\dot{\psi}_{1s'}\ddot{\psi}_{2s'}\ddot{\psi}_{\tau's}
=-\frac{\delta(\psi^{\dag}_{2s'}\psi_{1s'})}{\delta(\psi_{\tau's}^{\dag}\psi_{2s'}^{\dag})}
=[\psi^{\dag}_{2s'}\psi_{1s'},\psi_{2s'}\psi_{\tau's}]
=\{\psi^{\dag}_{2s'},\psi_{2s'}\}\psi_{\tau's}\psi_{1s'}
=\psi_{\tau's}\psi_{1s'},\\
	\end{aligned}
\end{equation}
where
\begin{equation} 
	\begin{aligned}
&2\mathcal{T}\dot{\psi}^{\dag}_{1s}\ddot{\psi}_{2s'}\ddot{\psi}_{\tau's}\dot{\psi}_{2s}
=\frac{2}{3}\psi_{2s'}\psi_{\tau's},\\
&-2\mathcal{T}\ddot{\psi}_{2s'}\ddot{\psi}_{\tau's}\dot{\psi}^{\dag}_{1s}\dot{\psi}_{2s}=\frac{2}{3}\psi_{2s'}\psi_{\tau's}
+2\psi_{2s'}\psi_{2s},\\
	\end{aligned}
\end{equation}

For $\Psi_{21}$:\\
\begin{equation} 
	\begin{aligned}
(1):\ &-\mathcal{T}\dot{\psi}_{1s'}\dot{\psi}^{\dag}_{2s'}\ddot{\psi}^{\dag}_{2s}\ddot{\psi}^{\dag}_{\tau' s'}
=-\mathcal{T}\dot{\psi}_{1s'}\ddot{\psi}^{\dag}_{2s}\ddot{\psi}^{\dag}_{\tau' s'}	\dot{\psi}^{\dag}_{2s'}
=-\frac{\delta(\psi_{1s'}\psi_{2s'}^{\dag})}{\delta(\psi_{\tau's'}\psi_{2s})}
=-[\psi_{1s'}\psi^{\dag}_{2s'},\psi^{\dag}_{2s}\psi^{\dag}_{\tau's'}]\\
&=[\psi^{\dag}_{2s}\psi^{\dag}_{\tau's'},\psi_{1s'}\psi^{\dag}_{2s'}]
=\psi^{\dag}_{2s}\{\psi^{\dag}_{\tau's'},\psi_{1s'}\}\psi^{\dag}_{2s'}
=2\mathcal{T}\dot{\psi}^{\dag}_{2s}\dot{\psi}^{\dag}_{\tau's'}\ddot{\psi}_{1s'}\ddot{\psi}^{\dag}_{2s'}
-\psi^{\dag}_{2s}\psi^{\dag}_{2s'}\\
&=-2\mathcal{T}\ddot{\psi}_{1s'}\ddot{\psi}^{\dag}_{2s'}
\dot{\psi}^{\dag}_{2s}\dot{\psi}^{\dag}_{\tau's'}
+\psi^{\dag}_{2s}\psi^{\dag}_{2s'}\\
&=-2\mathcal{T}
\ddot{\psi}_{1s'}\dot{\psi}^{\dag}_{2s}\dot{\psi}^{\dag}_{\tau's'}\ddot{\psi}^{\dag}_{2s'}
+\psi^{\dag}_{2s}\psi^{\dag}_{2s'}
=-\psi^{\dag}_{2s}\psi^{\dag}_{2s'},\\
%=-\psi_{2s'}^{\dag}\psi_{2s}+2\psi^{\dag}_{2s}\psi^{\dag}_{\tau's'},\\
		(2):\ &-\mathcal{T}\dot{\psi}_{2s}\dot{\psi}^{\dag}_{1s}\ddot{\psi}^{\dag}_{2s}\ddot{\psi}^{\dag}_{\tau' s'}
=-\mathcal{T}\dot{\psi}_{2s}\ddot{\psi}^{\dag}_{2s}\ddot{\psi}^{\dag}_{\tau' s'}	\dot{\psi}^{\dag}_{1s}
=-\frac{\delta(\psi_{2s}\psi_{1s}^{\dag})}{\delta(\psi_{\tau's'}\psi_{2s})}\\
&=[\psi^{\dag}_{2s}\psi^{\dag}_{\tau's'},\psi_{2s}\psi^{\dag}_{1s}]
=\{\psi^{\dag}_{2s},\psi_{2s}\}\psi^{\dag}_{1s}\psi^{\dag}_{\tau's'}
=\psi^{\dag}_{1s}\psi^{\dag}_{\tau's'}.
	\end{aligned}
\end{equation}
where
\begin{equation} 
	\begin{aligned}
&
\mathcal{T}
\ddot{\psi}_{1s'}\dot{\psi}^{\dag}_{2s}\dot{\psi}^{\dag}_{\tau's'}\ddot{\psi}^{\dag}_{2s'}=\psi^{\dag}_{2s}\psi^{\dag}_{2s'}.
	\end{aligned}
\end{equation}

For $\Psi_{22}=-\mathcal{T}\dot{\psi}_{1s'}\dot{\psi}^{\dag}_{2s'}\ddot{\psi}^{\dag}_{2s}\ddot{\psi}_{\tau' s}$ there are two cases:

$Case\ (i)$: $\psi^{\dag}_{2s}\psi_{\tau's}$ is the annihilational term,
\begin{equation} 
	\begin{aligned}
&
-\mathcal{T}\dot{\psi}_{2s}\dot{\psi}^{\dag}_{1s}\ddot{\psi}^{\dag}_{2s}\ddot{\psi}_{\tau' s}
=\mathcal{T}\dot{\psi}_{2s}\ddot{\psi}^{\dag}_{2s}\ddot{\psi}_{\tau's}\dot{\psi}_{1s}^{\dag}+\psi_{2s}\psi_{2s}^{\dag}
=\mathcal{T}\dot{\psi}_{2s}\ddot{\psi}^{\dag}_{2s}\ddot{\psi}_{\tau's}\dot{\psi}_{1s}^{\dag}\\
&
=\frac{\delta(\psi_{2s}\psi_{1s}^{\dag})}{\delta(\psi_{\tau's}^{\dag}\psi_{2s})}
=[\psi^{\dag}_{2s}\psi_{\tau's},\psi_{2s}\psi_{1s}^{\dag}]\\
&=\{\psi^{\dag}_{2s},\psi_{2s}\}\psi_{1s}^{\dag}\psi_{\tau's}
-\psi^{\dag}_{2s}\psi_{2s}\{\psi_{1s}^{\dag},\psi_{\tau's}\}
=\mathcal{T}\dot{\psi}_{2s}\ddot{\psi}^{\dag}_{2s}\dot{\psi}_{1s}^{\dag}\ddot{\psi}_{\tau's}
-\mathcal{T}\dot{\psi}^{\dag}_{2s}\ddot{\psi}_{2s}\dot{\psi}_{\tau's}\ddot{\psi}_{1s}^{\dag}\\
&=\mathcal{T}\dot{\psi}_{2s}\ddot{\psi}^{\dag}_{2s}\ddot{\psi}_{\tau's}\dot{\psi}_{1s}^{\dag}+\psi_{2s}\psi_{2s}^{\dag}
-\mathcal{T}\dot{\psi}^{\dag}_{2s}\ddot{\psi}_{2s}\dot{\psi}_{\tau's}\ddot{\psi}_{1s}^{\dag}\\
&=2\mathcal{T}\dot{\psi}_{2s}\ddot{\psi}^{\dag}_{2s}\ddot{\psi}_{\tau's}\dot{\psi}_{1s}^{\dag}+\psi_{2s}\psi_{2s}^{\dag}
-\psi_{\tau's}\psi_{1s}^{\dag}\\
&
=\psi_{\tau's}\psi_{1s}^{\dag}=-n_{\uparrow},\\
	\end{aligned}
\end{equation}
where
\begin{equation} 
	\begin{aligned}
		&
\mathcal{T}\dot{\psi}^{\dag}_{2s}\ddot{\psi}_{2s}\dot{\psi}_{\tau's}\ddot{\psi}_{1s}^{\dag}=0,\\
&
\mathcal{T}\dot{\psi}_{2s}\ddot{\psi}^{\dag}_{2s}\ddot{\psi}_{\tau's}\dot{\psi}_{1s}^{\dag}=
-\mathcal{T}\dot{\psi}_{2s}\dot{\psi}^{\dag}_{1s}\ddot{\psi}^{\dag}_{2s}\ddot{\psi}_{\tau' s}=\psi_{\tau's}\psi_{1s}^{\dag}.
	\end{aligned}
\end{equation}
For this case, combined with the result in Eqs.(\ref{645},\ref{646}),
the self-consistency relation is satisfied,
\begin{equation} 
	\begin{aligned}
&\Psi_{22}(\tau^{+},\tau';\tau)
=-\mathcal{T}\dot{\psi}_{1s'}\ddot{\psi}^{\dag}_{2s'}\ddot{\psi}^{\dag}_{2s}\dot{\psi}_{\tau's}
=\left[-\mathcal{T}\dot{\psi}_{1s'}\dot{\psi}^{\dag}_{2s'}\ddot{\psi}^{\dag}_{2s}\ddot{\psi}_{\tau's}\right]
+\left[-\mathcal{T}\dot{\psi}_{1s'}\ddot{\psi}^{\dag}_{2s'}\dot{\psi}^{\dag}_{2s}\ddot{\psi}_{\tau's}\right]
=-n_{\uparrow}+n_{\downarrow}.
	\end{aligned}
\end{equation}

$Case\ (ii)$: $\psi^{\dag}_{2s}\psi_{\tau's}$ is the creational term,
\begin{equation} 
	\begin{aligned}
		&
		-\mathcal{T}\dot{\psi}_{2s}\dot{\psi}^{\dag}_{1s}\ddot{\psi}^{\dag}_{2s}\ddot{\psi}_{\tau' s}
		=\mathcal{T}\dot{\psi}_{2s}\ddot{\psi}^{\dag}_{2s}\ddot{\psi}_{\tau's}\dot{\psi}_{1s}^{\dag}
		=\frac{\delta(\psi_{2s}\psi_{1s}^{\dag})}{\delta(\psi_{\tau's}^{\dag}\psi_{2s})}
		=[\psi_{2s}\psi_{1s}^{\dag},\psi^{\dag}_{2s}\psi_{\tau's}]\\
		&
		=\mathcal{T}\dot{\psi}^{\dag}_{2s}\ddot{\psi}_{2s}\dot{\psi}_{\tau's}\ddot{\psi}_{1s}^{\dag}-\mathcal{T}\dot{\psi}_{2s}\ddot{\psi}^{\dag}_{2s}\dot{\psi}_{1s}^{\dag}\ddot{\psi}_{\tau's}\\
		&=-\mathcal{T}\dot{\psi}_{2s}\ddot{\psi}^{\dag}_{2s}\ddot{\psi}_{\tau's}\dot{\psi}_{1s}^{\dag}-\psi_{2s}\psi_{2s}^{\dag}
		+\mathcal{T}\dot{\psi}^{\dag}_{2s}\ddot{\psi}_{2s}\dot{\psi}_{\tau's}\ddot{\psi}_{1s}^{\dag}
		=\frac{1}{3}\psi_{\tau's}\psi_{1s}^{\dag}=-\frac{1}{3}n_{\uparrow},\\
	\end{aligned}
\end{equation}
where
\begin{equation} 
	\begin{aligned}
		&
	\mathcal{T}\dot{\psi}^{\dag}_{2s}\ddot{\psi}_{2s}\dot{\psi}_{\tau's}\ddot{\psi}_{1s}^{\dag}=2\mathcal{T}\dot{\psi}_{2s}\ddot{\psi}^{\dag}_{2s}\ddot{\psi}_{\tau's}\dot{\psi}_{1s}^{\dag},\\
		&
	\mathcal{T}\dot{\psi}_{2s}\ddot{\psi}^{\dag}_{2s}\ddot{\psi}_{\tau's}\dot{\psi}_{1s}^{\dag}=\frac{1}{3}\psi_{\tau's}\psi_{1s}^{\dag}.
	\end{aligned}
\end{equation}
While for this case, 
the self-consistency relation is only possible to be satisfied in the following form (similar to the $\Psi_{11}$ and $\Psi_{12}$),
\begin{equation} 
	\begin{aligned}
		&\Psi_{22}(\tau^{+},\tau';\tau)
		=-\mathcal{T}\dot{\psi}_{1s'}\ddot{\psi}^{\dag}_{2s'}\ddot{\psi}^{\dag}_{2s}\dot{\psi}_{\tau's}
		=\left[-\mathcal{T}\dot{\psi}_{1s'}\dot{\psi}^{\dag}_{2s'}\ddot{\psi}^{\dag}_{2s}\ddot{\psi}_{\tau's}\right]
		-\left[-\mathcal{T}\dot{\psi}_{1s'}\ddot{\psi}^{\dag}_{2s'}\dot{\psi}^{\dag}_{2s}\ddot{\psi}_{\tau's}\right]\\
		&
		=-\frac{1}{3}n_{\uparrow}-n_{\downarrow}
		=-n_{\uparrow}+n_{\downarrow}.
	\end{aligned}
\end{equation}
 which enforce a fixed ratio between the up and down spin components
 $\frac{1}{3}n_{\uparrow}=n_{\downarrow}=\frac{1}{4}$.

For $Case\ (i)$ the other spin configurations read
\begin{equation}
	\begin{aligned}
		&
-\mathcal{T}\dot{\psi}_{1s'}\dot{\psi}^{\dag}_{2s'}\ddot{\psi}^{\dag}_{\tau's'}\ddot{\psi}_{2s'}\\
&= -\mathcal{T}\dot{\psi}_{1s'}\ddot{\psi}^{\dag}_{\tau's'}
\ddot{\psi}_{2s'}\dot{\psi}_{2s'}^{\dag}+\psi_{1s'}\psi^{\dag}_{\tau's'}
=-\frac{\delta(\psi_{1s'}\psi^{\dag}_{2s'})}{\delta(\psi^{\dag}_{2s'}\psi_{\tau's'})}+\psi_{1s'}\psi^{\dag}_{\tau's'}\\
&=\mathcal{T}\ddot{\psi}^{\dag}_{\tau's'}\dot{\psi}_{1s'}\dot{\psi}^{\dag}_{2s'}\ddot{\psi}_{2s'}-\psi^{\dag}_{2s'}\psi_{2s'}
=-\frac{\delta(\psi_{1s'}\psi^{\dag}_{2s'})}{\delta(\psi^{\dag}_{2s'}\psi_{\tau's'})}+\psi_{1s'}\psi^{\dag}_{\tau's'}
=-n_{\uparrow},\\
& -\mathcal{T}\dot{\psi}_{2s}\dot{\psi}^{\dag}_{1s}\ddot{\psi}^{\dag}_{\tau's'}\ddot{\psi}_{2s'}
= -\mathcal{T}\dot{\psi}_{2s}\ddot{\psi}^{\dag}_{\tau's'}
\ddot{\psi}_{2s'}\dot{\psi}_{1s}^{\dag}
= -\frac{\delta(\psi_{2s}\psi_{1s}^{\dag})}{\delta(\psi^{\dag}_{2s'}\psi_{\tau's'})}
=-n_{\uparrow},\\
&
-\mathcal{T}
\dot{\psi}_{1s'}\dot{\psi}^{\dag}_{2s'}\ddot{\psi}^{\dag}_{2s}\ddot{\psi}_{\tau's}
= -\mathcal{T}\dot{\psi}_{1s'}\ddot{\psi}^{\dag}_{2s}
\ddot{\psi}_{\tau's}\dot{\psi}_{2s'}^{\dag}
= -\frac{\delta(\psi_{1s'}\psi_{2s'}^{\dag})}{\delta(\psi^{\dag}_{\tau's}\psi_{2s})}
=-n_{\uparrow},
	\end{aligned}
\end{equation}
where
\begin{equation} 
	\begin{aligned}
		&
\mathcal{T}\dot{\psi}_{1s'}\ddot{\psi}^{\dag}_{\tau's'}
\ddot{\psi}_{2s'}\dot{\psi}_{2s'}^{\dag}=0,\\
&
\mathcal{T}\ddot{\psi}^{\dag}_{\tau's'}\dot{\psi}_{1s'}\dot{\psi}^{\dag}_{2s'}\ddot{\psi}_{2s'}=-n_{\uparrow},
	\end{aligned}
\end{equation}

\subsection{ $\dot{\psi}^{\dag}_{s}\ddot{\psi}_{s'}^{\dag}\ddot{\psi}_{s}\dot{\psi}_{s'}$}

For $\Psi_{11}$:
\begin{equation} 
	\begin{aligned}
&
\label{642}
 -\mathcal{T}\dot{\psi}^{\dag}_{1s}\ddot{\psi}_{2s}\dot{\psi}_{2s'}\ddot{\psi}^{\dag}_{\tau' s'}
=-\mathcal{T}\dot{\psi}^{\dag}_{1s}\ddot{\psi}_{2s}\ddot{\psi}^{\dag}_{\tau' s'}\dot{\psi}_{2s'}
+\psi_{1s}^{\dag}\psi_{2s}\\
&=[\psi^{\dag}_{1s}\psi_{2s'},\psi_{2s}\psi^{\dag}_{\tau's'}]
+\psi_{1s}^{\dag}\psi_{2s}\\
&=[\{\psi^{\dag}_{1s},\psi_{2s}\}\psi^{\dag}_{\tau's'}-\psi_{2s}\{\psi^{\dag}_{\tau's'},\psi^{\dag}_{1s}\}]\psi_{2s'}
+\psi^{\dag}_{1s}
[\{\psi_{2s'},\psi_{2s}\}\psi^{\dag}_{\tau's'}-\psi_{2s}\{\psi^{\dag}_{\tau's'},
\psi_{2s'}\}]
+\psi_{1s}^{\dag}\psi_{2s}\\
&=[\{\psi^{\dag}_{1s},\psi_{2s}\}\psi^{\dag}_{\tau's'}\psi_{2s'}
-\psi^{\dag}_{1s}
\psi_{2s}\{\psi^{\dag}_{\tau's'},
\psi_{2s'}\}]
+\psi_{1s}^{\dag}\psi_{2s}\\
&=[\mathcal{T}\dot{\psi}_{2s}\ddot{\psi}^{\dag}_{1s}\dot{\psi}^{\dag}_{\tau's'}\ddot{\psi}_{2s'}
-\mathcal{T}\dot{\psi}^{\dag}_{1s}\ddot{\psi}_{2s}\dot{\psi}_{2s'}\ddot{\psi}^{\dag}_{\tau's'}]
+\psi_{1s}^{\dag}\psi_{2s}
=\psi_{1s}^{\dag}\psi_{2s}=n_{\uparrow}.
	\end{aligned}
\end{equation}

For $\Psi_{12}$:
\begin{equation} 
	\begin{aligned}
&
 -\mathcal{T}\dot{\psi}^{\dag}_{1s}\ddot{\psi}_{2s}\dot{\psi}_{2s'}\ddot{\psi}_{\tau' s}
= \mathcal{T}\dot{\psi}^{\dag}_{1s}\ddot{\psi}_{2s}\ddot{\psi}_{\tau' s}\dot{\psi}_{2s'}\\
&=[\psi_{2s}\psi_{\tau' s},\psi^{\dag}_{1s}\psi_{2s'}]
=\{\psi_{2s},\psi^{\dag}_{1s}\}\psi_{2s'}\psi_{\tau's}
+\psi_{2s}\{\psi_{\tau' s},\psi^{\dag}_{1s}\}
\psi_{2s'}\\
&=\mathcal{T}\dot{\psi}_{1s}^{\dag}\ddot{\psi}_{2s}\dot{\psi}_{2s'}\ddot{\psi}_{\tau's}
+\mathcal{T}\ddot{\psi}_{2s}\ddot{\psi}_{\tau's}\dot{\psi}^{\dag}_{1s}\dot{\psi}_{2s'}\\
&=
-\mathcal{T}\ddot{\psi}_{2s}\ddot{\psi}_{\tau's}\dot{\psi}^{\dag}_{1s}\dot{\psi}_{2s'}-\psi_{2s}\psi_{2s'}+\psi_{2s'}\psi_{\tau's}
+\mathcal{T}\ddot{\psi}_{2s}\ddot{\psi}_{\tau's}\dot{\psi}^{\dag}_{1s}\dot{\psi}_{2s'}\\
&=-\psi_{2s}\psi_{2s'}+\psi_{2s'}\psi_{\tau's}.
	\end{aligned}
\end{equation}

For $\Psi_{21}$:
\begin{equation} 
	\begin{aligned}
&-\mathcal{T}\dot{\psi}_{1s'}\ddot{\psi}^{\dag}_{2s'}\dot{\psi}^{\dag}_{2s}\ddot{\psi}^{\dag}_{\tau' s'}
=\mathcal{T}\dot{\psi}_{1s'}\ddot{\psi}^{\dag}_{2s'}\ddot{\psi}^{\dag}_{\tau' s'}\dot{\psi}^{\dag}_{2s}
=\frac{\delta(\psi_{1s'}\psi_{2s}^{\dag})}{\delta(\psi_{\tau's'}\psi_{2s'})}
=[\psi_{1s'}\psi^{\dag}_{2s},\psi^{\dag}_{2s'}\psi^{\dag}_{\tau' s'}]\\
&=
\{\psi_{1s'},\psi^{\dag}_{2s'}\}\psi^{\dag}_{\tau's'}\psi^{\dag}_{2s}
-\psi^{\dag}_{2s'}\{\psi^{\dag}_{\tau's'},\psi_{1s'}\}\psi^{\dag}_{2s}\\
&=\mathcal{T}\ddot{\psi}_{1s'}\dot{\psi}^{\dag}_{2s'}
\dot{\psi}^{\dag}_{\tau's'}\ddot{\psi}^{\dag}_{2s}
-\mathcal{T}\ddot{\psi}_{2s'}^{\dag}
\ddot{\psi}_{\tau's'}^{\dag}\dot{\psi}_{1s'}\dot{\psi}_{2s}^{\dag}
=-\psi^{\dag}_{2s'}\psi^{\dag}_{2s}-\psi^{\dag}_{\tau's'}\psi^{\dag}_{2s}.\\
	\end{aligned}
\end{equation}

For $\Psi_{22}$:
\begin{equation} 
	\begin{aligned}
		\label{646}
& -\mathcal{T}\dot{\psi}_{1s'}\ddot{\psi}^{\dag}_{2s'}\dot{\psi}^{\dag}_{2s}\ddot{\psi}_{\tau' s}
= -\mathcal{T}\dot{\psi}_{1s'}\ddot{\psi}^{\dag}_{2s'}\ddot{\psi}_{\tau' s}\dot{\psi}^{\dag}_{2s}
-\psi_{1s'}\psi_{2s'}^{\dag}\\
&=[\psi_{1s'}\psi_{2s}^{\dag},\psi_{2s'}^{\dag}\psi_{\tau's}]
-\psi_{1s'}\psi_{2s'}^{\dag}\\
&=\{\psi_{1s'},\psi_{2s'}^{\dag}\}\psi_{\tau's}\psi_{2s}^{\dag}
-\psi_{1s'}\psi_{2s'}^{\dag}\{\psi_{\tau's},\psi_{2s}^{\dag}\}
-\psi_{1s'}\psi_{2s'}^{\dag}\\
&=\mathcal{T}\dot{\psi}_{2s'}^{\dag}\ddot{\psi}_{1s'}\dot{\psi}_{\tau's}\ddot{\psi}_{2s}^{\dag}
-\mathcal{T}\dot{\psi}_{1s'}\ddot{\psi}_{2s'}^{\dag}\dot{\psi}_{2s}^{\dag}\ddot{\psi}_{\tau's}
-\psi_{1s'}\psi_{2s'}^{\dag}\\
&=
\psi_{\tau's}\psi_{2s}^{\dag}
-\psi_{1s'}\psi_{2s'}^{\dag}
-\psi_{1s'}\psi_{2s'}^{\dag}
=
-\psi^{\dag}_{2s'}\psi_{1s'}
+\psi_{2s}^{\dag}\psi_{\tau's}
-\psi_{1s'}\psi_{2s'}^{\dag}
=-\psi_{1s'}\psi_{2s'}^{\dag}=n_{\downarrow},
	\end{aligned}
\end{equation}

%\subsection{Final result obtained from above subsections}

Then we obtain that the relation between different decompositions for each element
(we only consider the normal case here where
$\psi_{\tau's}\psi_{2s}^{\dag}$
and $\psi^{\dag}_{2s}\psi_{\tau's}$
be the annihilational termfor $\Psi_{11}$
and $\Psi_{22}$, respectively)
\begin{equation} 
	\begin{aligned}
&\Psi_{11}(\tau^{+},\tau';\tau)=
-\mathcal{T}\dot{\psi}^{\dag}_{1s}\ddot{\psi}_{2s}\ddot{\psi}_{2s'}\dot{\psi}^{\dag}_{\tau's'}
=
\left[-\mathcal{T}
\dot{\psi}^{\dag}_{1s}\dot{\psi}_{2s}\ddot{\psi}_{2s'}\ddot{\psi}^{\dag}_{\tau's'}\right]-
 \left[-\mathcal{T}\dot{\psi}^{\dag}_{1s}\ddot{\psi}_{2s}\dot{\psi}_{2s'}\ddot{\psi}^{\dag}_{\tau' s'}
\right],\\
&\Psi_{12}(\tau^{+},\tau';\tau)
=-\mathcal{T}\dot{\psi}_{1s'}\ddot{\psi}^{\dag}_{2s'}\ddot{\psi}^{\dag}_{2s}\dot{\psi}^{\dag}_{\tau' s'}
=\left[-\mathcal{T}\dot{\psi}_{1s'}\dot{\psi}^{\dag}_{2s'}\ddot{\psi}^{\dag}_{2s}\ddot{\psi}^{\dag}_{\tau' s'}\right]
-
\left[-\mathcal{T}\dot{\psi}_{1s'}\ddot{\psi}^{\dag}_{2s'}\dot{\psi}^{\dag}_{2s}\ddot{\psi}^{\dag}_{\tau' s'}\right],\\
&\Psi_{21}(\tau^{+},\tau';\tau)
=-\mathcal{T}\dot{\psi}_{1s'}\ddot{\psi}^{\dag}_{2s'}\ddot{\psi}^{\dag}_{2s}\dot{\psi}^{\dag}_{\tau' s'}
=
\left[-\mathcal{T}\dot{\psi}_{1s'}\dot{\psi}^{\dag}_{2s'}\ddot{\psi}^{\dag}_{2s}\ddot{\psi}^{\dag}_{\tau' s'}\right]
+
\left[-\mathcal{T}\dot{\psi}_{1s'}\ddot{\psi}^{\dag}_{2s'}\dot{\psi}^{\dag}_{2s}\ddot{\psi}^{\dag}_{\tau' s'}\right],\\
&\Psi_{22}(\tau^{+},\tau';\tau)
=-\mathcal{T}\dot{\psi}_{1s'}\ddot{\psi}^{\dag}_{2s'}\ddot{\psi}^{\dag}_{2s}\dot{\psi}_{\tau's}
=\left[-\mathcal{T}\dot{\psi}_{1s'}\dot{\psi}^{\dag}_{2s'}\ddot{\psi}^{\dag}_{2s}\ddot{\psi}_{\tau's}\right]
+\left[-\mathcal{T}\dot{\psi}_{1s'}\ddot{\psi}^{\dag}_{2s'}\dot{\psi}^{\dag}_{2s}\ddot{\psi}_{\tau's}\right].
	\end{aligned}
\end{equation}

\section{Appendix.B: Condition for the cutoff at second order derivative}

For the propagator labelled by $(\tau_{1},\tau_{2})$, to make sure its derivative with $\tau_{1}$ be consistent with the decomposed four-point functions in $\pmb{\Psi}(\tau^{+},\tau';\tau)$,
a broken U(1)-symmetry is needed in the first order derivative of the $\tau_{1}$-dependent operator.
Thus $\tau_{1}$-dependence can be regarded as a new kinds of source field
which is responsible for the generation of ill-defined normal propagator,
and the fermion bilinears appear here that each operator has been cassified into $\alpha$ or $\beta$ according to their variational dependence on the orthonormal set, analogous to the off-diagonal propagators appear in the Bethe-Salpeter equation for the three-point susceptibility (source field functional derivative of normal propagator),
as far as $\alpha\neq\beta$.

Next we see what condition is required to realize the above-mentioned
cutoff at the second order time derivative for the operators.
Here, we take $\psi^{\dag}_{s}(\tau_{1})$ as an example,
\begin{equation} 
	\begin{aligned}
		\frac{\partial \psi_{s}^{\dag}(\tau_{1})}{\partial \tau_{1}}
		&=
		\hat{H}^{*}_{1}\psi_{s}^{\dag}(\tau_{1})
		+
		\int dr_{2}' U_{r_{1}-r_{2}'}    
		\psi^{\dag}_{s}(\tau_{1}) \psi^{\dag}_{s'}(\tau_{2}')\psi_{s'}(\tau_{2}')
		-
		\int dr_{2} U_{r_{2}-r_{1}}   
		\psi^{\dag}_{s}(\tau_{2}) \psi^{\dag}_{s'}(\tau_{1})\psi_{s}(\tau_{2})\\
				&=
		\hat{H}^{*}_{1}\psi_{s}^{\dag}(\tau_{1})
		+
		\int dr_{2}' U_{r_{1}-r_{2}'}    
		\psi^{\dag}_{s}(\tau_{1}) \psi^{\dag}_{s'}(\tau_{2}')\psi_{s'}(\tau_{2}')
		+
		\int dr_{2} U_{r_{2}-r_{1}}   
	\psi^{\dag}_{s'}(\tau_{1}) 	\psi^{\dag}_{s}(\tau_{2}) \psi_{s}(\tau_{2})\\
	\end{aligned}
\end{equation}

\begin{equation} 
	\begin{aligned}
		\frac{\partial ^{(2)}\psi_{s}^{\dag}(\tau_{1})}{\partial \tau_{1}^{(2)}}
		&=
		\hat{H}^{*}_{1}\frac{\partial \psi_{s}^{\dag}(\tau_{1})}{\partial \tau_{1}}
+
		\frac{\partial }{\partial \tau_{1}}
		\left[	\psi^{\dag}_{s}(\tau_{1})	\int dr_{2}' U_{r_{1}-r_{2}'}    
	 \psi^{\dag}_{s'}(\tau_{2}')\psi_{s'}(\tau_{2}')
		+
		\psi^{\dag}_{s'}(\tau_{1}) 	\int dr_{2} U_{r_{2}-r_{1}}   
		\psi^{\dag}_{s}(\tau_{2}) \psi_{s}(\tau_{2})\right].
	\end{aligned}
\end{equation}
As we discuss in above,
an infinitely small shifment on imaginary time in nonstring form
is required for a well-performed four-point function decomposition,
and this shifment cannot be seen from the time components in string form,
but can be observed in terms of the distinct dependences on the corresponding orthogonal basis for the operators within a fermion bilinear 
(e.g., with same spin and at the same time in string form, like $\psi^{\dag}_{s}(t_{1}')\psi_{s}(t_{1}')$).
Such distinct dependence, or distinct (although related to each other as enforced by the correlations between two spin basis)
corresponding coefficient sets,
breaks the U(1) symmetry of this fermion bilinear
and we can see this effect by rewriting the above bilinear as
$\psi^{\dag}_{s}(t_{1}';\alpha)\psi_{s}(t_{1}';\beta)$.
We can thus write the integral over $\tau_{2}'(\neq\tau_{1})$ for the normal propagator that appears in above second order derivative as (this form will appear again, see Eq.(\ref{711}))
\begin{equation} 
	\begin{aligned}
		\label{6181}
\int dr_{2}' U_{r_{1}-r_{2}'}    
\psi^{\dag}_{s'}(\tau_{2}';\alpha)\psi_{s'}(\tau_{2}';\beta)=
\int dr_{2}' U_{r_{1}-r_{2}'}    [
\psi^{\dag}_{s'}(\tau_{2}';\alpha)\psi_{s'}(\tau_{2}';\alpha)
+\psi^{\dag}_{s'}(\tau_{2}';\alpha)\psi_{s'}(\tau_{2}';\alpha)(e^{u(\alpha-\beta)}-1)],
	\end{aligned}
\end{equation}
where we use the conformal approximation 
$\psi_{s'}(\tau_{2}';\beta)=\psi_{s'}(\tau_{2}';\alpha)e^{u(\alpha-\beta)}$,
with the conserved energy density $u(\alpha-\beta)$ to overcome the scaling
invariance of conformal states.
Then we have the mutually independent fermion bilinears in the first term
where each one of them is a local observable (local or quasilocal conserved quantities of an integrable system).
The integral in Eq.(\ref{6181}) can be viewed as a chaotic non-fermi-liquid system without the lowest energy state,
which containing two correlated terms
$\int dr_{2}' U_{r_{1}-r_{2}'}   
\psi^{\dag}_{s'}(\tau_{2}';\alpha)\psi_{s'}(\tau_{2}';\alpha)$
and $\int dr_{2}' U_{r_{1}-r_{2}'}   
\psi^{\dag}_{s'}(\tau_{2}';\alpha)\psi_{s'}(\tau_{2}';\alpha)(e^{u(\alpha-\beta)}-1)$.

\subsection{Self-consistency in terms of Bethe-Salpeter-type expression}

Using the functional derivative scheme and base on the above-mentioned energy density $u(\alpha-\beta)$,
we can write the derivative with respect to $\tau_{1}$ in the following form,
which is necessary for the following step (see Eq.(\ref{6268})) to cast it into the form of Bethe-Salpeter equation 
\begin{equation} 
	\begin{aligned}
		\label{61821}
&\frac{\partial}{\partial\tau_{1}}
		\int dr_{2}' U_{r_{1}-r_{2}'}    
		\psi^{\dag}_{s'}(\tau_{2}';\alpha)\psi_{s'}(\tau_{2}';\beta)\\
		&=
\frac{\delta}{\delta 	\psi^{\dag}_{s'}(\tau_{2}';\alpha)\psi_{s'}(\tau_{2}';\alpha)}
\int dr_{2}' U_{r_{1}-r_{2}'}    
\psi^{\dag}_{s'}(\tau_{2}';\alpha)\psi_{s'}(\tau_{2}';\beta)\bigg|_{u(\alpha-\beta)=0}
\frac{\delta 	\psi^{\dag}_{s'}(\tau_{2}';\alpha)\psi_{s'}(\tau_{2}';\beta)}{\delta 	\tau_{1}}
\bigg|_{u(\alpha-\beta)=0},\\
&
\frac{\partial}{\partial \tau_{1}}	\int dr_{2} U_{r_{2}-r_{1}}   
\psi^{\dag}_{s}(\tau_{2};\alpha) \psi_{s}(\tau_{2};\beta)\\
		&=
\frac{\delta}{\delta 	\psi^{\dag}_{s}(\tau_{2};\alpha)\psi_{s}(\tau_{2};\alpha)}
\int dr_{2} U_{r_{2}-r_{1}}    
\psi^{\dag}_{s}(\tau_{2};\alpha)\psi_{s}(\tau_{2};\beta)\bigg|_{u(\alpha-\beta)=0}
\frac{\delta 	\psi^{\dag}_{s}(\tau_{2};\alpha)\psi_{s}(\tau_{2};\beta)}{\delta 	\tau_{1}}
\bigg|_{u(\alpha-\beta)=0}.
	\end{aligned}
\end{equation}
For notational simplicity, hereafter we define
\begin{equation} 
	\begin{aligned}
		&	\int dr_{2}' U_{r_{1}-r_{2}'}    
		\psi^{\dag}_{s'}(\tau_{2}';\alpha)\psi_{s'}(\tau_{2}';\beta):=\digamma_{s'}(\alpha,\beta),\\
		&
			\int dr_{2} U_{r_{2}-r_{1}}    
		\psi^{\dag}_{s}(\tau_{2};\alpha)\psi_{s}(\tau_{2};\beta):=\digamma_{s}(\alpha,\beta),\\
		&
		\psi^{\dag}_{s'}(\tau_{2}';\alpha)\psi_{s'}(\tau_{2}';\alpha):=\mathcal{N}_{s'}(\tau_{2}';\alpha),\\
		&
		\psi^{\dag}_{s}(\tau_{2};\alpha)\psi_{s}(\tau_{2};\alpha):=\mathcal{N}_{s}(\tau_{2};\alpha).
			\end{aligned}
	\end{equation}
Analogous to Bethe-Salpeter equation,
where the product of self-energy and Green function equivalents to the product of irreducible vertex and susceptibility,
the Eq.(\ref{61821}) can be decomposed into two terms,
where the first term reads
\begin{equation} 
	\begin{aligned}
		\label{6182}
		&
		\frac{\delta \digamma_{s'}(\alpha,\beta)}{\delta 	\mathcal{N}_{s'}(\tau_{2}';\alpha)}
=		\frac{\delta \digamma_{s'}(\alpha,\beta)}{\delta {\rm ln}
\mathcal{N}_{s'}(\tau_{2}';\alpha)}
[\mathcal{N}_{s'}(\tau_{2}';\alpha)]^{-1}
=		\frac{\delta \hat{\Sigma}_{s'}(\alpha,\beta)}{\delta {\rm ln}
\mathcal{N}_{s'}(\tau_{2}';\alpha)}
,\\
		&
		\frac{\delta \digamma_{s}(\alpha,\beta)}{\delta 	\mathcal{N}_{s}(\tau_{2};\alpha)}
=		\frac{\delta \digamma_{s}(\alpha,\beta)}{\delta  {\rm ln}
\mathcal{N}_{s}(\tau_{2};\alpha)}
[\mathcal{N}_{s}(\tau_{2};\alpha)]^{-1}
=		\frac{\delta \hat{\Sigma}_{s}(\alpha,\beta)}{\delta  {\rm ln}
\mathcal{N}_{s}(\tau_{2};\alpha)}
,
	\end{aligned}
\end{equation}
with $\hat{\Sigma}_{s'}(\alpha,\beta)$ (off-diagonal type) the effective self-energy
related to the exchange between $\alpha$ and $\beta$,
and the off-diagonal type self-energy is nonzero as long as the conserved energy density $u(\alpha-\beta)$ is nonzero.
In terms of the Bethe-Salpeter equation,
we express the effective self-energies as
\begin{equation} 
	\begin{aligned}
		\hat{\Sigma}_{s'}(\alpha,\beta)
		&=\digamma_{s'}(\alpha,\beta)
		\digamma[{\rm ln}\mathcal{N}_{s'}(\tau_{2}';\alpha)]\\
		&
		=\digamma_{s'}(\alpha,\beta)\left\{
		[\mathcal{N}_{s'}(\tau_{2}';\alpha)]^{-1}
		-
	\left[
			\frac{\delta \digamma_{s'}(\alpha,\beta)}{\delta {\rm ln}
			\mathcal{N}_{s'}(\tau_{2}';\alpha)}
	\right]^{-1}
	\digamma_{s'}(\alpha,\beta)
	\frac{\delta \digamma[{\rm ln}\mathcal{N}_{s'}(\tau_{2}';\alpha)]}
		{\delta {\rm ln}\mathcal{N}_{s'}(\tau_{2}';\alpha)}\right\},\\
\hat{\Sigma}_{s}(\alpha,\beta)
		&=\digamma_{s}(\alpha,\beta)
\digamma[{\rm ln}\mathcal{N}_{s}(\tau_{2};\alpha)]\\
		&
=\digamma_{s}(\alpha,\beta)\left\{
[\mathcal{N}_{s}(\tau_{2};\alpha)]^{-1}
-
\left[
\frac{\delta \digamma_{s}(\alpha,\beta)}{\delta {\rm ln}
\mathcal{N}_{s}(\tau_{2};\alpha)}
\right]^{-1}
\digamma_{s}(\alpha,\beta)
\frac{\delta \digamma[{\rm ln}\mathcal{N}_{s}(\tau_{2};\alpha)]}
{\delta {\rm ln}\mathcal{N}_{s}(\tau_{2};\alpha)}\right\},
	\end{aligned}
\end{equation}
and according to the relation between irreducible self energy and susceptibility, we have
(through the standard operator identities)
\begin{equation} 
	\begin{aligned}
		\label{6268}
		&
	\mathcal{G}^{-1}_{L;s'}	\frac{\delta
		\digamma_{s'}(\alpha,\beta)
	}{\delta \mathcal{N}_{s'}(\tau_{2}';\alpha)}
	\bigg|_{u(\alpha-\beta)=0}
		\frac{\delta 	\psi^{\dag}_{s'}(\tau_{2}';\alpha)\psi_{s'}(\tau_{2}';\beta)}{\delta 	\tau_{1}}
		\bigg|_{u(\alpha-\beta)=0}	\mathcal{G}^{-1}_{R;s'}	
		= 	\frac{\delta 	\psi^{\dag}_{s'}(\tau_{2}';\alpha)\psi_{s'}(\tau_{2}';\beta)}{\delta 	\tau_{1}}
		\bigg|_{u(\alpha-\beta)=0}	
		- \mathcal{G}^{-1}_{L;s'}		\mathcal{G}^{-1}_{R;s'}	,\\
		&
	\mathcal{G}^{-1}_{L;s}	
		\frac{\delta   \digamma_{s}(\alpha,\beta)
		}{\delta \mathcal{N}_{s}(\tau_{2};\alpha)}
	\bigg|_{u(\alpha-\beta)=0}
		\frac{\delta 	\psi^{\dag}_{s}(\tau_{2};\alpha)\psi_{s}(\tau_{2};\beta)}{\delta 	\tau_{1}}
		\bigg|_{u(\alpha-\beta)=0}	\mathcal{G}^{-1}_{R;s}	
		= \frac{\delta 	\psi^{\dag}_{s}(\tau_{2};\alpha)\psi_{s}(\tau_{2};\beta)}{\delta 	\tau_{1}}
		\bigg|_{u(\alpha-\beta)=0}- 
\mathcal{G}^{-1}_{L;s}		\mathcal{G}^{-1}_{R;s}	,\\
	\end{aligned}
\end{equation}
which also implies
\begin{equation} 
	\begin{aligned}
		\label{6225}
		& 
\mathcal{G}_{L;s'}		\frac{\delta 	\psi^{\dag}_{s'}(\tau_{2}';\alpha)\psi_{s'}(\tau_{2}';\beta)}{\delta 	\tau_{1}}
		\bigg|_{u(\alpha-\beta)=0}		\mathcal{G}_{R;s'}	
=\left[\frac{\delta
\digamma_{s'}(\alpha,\beta)
}{\delta  \mathcal{N}_{s'}(\tau_{2}';\alpha)}
\bigg|_{u(\alpha-\beta)=0}
\frac{\delta 	\psi^{\dag}_{s'}(\tau_{2}';\alpha)\psi_{s'}(\tau_{2}';\beta)}{\delta 	\tau_{1}}
\bigg|_{u(\alpha-\beta)=0}
+1\right],\\
		& 
	\mathcal{G}_{L;s}		\frac{\delta 	\psi^{\dag}_{s}(\tau_{2};\alpha)\psi_{s}(\tau_{2};\beta)}{\delta 	\tau_{1}}
		\bigg|_{u(\alpha-\beta)=0}	\mathcal{G}_{R;s}	
=\left[\frac{\delta \digamma_{s}(\alpha,\beta)
}{\delta \mathcal{N}_{s}(\tau_{2};\alpha)}
\bigg|_{u(\alpha-\beta)=0}
\frac{\delta 	\psi^{\dag}_{s}(\tau_{2};\alpha)\psi_{s}(\tau_{2};\beta)}{\delta 	\tau_{1}}
\bigg|_{u(\alpha-\beta)=0}
+1\right].
	\end{aligned}
\end{equation}
where $\mathcal{G}_{R;s'}=e^{-{\rm ln}\mathcal{G}^{-1}_{R;s'}}=e^{-{\bf x}{\bf A}}$,
and we use the identity
$\mathcal{G}_{L;s'}\mathcal{G}^{-1}_{L;s'}\mathcal{G}^{-1}_{R;s'}\mathcal{G}_{R;s'}=1$ that can be obtained both diagrammatically (reducible) or functionally 
(irreducible; e.g., using the self-energy in Dyson function form).
However,
as shown below, this identity will be invalidated
in the case that we approximate,
for $\mathcal{G}_{L;s'}\hat{O}(\tau_{2}')\mathcal{G}_{R;s'}$,
the effect of identity operator
is just replacing the observable $\hat{O}(\tau_{2}')$ inserted in the middle, by that at $\tau_{1}$, i.e., $\hat{O}(\tau_{1})$.
Dividing by the susceptibility (functional derivative with respect to $\tau_{1}$ instead of the source field here) in both sides of above equation,
we arrive at
\begin{equation} 
	\begin{aligned}
		\label{6267}
		&
	\frac{	\mathcal{G}_{L;s'}		\frac{\delta 	\psi^{\dag}_{s'}(\tau_{2}';\alpha)\psi_{s'}(\tau_{2}';\beta)}{\delta 	\tau_{1}}
		\bigg|_{u(\alpha-\beta)=0}		\mathcal{G}_{R;s'}	}
	{	\frac{\delta 	\psi^{\dag}_{s'}(\tau_{2}';\alpha)\psi_{s'}(\tau_{2}';\beta)}{\delta 	\tau_{1}}
		\bigg|_{u(\alpha-\beta)=0}	}
		=
\frac{\delta
		\digamma_{s'}(\alpha,\beta)
		}{\delta \mathcal{N}_{s'}(\tau'_{2};\alpha)}
		\bigg|_{u(\alpha-\beta)=0}
+\left[\frac{\delta 	\psi^{\dag}_{s'}(\tau_{2}';\alpha)\psi_{s'}(\tau_{2}';\beta)}{\delta 	\tau_{1}}
\bigg|_{u(\alpha-\beta)=0}\right]^{-1}\\
&
=\frac{\mathcal{G}_{L;s'}}{\mathcal{G}_{R;s'}^{-1}}
\left[1+\sum_{n}\frac{{\bf x}^{n}}{n!}
\left(
\frac{\delta 	\psi^{\dag}_{s'}(\tau_{2}';\alpha)\psi_{s'}(\tau_{2}';\beta)}{\delta 	\tau_{1}}
\bigg|_{u(\alpha-\beta)=0}\right)^{(n)}
\left[\frac{\delta 	\psi^{\dag}_{s'}(\tau_{2}';\alpha)\psi_{s'}(\tau_{2}';\beta)}{\delta 	\tau_{1}}
\bigg|_{u(\alpha-\beta)=0}\right]^{-1}
\right]\\
&
=\frac{\mathcal{G}_{L;s'}}{\mathcal{G}_{R;s'}^{-1}}
\left[1+\sum_{n}\frac{{\bf x}^{n}}{n!}
\left({\rm ln}
\frac{\delta 	\psi^{\dag}_{s'}(\tau_{2}';\alpha)\psi_{s'}(\tau_{2}';\beta)}{\delta 	\tau_{1}}
\bigg|_{u(\alpha-\beta)=0}\right)^{(n)}
\right],\\
		&
		\frac{	\mathcal{G}_{L;s}		\frac{\delta 	\psi^{\dag}_{s}(\tau_{2};\alpha)\psi_{s}(\tau_{2};\beta)}{\delta 	\tau_{1}}
			\bigg|_{u(\alpha-\beta)=0}	\mathcal{G}_{R;s}	}
		{	\frac{\delta 	\psi^{\dag}_{s}(\tau_{2};\alpha)\psi_{s}(\tau_{2};\beta)}{\delta 	\tau_{1}}
			\bigg|_{u(\alpha-\beta)=0}}
=
	\frac{\delta \digamma_{s}(\alpha,\beta)
		}{\delta \mathcal{N}_{s}(\tau_{2};\alpha)}
		\bigg|_{u(\alpha-\beta)=0}
+\left[\frac{\delta 	\psi^{\dag}_{s}(\tau_{2};\alpha)\psi_{s}(\tau_{2};\beta)}{\delta 	\tau_{1}}
\bigg|_{u(\alpha-\beta)=0}\right]^{-1}\\
&=\frac{\mathcal{G}_{L;s}}{\mathcal{G}_{R;s}^{-1}}
\left[1+\sum_{n}\frac{{\bf x}^{n}}{n!}
\left(
\frac{\delta 	\psi^{\dag}_{s}(\tau_{2};\alpha)\psi_{s}(\tau_{2};\beta)}{\delta 	\tau_{1}}
	\bigg|_{u(\alpha-\beta)=0}\right)^{(n)}
	\left[\frac{\delta 	\psi^{\dag}_{s}(\tau_{2};\alpha)\psi_{s}(\tau_{2};\beta)}{\delta 	\tau_{1}}
	\bigg|_{u(\alpha-\beta)=0}\right]^{-1}
\right]\\
&=\frac{\mathcal{G}_{L;s}}{\mathcal{G}_{R;s}^{-1}}
\left[1+\sum_{n}\frac{{\bf x}^{n}}{n!}
\left({\rm ln}
\frac{\delta 	\psi^{\dag}_{s}(\tau_{2};\alpha)\psi_{s}(\tau_{2};\beta)}{\delta 	\tau_{1}}
\bigg|_{u(\alpha-\beta)=0}\right)^{(n)}
\right]\\
	\end{aligned}
\end{equation}
where $(\cdot)^{(n)}$ denotes the functional derivative with respect to ${\bf A}^{\dag}$ to $n$-th order,
and the last line is satisfied under the conditions
\begin{equation} 
	\begin{aligned}
	&	\left[	{\bf A},\left[\frac{\delta 	\psi^{\dag}_{s}(\tau_{2};\alpha)\psi_{s}(\tau_{2};\beta)}{\delta 	\tau_{1}}
		\bigg|_{u(\alpha-\beta)=0}\right]^{-1}\right]=0.
	\end{aligned}
\end{equation}
In language of mean-field approximation as well as the random-phase approximation,
the susceptibility here and the Eq.(\ref{6267})
are related to the noninteracting and interacting density-density response.

Eq.(\ref{61821}) then become
\begin{equation} 
	\begin{aligned}
		\label{6183}
		&\frac{\partial}{\partial\tau_{1}}
	\digamma_{s'}(\alpha,\beta)
		=[\psi^{\dag}_{s}(\tau_{1})]^{-1}\left[
\psi^{\dag}_{s}(\tau_{1})\mathcal{G}_{L;s'}
		\frac{\delta 	\psi^{\dag}_{s'}(\tau_{2}';\alpha)\psi_{s'}(\tau_{2}';\beta)}{\delta 	\tau_{1}}
		\bigg|_{u(\alpha-\beta)=0}\mathcal{G}_{R;s'}-\bf{I}[\psi_{s}^{\dag}(\tau_{1})]\right],\\
		&
		\frac{\partial}{\partial \tau_{1}}		\digamma_{s}(\alpha,\beta)
		=[\psi^{\dag}_{s'}(\tau_{1})]^{-1}\left[
\psi^{\dag}_{s'}(\tau_{1})\mathcal{G}_{L;s}
		\frac{\delta 	\psi^{\dag}_{s}(\tau_{2};\alpha)\psi_{s}(\tau_{2};\beta)}{\delta 	\tau_{1}}
		\bigg|_{u(\alpha-\beta)=0}\mathcal{G}_{R;s}-\bf{I}[\psi_{s'}^{\dag}(\tau_{1})]\right],\\
	\end{aligned}
\end{equation}
with
$\psi_{s}^{\dag}(\tau_{1})\mathcal{G}_{L;s'}\mathcal{G}^{-1}_{L;s'}\mathcal{G}^{-1}_{R;s'}\mathcal{G}_{R;s'}=\bf{I}[\psi_{s}^{\dag}(\tau_{1})]$ .
Substituting Eq.(\ref{6183}) into the last two terms in the expression of second order derivative,
\begin{equation} 
	\begin{aligned}
		\label{6212}
		\frac{\partial ^{(2)}\psi_{s}^{\dag}(\tau_{1})}{\partial \tau_{1}^{(2)}}
		&=
		\hat{H}^{*}_{1}\frac{\partial \psi_{s}^{\dag}(\tau_{1})}{\partial \tau_{1}}
		+
		\frac{\partial \psi^{\dag}_{s}(\tau_{1})}{\partial \tau_{1}}
			\int dr_{2}' U_{r_{1}-r_{2}'}    
		\psi^{\dag}_{s'}(\tau_{2}')\psi_{s'}(\tau_{2}')
		+
		\frac{\partial	\psi^{\dag}_{s'}(\tau_{1}) }{\partial \tau_{1}}
			\int dr_{2} U_{r_{2}-r_{1}}   
		\psi^{\dag}_{s}(\tau_{2}) \psi_{s}(\tau_{2})\\
	&			+
 \psi^{\dag}_{s}(\tau_{1})
	\frac{\partial}{\partial \tau_{1}}	\int dr_{2}' U_{r_{1}-r_{2}'}    
	\psi^{\dag}_{s'}(\tau_{2}')\psi_{s'}(\tau_{2}')
	+
	\psi^{\dag}_{s'}(\tau_{1}) 
	\frac{\partial}{\partial \tau_{1}}	\int dr_{2} U_{r_{2}-r_{1}}   
	\psi^{\dag}_{s}(\tau_{2}) \psi_{s}(\tau_{2}),
		\end{aligned}
\end{equation}
we obtain
	\begin{equation} 
		\begin{aligned}
			\label{6213}
	&			
\psi^{\dag}_{s}(\tau_{1})
\frac{\partial}{\partial \tau_{1}}\digamma_{s'}(\alpha,\beta)
=\int d\tau_{1}' \psi^{\dag}_{s}(\tau_{1}')
	\frac{\partial \psi^{\dag}_{s'}(\tau_{1}';\alpha)\psi_{s'}(\tau_{1}';\beta)}{\partial \tau'_{1}}	
	\bigg|_{u(\alpha-\beta)=0}-{\bf I}[\psi^{\dag}_{s}(\tau_{1})],\\
&
\psi^{\dag}_{s'}(\tau_{1}) 
\frac{\partial}{\partial \tau_{1}}	\digamma_{s}(\alpha,\beta)
=\int d\tau_{1}'\psi^{\dag}_{s'}(\tau_{1}') 
\frac{\partial \psi^{\dag}_{s}(\tau_{1}';\alpha) \psi_{s}(\tau_{1}';\beta)}
{\partial \tau'_{1}}		
\bigg|_{u(\alpha-\beta)=0}-{\bf I}[\psi^{\dag}_{s'}(\tau_{1})]
,
	\end{aligned}
\end{equation}
where the identity operator $\mathcal{G}_{R;s'}^{-1}\mathcal{G}_{R;s'}$ endows the susceptibility
($\frac{\delta 	\psi^{\dag}_{s'}(\tau_{2}';\alpha)\psi_{s'}(\tau_{2}';\beta)}{\delta 	\tau_{1}}$)
with the dependence on 
$\frac{\partial \psi^{\dag}(\tau_{1}')}{\partial \tau_{1}'}$
and the three-operator product
$\psi^{\dag}_{s'}(\tau_{1}') 
\psi^{\dag}_{s}(\tau_{1}';\alpha) \psi_{s}(\tau_{1}';\beta)$ 
(in terms of the Baker-Hausdorff lemma).

Eq.(\ref{6183}) and Eq.(\ref{6213}) are related by the following property
\begin{equation} 
	\begin{aligned}
		\label{6221}
&\mathcal{G}_{L;s'}
		\frac{\delta 	\psi^{\dag}_{s'}(\tau_{2}';\alpha)\psi_{s'}(\tau_{2}';\beta)}{\delta 	\tau_{1}}
		\bigg|_{u(\alpha-\beta)=0}\mathcal{G}_{R;s'}
=\left[\frac{\delta
			\digamma_{s'}(\alpha,\beta)
		}{\delta  \mathcal{N}_{s'}(\tau_{2}';\alpha)}
		\bigg|_{u(\alpha-\beta)=0}
		\frac{\delta 	\psi^{\dag}_{s'}(\tau_{2}';\alpha)\psi_{s'}(\tau_{2}';\beta)}{\delta 	\tau_{1}}
		\bigg|_{u(\alpha-\beta)=0}
		+\underline{1}\right]\\
		&
=[\psi_{s}^{\dag}(\tau_{1})]^{-1}\int d\tau_{1}' \psi^{\dag}_{s}(\tau_{1}')
\frac{\partial \psi^{\dag}_{s'}(\tau_{1}';\alpha)\psi_{s'}(\tau_{1}';\beta)}{\partial \tau'_{1}}	
\bigg|_{u(\alpha-\beta)=0}\\
		&
\stackrel{\boldsymbol{\cdot}}{=}
\frac{\partial \psi^{\dag}_{s'}(\tau_{1};\alpha)\psi_{s'}(\tau_{1};\beta)}{\partial \tau_{1}}	
\bigg|_{u(\alpha-\beta)=0},\\
		&
	\mathcal{G}_{L;s}
		\frac{\delta 	\psi^{\dag}_{s}(\tau_{2};\alpha)\psi_{s}(\tau_{2};\beta)}{\delta 	\tau_{1}}
		\bigg|_{u(\alpha-\beta)=0}\mathcal{G}_{R;s}
		=\left[\frac{\delta \digamma_{s}(\alpha,\beta)
		}{\delta \mathcal{N}_{s}(\tau_{2};\alpha)}
		\bigg|_{u(\alpha-\beta)=0}
		\frac{\delta 	\psi^{\dag}_{s}(\tau_{2};\alpha)\psi_{s}(\tau_{2};\beta)}{\delta 	\tau_{1}}
		\bigg|_{u(\alpha-\beta)=0}
		+\underline{1}\right]\\
		&=[	\psi^{\dag}_{s'}(\tau_{1})]^{-1}
\int d\tau_{1}'\psi^{\dag}_{s'}(\tau_{1}') 
\frac{\partial \psi^{\dag}_{s}(\tau_{1}';\alpha) \psi_{s}(\tau_{1}';\beta)}
{\partial \tau'_{1}}		
\bigg|_{u(\alpha-\beta)=0}\\
		&\stackrel{\boldsymbol{\cdot}}{=}
\frac{\partial \psi^{\dag}_{s}(\tau_{1};\alpha) \psi_{s}(\tau_{1};\beta)}
{\partial \tau_{1}}		
\bigg|_{u(\alpha-\beta)=0}.
	\end{aligned}
\end{equation}
Here the constant 1 with underline indicates the role of the above-mentioned
invalidated identity,
and we simply treat the integral over $\tau_{1}'$ (over all possible fluctuations around $\tau_{1}$) as $\tau_{1}'=\tau_{1}$,
which is conditionally allowed here, as will be clarified below,
and we use an equal sign with a bolded dot above it to denote this process.
Performing the integral over $\tau_{1}$ on above equations, we obtain
\begin{equation} 
	\begin{aligned}
		\label{6222}
		&\mathcal{G}_{L;s'}\int d\tau_{1}
		\frac{\delta 	\psi^{\dag}_{s'}(\tau_{2}';\alpha)\psi_{s'}(\tau_{2}';\beta)}{\delta 	\tau_{1}}
		\bigg|_{u(\alpha-\beta)=0}\mathcal{G}_{R;s'}\\
		&=\int d\tau_{1} \left[\frac{\delta
			\digamma_{s'}(\alpha,\beta)
		}{\delta  \mathcal{N}_{s'}(\tau_{2}';\alpha)}
		\bigg|_{u(\alpha-\beta)=0}
		\frac{\delta 	\psi^{\dag}_{s'}(\tau_{2}';\alpha)\psi_{s'}(\tau_{2}';\beta)}{\delta 	\tau_{1}}
		\bigg|_{u(\alpha-\beta)=0}
		+\underline{1}\right]
		=\int d\tau_{1}
		\frac{\partial \psi^{\dag}_{s'}(\tau_{1};\alpha)\psi_{s'}(\tau_{1};\beta)}{\partial \tau_{1}}	
		\bigg|_{u(\alpha-\beta)=0}\\
		&=\digamma_{s'}(\alpha,\beta)+\int d\tau_{1}[\psi^{\dag}_{s}(\tau_{1})]^{-1}{\bf I}[\psi^{\dag}_{s}(\tau_{1})]
		,\\
		&	
			\mathcal{G}_{L;s}\int d\tau_{1}
		\frac{\delta 	\psi^{\dag}_{s}(\tau_{2};\alpha)\psi_{s}(\tau_{2};\beta)}{\delta 	\tau_{1}}
		\bigg|_{u(\alpha-\beta)=0}\mathcal{G}_{R;s}\\
		&=	\int d\tau_{1} \left[\frac{\delta \digamma_{s}(\alpha,\beta)
		}{\delta \mathcal{N}_{s}(\tau_{2};\alpha)}
		\bigg|_{u(\alpha-\beta)=0}
		\frac{\delta 	\psi^{\dag}_{s}(\tau_{2};\alpha)\psi_{s}(\tau_{2};\beta)}{\delta 	\tau_{1}}
		\bigg|_{u(\alpha-\beta)=0}
		+\underline{1}\right]
		=
	\int d\tau_{1}	\frac{\partial \psi^{\dag}_{s}(\tau_{1};\alpha) \psi_{s}(\tau_{1};\beta)}
		{\partial \tau_{1}}		
		\bigg|_{u(\alpha-\beta)=0}\\
		&=\digamma_{s}(\alpha,\beta)+
		\int d\tau_{1}[\psi^{\dag}_{s'}(\tau_{1})]^{-1}{\bf I}[\psi^{\dag}_{s'}(\tau_{1})].
	\end{aligned}
\end{equation}

Eq.(\ref{6221}) can be equivalently rewritten as
\begin{equation} 
	\begin{aligned}
	\mathcal{G}_{R;s'}^{-1}
		\frac{\delta 	\psi^{\dag}_{s'}(\tau_{2}';\alpha)\psi_{s'}(\tau_{2}';\beta)}{\delta 	\tau_{1}}
		\bigg|_{u(\alpha-\beta)=0}\mathcal{G}_{R;s'}
		&=\mathcal{G}_{R;s'}^{-1}
	[\psi^{\dag}_{s}(\tau_{1})]^{-1}
	\mathcal{G}_{L;s'}^{-1}
	\int d\tau_{1}' \psi^{\dag}_{s}(\tau_{1}')
		\frac{\partial \psi^{\dag}_{s'}(\tau_{1}';\alpha)\psi_{s'}(\tau_{1}';\beta)}{\partial \tau'_{1}}	
		\bigg|_{u(\alpha-\beta)=0}\\
		&
				=\mathcal{G}_{R;s'}^{-1}
		\mathcal{G}_{L;s'}^{-1}
		\frac{\partial \psi^{\dag}_{s'}(\tau_{1};\alpha)\psi_{s'}(\tau_{1};\beta)}{\partial \tau_{1}}	
		\bigg|_{u(\alpha-\beta)=0}\\
		&
=\mathcal{G}_{R;s'}^{-1}
		\frac{\partial \psi^{\dag}_{s'}(\tau_{2}';\alpha)\psi_{s'}(\tau_{2}';\beta)}{\partial \tau_{1}}	
		\bigg|_{u(\alpha-\beta)=0}	\mathcal{G}_{R;s'},\\
\mathcal{G}_{R;s}^{-1}
		\frac{\delta 	\psi^{\dag}_{s}(\tau_{2};\alpha)\psi_{s}(\tau_{2};\beta)}{\delta 	\tau_{1}}
		\bigg|_{u(\alpha-\beta)=0}\mathcal{G}_{R;s}
		&=\mathcal{G}_{R;s}^{-1}
			[\psi^{\dag}_{s'}(\tau_{1})]^{-1}\mathcal{G}^{-1}_{L;s}
		\int d\tau_{1}'\psi^{\dag}_{s'}(\tau_{1}') 
		\frac{\partial \psi^{\dag}_{s}(\tau_{1}';\alpha) \psi_{s}(\tau_{1}';\beta)}
		{\partial \tau'_{1}}		
		\bigg|_{u(\alpha-\beta)=0}\\
		&		=\mathcal{G}_{R;s}^{-1}
\mathcal{G}^{-1}_{L;s}
		\frac{\partial \psi^{\dag}_{s}(\tau_{1};\alpha) \psi_{s}(\tau_{1};\beta)}
		{\partial \tau_{1}}		
		\bigg|_{u(\alpha-\beta)=0}\\
		&
					=\mathcal{G}_{R;s}^{-1}
		\frac{\partial \psi^{\dag}_{s}(\tau_{2};\alpha) \psi_{s}(\tau_{2};\beta)}
		{\partial \tau_{1}}		
		\bigg|_{u(\alpha-\beta)=0}		\mathcal{G}_{R;s}.
	\end{aligned}
\end{equation}

Next we substituting the Eq.(\ref{6183}) integrated over $\tau_{1}$
into the second and third terms of $\frac{\partial ^{(2)}\psi_{s}^{\dag}(\tau_{1})}{\partial \tau_{1}^{(2)}}$,
\begin{equation} 
	\begin{aligned}
		\label{6211}
		&	
		\frac{\partial \psi^{\dag}_{s}(\tau_{1})}{\partial \tau_{1}}
	\digamma_{s'}(\alpha,\beta)
		=\int d\tau_{1}' 
		\frac{\partial \psi^{\dag}_{s}(\tau_{1}')}{\partial \tau'_{1}}
		\psi^{\dag}_{s'}(\tau_{1}';\alpha)\psi_{s'}(\tau_{1}';\beta)
+{\bf I}[\psi^{\dag}_{s}(\tau_{1})],\\
&
		\frac{\partial	\psi^{\dag}_{s'}(\tau_{1}) }{\partial \tau_{1}}
		\digamma_{s}(\alpha,\beta)
=\int d\tau_{1}'
		\frac{\partial \psi^{\dag}_{s'}(\tau_{1}') }		{\partial \tau'_{1}}		
			\psi^{\dag}_{s}(\tau_{1}';\alpha) \psi_{s}(\tau_{1}';\beta)
+{\bf I}[\psi^{\dag}_{s'}(\tau_{1})],
	\end{aligned}
\end{equation}
where
\begin{equation} 
	\begin{aligned}
				&	
				[\psi^{\dag}_{s}(\tau_{1})]^{-1}
				\psi^{\dag}_{s}(\tau_{1})\mathcal{G}_{L;s'}
				\frac{\delta 	\psi^{\dag}_{s'}(\tau_{2}';\alpha)\psi_{s'}(\tau_{2}';\beta)}{\delta 	\tau_{1}}
				\bigg|_{u(\alpha-\beta)=0}\mathcal{G}_{R;s'}\\
				&=
				\frac{\partial}{\partial \tau_{1}}\left[
\left(	\frac{\partial \psi^{\dag}_{s}(\tau_{1})}{\partial \tau_{1}}\right)^{-1}
\int d\tau_{1}' 
		\frac{\partial \psi^{\dag}_{s}(\tau_{1}')}{\partial \tau'_{1}}
		\psi^{\dag}_{s'}(\tau_{1}';\alpha)\psi_{s'}(\tau_{1}';\beta)
\right],\\
&
[\psi^{\dag}_{s'}(\tau_{1})]^{-1}
\psi^{\dag}_{s'}(\tau_{1})\mathcal{G}_{L;s}
\frac{\delta 	\psi^{\dag}_{s}(\tau_{2};\alpha)\psi_{s}(\tau_{2};\beta)}{\delta 	\tau_{1}}
\bigg|_{u(\alpha-\beta)=0}\mathcal{G}_{R;s}\\
&=
\frac{\partial}{\partial \tau_{1}}\left[
\left(	\frac{\partial	\psi^{\dag}_{s'}(\tau_{1}) }{\partial \tau_{1}}\right)^{-1}
\int d\tau_{1}'
		\frac{\partial \psi^{\dag}_{s'}(\tau_{1}') }		{\partial \tau'_{1}}		
		\psi^{\dag}_{s}(\tau_{1}';\alpha) \psi_{s}(\tau_{1}';\beta)
\right],\\
		&	
		[\psi^{\dag}_{s}(\tau_{1})]^{-1}
{\bf I}[\psi^{\dag}_{s}(\tau_{1})]=\frac{\partial}{\partial \tau_{1}}
	\left[\left(	\frac{\partial \psi^{\dag}_{s}(\tau_{1})}{\partial \tau_{1}}\right)^{-1}{\bf I}[\psi^{\dag}_{s}(\tau_{1})]\right],\\
		&		[\psi^{\dag}_{s'}(\tau_{1})]^{-1}
{\bf I}[\psi^{\dag}_{s'}(\tau_{1})]
=\frac{\partial}{\partial \tau_{1}}
\left[\left(	\frac{\partial	\psi^{\dag}_{s'}(\tau_{1}) }{\partial \tau_{1}}\right)^{-1}{\bf I}[\psi^{\dag}_{s'}(\tau_{1})]
\right],
	\end{aligned}
\end{equation}

Combining Eqs.(\ref{6183},\ref{6211}),
\begin{equation} 
	\begin{aligned}
		\label{6214}
		\digamma_{s'}(\alpha,\beta)
		&=\int d\tau_{1} 
		[\psi^{\dag}_{s}(\tau_{1})]^{-1}\left[
		\psi^{\dag}_{s}(\tau_{1})\mathcal{G}_{L;s'}
		\frac{\delta 	\psi^{\dag}_{s'}(\tau_{2}';\alpha)\psi_{s'}(\tau_{2}';\beta)}{\delta 	\tau_{1}}
		\bigg|_{u(\alpha-\beta)=0}\mathcal{G}_{R;s'}-\bf{I}[\psi_{s}^{\dag}(\tau_{1})]\right]\\
		&=\int d\tau_{1} [\psi_{s}^{\dag}(\tau_{1})]^{-1}\left(\int d\tau_{1}' \psi^{\dag}_{s}(\tau_{1}')
		\frac{\partial \psi^{\dag}_{s'}(\tau_{1}';\alpha)\psi_{s'}(\tau_{1}';\beta)}{\partial \tau'_{1}}	
		\bigg|_{u(\alpha-\beta)=0}-{\bf I}[\psi^{\dag}_{s}(\tau_{1})]\right)\\		
		&=
\left[	\frac{\partial \psi^{\dag}_{s}(\tau_{1})}{\partial \tau_{1}}
\right]^{-1}\int d\tau_{1}' 
		\frac{\partial \psi^{\dag}_{s}(\tau_{1}')}{\partial \tau'_{1}}
		\psi^{\dag}_{s'}(\tau_{1}';\alpha)\psi_{s'}(\tau_{1}';\beta)
		+\left[	\frac{\partial \psi^{\dag}_{s}(\tau_{1})}{\partial \tau_{1}}
		\right]^{-1}
		{\bf I}[\psi^{\dag}_{s}(\tau_{1})]\\		
		&=\psi^{\dag}_{s'}(\tau_{1};\alpha)\psi_{s'}(\tau_{1};\alpha),\\
		\digamma_{s}(\alpha,\beta)
		&=\int d\tau_{1} 
		[\psi^{\dag}_{s'}(\tau_{1})]^{-1}\left[
		\psi^{\dag}_{s'}(\tau_{1})\mathcal{G}_{L;s}
		\frac{\delta 	\psi^{\dag}_{s}(\tau_{2};\alpha)\psi_{s}(\tau_{2};\beta)}{\delta 	\tau_{1}}
		\bigg|_{u(\alpha-\beta)=0}\mathcal{G}_{R;s}-\bf{I}[\psi_{s'}^{\dag}(\tau_{1})]\right]\\
		&=\int d\tau_{1}[\psi_{s'}^{\dag}(\tau_{1})]^{-1}\left(\int d\tau_{1}'\psi^{\dag}_{s'}(\tau_{1}') 
		\frac{\partial \psi^{\dag}_{s}(\tau_{1}';\alpha) \psi_{s}(\tau_{1}';\beta)}
		{\partial \tau'_{1}}		
		\bigg|_{u(\alpha-\beta)=0}-{\bf I}[\psi^{\dag}_{s'}(\tau_{1})]\right)\\
		&=	\left[	\frac{\partial	\psi^{\dag}_{s'}(\tau_{1}) }{\partial \tau_{1}}
\right]^{-1}\int d\tau_{1}'
		\frac{\partial \psi^{\dag}_{s'}(\tau_{1}') }		{\partial \tau'_{1}}		
		\psi^{\dag}_{s}(\tau_{1}';\alpha) \psi_{s}(\tau_{1}';\beta)
		+\left[	\frac{\partial	\psi^{\dag}_{s'}(\tau_{1}) }{\partial \tau_{1}}
		\right]^{-1}{\bf I}[\psi^{\dag}_{s'}(\tau_{1})]\\
		&=\psi^{\dag}_{s}(\tau_{1};\alpha)\psi_{s}(\tau_{1};\alpha).
	\end{aligned}
\end{equation}
Follow the above treatment for the integral over $\tau_{1}'$,
\begin{equation} 
	\begin{aligned}
		\label{6272}
		\digamma_{s'}(\alpha,\beta)
&\stackrel{\boldsymbol{\cdot}}{=}
		\psi^{\dag}_{s'}(\tau_{1};\alpha)\psi_{s'}(\tau_{1};\beta)
		+\left[	\frac{\partial \psi^{\dag}_{s}(\tau_{1})}{\partial \tau_{1}}
		\right]^{-1}
		{\bf I}[\psi^{\dag}_{s}(\tau_{1})]
		=\psi^{\dag}_{s'}(\tau_{1};\alpha)\psi_{s'}(\tau_{1};\alpha),\\
		\digamma_{s}(\alpha,\beta)
		&\stackrel{\boldsymbol{\cdot}}{=}
		\psi^{\dag}_{s}(\tau_{1};\alpha) \psi_{s}(\tau_{1};\beta)
		+\left[	\frac{\partial	\psi^{\dag}_{s'}(\tau_{1}) }{\partial \tau_{1}}
		\right]^{-1}{\bf I}[\psi^{\dag}_{s'}(\tau_{1})]
=\psi^{\dag}_{s}(\tau_{1};\alpha)\psi_{s}(\tau_{1};\alpha).
	\end{aligned}
\end{equation}

Also, the Eq.(\ref{6222}) provides the alternative expressions
\begin{equation} 
	\begin{aligned}
		\label{6223}
		\digamma_{s'}(\alpha,\beta)
		&=\int dr_{2}' U_{r_{1}-r_{2}'}\psi^{\dag}_{s'}(\tau_{2}';\alpha)\psi_{s'}(\tau_{2}';\beta)\\
	&=\psi^{\dag}_{s}(\tau_{1};\alpha)\psi_{s}(\tau_{1};\alpha)
=\int d\tau_{1}
	\frac{\partial \psi^{\dag}_{s'}(\tau_{1};\alpha)\psi_{s'}(\tau_{1};\beta)}{\partial \tau_{1}}	
	\bigg|_{u(\alpha-\beta)=0}-\int d\tau_{1}[\psi^{\dag}_{s}(\tau_{1})]^{-1}{\bf I}[\psi^{\dag}_{s}(\tau_{1})],\\
		\digamma_{s}(\alpha,\beta)
=&\int dr_{2}U_{r_{2}-r_{1}}\psi^{\dag}_{s}(\tau_{2};\alpha)\psi_{s}(\tau_{2};\beta)\\
&		=\psi^{\dag}_{s}(\tau_{1};\alpha)\psi_{s}(\tau_{1};\alpha)
	=	\int d\tau_{1}	\frac{\partial \psi^{\dag}_{s}(\tau_{1};\alpha) \psi_{s}(\tau_{1};\beta)}
		{\partial \tau_{1}}		
		\bigg|_{u(\alpha-\beta)=0}	-
		\int d\tau_{1}[\psi^{\dag}_{s'}(\tau_{1})]^{-1}{\bf I}[\psi^{\dag}_{s'}(\tau_{1})].
	\end{aligned}
\end{equation}
The reason that the integral over $\tau_{2}'$ (or $\tau_{2}$) counterintuitively
results in a well-defined bilinear,
is due to the nonlocal conserved mode about the coupled operator of $\alpha$ and $\beta$.
Ideally, we suppose such conservation forms a fluctuation around $\tau_{1}$, which can be collect by the dynamics at $\tau_{1}^{+}$,
\begin{equation} 
	\begin{aligned}
		\label{711}
	&	\digamma_{s'}(\alpha,\beta)=\int dr_{2}' U_{r_{1}-r_{2}'}\psi^{\dag}_{s'}(\tau_{2}';\alpha)\psi_{s'}(\tau_{2}';\beta)
=\psi^{\dag}_{s'}(\tau_{1};\alpha)\psi_{s'}(\tau_{1};\beta)
+\psi^{\dag}_{s'}(\tau_{1}^{+};\alpha)\psi_{s'}(\tau_{1}^{+};\beta)
=\psi^{\dag}_{s}(\tau_{1};\alpha)\psi_{s}(\tau_{1};\alpha),\\
&
\digamma_{s}(\alpha,\beta)=	\int dr_{2} U_{r_{2}-r_{1}}\psi^{\dag}_{s'}(\tau_{2}';\alpha)\psi_{s'}(\tau_{2}';\beta)
=\psi^{\dag}_{s}(\tau_{1};\alpha)\psi_{s}(\tau_{1};\beta)
+\psi^{\dag}_{s}(\tau_{1}^{+};\alpha)\psi_{s}(\tau_{1}^{+};\beta)
=\psi^{\dag}_{s}(\tau_{1};\alpha)\psi_{s}(\tau_{1};\alpha).
\end{aligned}
\end{equation}
That is why the function $\digamma_{s'(s)}(\alpha,\beta)=\mathcal{N}_{s'(s)}(\alpha,\beta)$ is still a function of both the $\alpha$ and $\beta$,
instead of only $\alpha$.
That also implies the stubborn correlation between the observables of the $\alpha$ and that of $\beta$.
In this way,
the fluctuation around $\tau_{1}$ can be easily counted in
by simply consider two points, $\tau_{1}$ and $\tau^{+}_{1}$,
which is benefit by the U(1) conservation in terms of the well-defined
fermion bilinear at $\tau_{1}$.
And from the Eq.(\ref{6272}),
we can obtain the fluctuational part,
\begin{equation} 
	\begin{aligned}
&
\psi^{\dag}_{s'}(\tau_{1}^{+};\alpha)\psi_{s'}(\tau_{1}^{+};\beta)
=
\left[	\frac{\partial \psi^{\dag}_{s}(\tau_{1})}{\partial \tau_{1}}
\right]^{-1}
{\bf I}[\psi^{\dag}_{s}(\tau_{1})],\\
		&
\psi^{\dag}_{s}(\tau_{1}^{+};\alpha)\psi_{s}(\tau_{1}^{+};\beta)
	=\left[	\frac{\partial	\psi^{\dag}_{s'}(\tau_{1}) }{\partial \tau_{1}}
	\right]^{-1}{\bf I}[\psi^{\dag}_{s'}(\tau_{1})].
	\end{aligned}
\end{equation}

\section{Appendix.C: Effective susceptibility and the functional derivative for effective self-energy with respect to normal propagator}

Comparing Eq.(\ref{6272}) and Eq.(\ref{6223}),
we can obtain
\begin{equation} 
	\begin{aligned}
		&
\frac{\partial \psi^{\dag}_{s'}(\tau_{1};\alpha)\psi_{s'}(\tau_{1};\beta)}{\partial \tau_{1}}-
\frac{\partial \psi^{\dag}_{s'}(\tau_{1};\alpha)\psi_{s'}(\tau_{1};\beta)}{\partial \tau_{1}}
\bigg|_{u(\alpha-\beta)=0}
=-\frac{\partial}{\partial\tau_{1}} 
\left[	\frac{\partial \psi^{\dag}_{s}(\tau_{1})}{\partial \tau_{1}}
\right]^{-1}
{\bf I}[\psi^{\dag}_{s}(\tau_{1})]
-[\psi^{\dag}_{s}(\tau_{1})]^{-1}{\bf I}[\psi^{\dag}_{s}(\tau_{1})],\\
&
\frac{\partial \psi^{\dag}_{s}(\tau_{1};\alpha) \psi_{s}(\tau_{1};\beta)}
{\partial \tau_{1}}-
\frac{\partial \psi^{\dag}_{s}(\tau_{1};\alpha) \psi_{s}(\tau_{1};\beta)}
{\partial \tau_{1}}		
\bigg|_{u(\alpha-\beta)=0}	=
-\frac{\partial}{\partial\tau_{1}} 
\left[	\frac{\partial	\psi^{\dag}_{s'}(\tau_{1}) }{\partial \tau_{1}}
\right]^{-1}{\bf I}[\psi^{\dag}_{s'}(\tau_{1})]
-[\psi^{\dag}_{s'}(\tau_{1})]^{-1}{\bf I}[\psi^{\dag}_{s'}(\tau_{1})].
\end{aligned}
\end{equation}

Performing the derivative with $\tau_{1}$ in both sides of Eq.(\ref{6223}),
\begin{equation} 
\begin{aligned}
	\label{6226}
\frac{\partial}{\partial \tau_{1}}	\digamma_{s'}(\alpha,\beta)
	&=\frac{\partial}{\partial \tau_{1}}	
	\psi^{\dag}_{s}(\tau_{1};\alpha)\psi_{s}(\tau_{1};\alpha)
	=
	\frac{\partial \psi^{\dag}_{s'}(\tau_{1};\alpha)\psi_{s'}(\tau_{1};\beta)}{\partial \tau_{1}}	
	\bigg|_{u(\alpha-\beta)=0}
	-\frac{\partial}{\partial \tau_{1}}	
	\int d\tau_{1}[\psi^{\dag}_{s}(\tau_{1})]^{-1}{\bf I}[\psi^{\dag}_{s}(\tau_{1})],\\
\frac{\partial}{\partial \tau_{1}}		\digamma_{s}(\alpha,\beta)
	&		=\frac{\partial}{\partial \tau_{1}}	
	\psi^{\dag}_{s}(\tau_{1};\alpha)\psi_{s}(\tau_{1};\alpha)
	=\frac{\partial \psi^{\dag}_{s}(\tau_{1};\alpha) \psi_{s}(\tau_{1};\beta)}
	{\partial \tau_{1}}		
	\bigg|_{u(\alpha-\beta)=0}	-\frac{\partial}{\partial \tau_{1}}	
	\int d\tau_{1}[\psi^{\dag}_{s'}(\tau_{1})]^{-1}{\bf I}[\psi^{\dag}_{s'}(\tau_{1})],
\end{aligned}
\end{equation}
thus the nonzero value of the last terms of the above two expressions,
or the invalidated identity mentioned-above is directly related to the differece
\begin{equation} 
	\begin{aligned}
		\label{6224}
	&	\frac{\partial \psi^{\dag}_{s'}(\tau_{1};\alpha)\psi_{s'}(\tau_{1};\beta)}{\partial \tau_{1}}	
		\bigg|_{u(\alpha-\beta)=0}
		-\frac{\partial}{\partial \tau_{1}}	
		\psi^{\dag}_{s}(\tau_{1};\alpha)\psi_{s}(\tau_{1};\alpha)=\frac{\partial \psi^{\dag}_{s'}(\tau_{1};\alpha)\psi_{s'}(\tau_{1};\beta)}{\partial \tau_{1}}	
\bigg|_{u(\alpha-\beta)=0}
-\frac{\partial 
\psi^{\dag}_{s}(\tau_{1};\alpha)\psi_{s}(\tau_{1};\beta)\bigg|_{u(\alpha-\beta)=0}}{\partial \tau_{1}}	
\\
	&	=
		\frac{\partial}{\partial \tau_{1}}	
		\int d\tau_{1}[\psi^{\dag}_{s}(\tau_{1})]^{-1}{\bf I}[\psi^{\dag}_{s}(\tau_{1})],\\
	&
\frac{\partial \psi^{\dag}_{s}(\tau_{1};\alpha) \psi_{s}(\tau_{1};\beta)}
		{\partial \tau_{1}}		
		\bigg|_{u(\alpha-\beta)=0}	-
			\frac{\partial}{\partial \tau_{1}}	
		\psi^{\dag}_{s}(\tau_{1};\alpha)\psi_{s}(\tau_{1};\alpha)
	=	\frac{\partial \psi^{\dag}_{s}(\tau_{1};\alpha) \psi_{s}(\tau_{1};\beta)}
		{\partial \tau_{1}}		
		\bigg|_{u(\alpha-\beta)=0}	-
		\frac{\partial
			\psi^{\dag}_{s}(\tau_{1};\alpha)\psi_{s}(\tau_{1};\beta)\bigg|_{u(\alpha-\beta)=0}}{\partial \tau_{1}}	
	\\
	&	=\frac{\partial}{\partial \tau_{1}}	
		\int d\tau_{1}[\psi^{\dag}_{s'}(\tau_{1})]^{-1}{\bf I}[\psi^{\dag}_{s'}(\tau_{1})].
	\end{aligned}
\end{equation}
As long as Eq.(\ref{6224}) is nonzero,
the identity ${\bf I}[\psi^{\dag}_{s'}(\tau_{1})]$ satisfies
\begin{equation} 
	\begin{aligned}
&
[\psi^{\dag}_{s}(\tau_{1})]^{-1}{\bf I}[\psi^{\dag}_{s}(\tau_{1})]
=[\psi^{\dag}_{s}(\tau_{1})]^{-1}(\psi^{\dag}_{s}(\tau_{1})
\mathcal{G}_{L;s'}\mathcal{G}^{-1}_{L;s'}\mathcal{G}^{-1}_{R;s'}\mathcal{G}_{R;s'})
\neq ([\psi^{\dag}_{s}(\tau_{1})]^{-1}\psi^{\dag}_{s}(\tau_{1}))
\mathcal{G}_{L;s'}\mathcal{G}^{-1}_{L;s'}\mathcal{G}^{-1}_{R;s'}\mathcal{G}_{R;s'}
= 1,\\
&
[\psi^{\dag}_{s'}(\tau_{1})]^{-1}{\bf I}[\psi^{\dag}_{s'}(\tau_{1})]
=[\psi^{\dag}_{s'}(\tau_{1})]^{-1}(\psi^{\dag}_{s'}(\tau_{1})
\mathcal{G}_{L;s}\mathcal{G}^{-1}_{L;s}\mathcal{G}^{-1}_{R;s}\mathcal{G}_{R;s})
\neq ([\psi^{\dag}_{s'}(\tau_{1})]^{-1}(\psi^{\dag}_{s'}(\tau_{1}))
\mathcal{G}_{L;s}\mathcal{G}^{-1}_{L;s}\mathcal{G}^{-1}_{R;s}\mathcal{G}_{R;s})
= 1.
	\end{aligned}
\end{equation}

Next we turn back to Eq.(\ref{6182}), which is equivalent to
the functional derivative of an effective self-energy with respect to a
trivial propagator (not at the time $\tau_{1}$)
\begin{equation} 
	\begin{aligned}
		\label{715}
		&
		\frac{\delta \digamma_{s'}(\alpha,\beta)}{\delta 	\mathcal{N}_{s'}(\tau_{2}';\alpha)}
		=		\frac{\delta \digamma_{s'}(\alpha,\beta)}{\delta {\rm ln}
			\mathcal{N}_{s'}(\tau_{2}';\alpha)}
		[\mathcal{N}_{s'}(\tau_{2}';\alpha)]^{-1}
				=	
						[\mathcal{N}_{s'}(\tau_{1};\alpha)]^{-1}
							\frac{\delta \digamma_{s'}(\alpha,\beta)}{\delta {\rm ln}
			\mathcal{N}_{s'}(\tau_{2}';\alpha)}
,\\
		&
		\frac{\delta \digamma_{s}(\alpha,\beta)}{\delta 	\mathcal{N}_{s}(\tau_{2};\alpha)}
		=		\frac{\delta \digamma_{s}(\alpha,\beta)}{\delta  {\rm ln}
			\mathcal{N}_{s}(\tau_{2};\alpha)}
		[\mathcal{N}_{s}(\tau_{2};\alpha)]^{-1}
				=		[\mathcal{N}_{s}(\tau_{1};\alpha)]^{-1}
					\frac{\delta \digamma_{s}(\alpha,\beta)}{\delta  {\rm ln}
			\mathcal{N}_{s}(\tau_{2};\alpha)}
		,
	\end{aligned}
\end{equation}
which imply the following commutation relations (related to the difference between normal propagators at $\tau_{1}$ and $\tau_{2}$)
\begin{equation} 
	\begin{aligned}
		&
\left[	\frac{\delta \digamma_{s'}(\alpha,\beta)}{\delta {\rm ln}
			\mathcal{N}_{s'}(\tau_{2}';\alpha)},
		[\mathcal{N}_{s'}(\tau_{2}';\alpha)]^{-1}\right]
		=	\left(
		[\mathcal{N}_{s'}(\tau_{1};\alpha)]^{-1}-		[\mathcal{N}_{s'}(\tau_{2}';\alpha)]^{-1}\right)
	\frac{\delta \digamma_{s'}(\alpha,\beta)}{\delta {\rm ln}
			\mathcal{N}_{s'}(\tau_{2}';\alpha)}
		\\
		&		=	\left(
	\digamma^{-1}_{s'}(\alpha,\beta)-		[\mathcal{N}_{s'}(\tau_{2}';\alpha)]^{-1}\right)
		\frac{\delta \digamma_{s'}(\alpha,\beta)}{\delta {\rm ln}
			\mathcal{N}_{s'}(\tau_{2}';\alpha)}
		\\
		&
\left[\frac{\delta \digamma_{s}(\alpha,\beta)}{\delta  {\rm ln}
			\mathcal{N}_{s}(\tau_{2};\alpha)},
		[\mathcal{N}_{s}(\tau_{2};\alpha)]^{-1}\right]
		=	\left(	[\mathcal{N}_{s}(\tau_{1};\alpha)]^{-1}-	[\mathcal{N}_{s}(\tau_{2};\alpha)]^{-1}\right)
		\frac{\delta \digamma_{s}(\alpha,\beta)}{\delta  {\rm ln}
			\mathcal{N}_{s}(\tau_{2};\alpha)}\\
		&		=	\left(\digamma^{-1}_{s}(\alpha,\beta)
		-	[\mathcal{N}_{s}(\tau_{2};\alpha)]^{-1}\right)
		\frac{\delta \digamma_{s}(\alpha,\beta)}{\delta  {\rm ln}
			\mathcal{N}_{s}(\tau_{2};\alpha)}
		,
	\end{aligned}
\end{equation}
and the corresponding self-energies satisfy (conditionally)
\begin{equation} 
	\begin{aligned}
		\hat{\Sigma}_{s'}(\alpha,\beta)
		&={\rm ln}\digamma_{s'}(\alpha,\beta),\\
		\hat{\Sigma}_{s}(\alpha,\beta)
		&={\rm ln}\digamma_{s}(\alpha,\beta),
	\end{aligned}
\end{equation}
with
\begin{equation} 
	\begin{aligned}
		&
[\mathcal{N}_{s'}(\tau_{2}';\alpha)]^{-1}=
\frac{1}{\digamma_{s'}(\alpha,\beta)}\sum^{\digamma_{s'}(\alpha,\beta)}_{\flat=1}\frac{1}{\flat}, \ 
	\left[	\frac{\delta \digamma_{s'}(\alpha,\beta)}{\delta {\rm ln}
			\mathcal{N}_{s'}(\tau_{2}';\alpha)}
		\right]^{-1}
		\digamma_{s'}(\alpha,\beta)
		\frac{\delta \digamma[{\rm ln}\mathcal{N}_{s'}(\tau_{2}';\alpha)]}
		{\delta {\rm ln}\mathcal{N}_{s'}(\tau_{2}';\alpha)}
		=\frac{\upgamma}{\digamma_{s'}(\alpha,\beta)},\\
&
	[\mathcal{N}_{s}(\tau_{2};\alpha)]^{-1}=\frac{1}{\digamma_{s}(\alpha,\beta)}\sum^{\digamma_{s}(\alpha,\beta)}_{\flat=1}\frac{1}{\flat}, \  
		\left[
		\frac{\delta \digamma_{s}(\alpha,\beta)}{\delta {\rm ln}
			\mathcal{N}_{s}(\tau_{2};\alpha)}
		\right]^{-1}
		\digamma_{s}(\alpha,\beta)
		\frac{\delta \digamma[{\rm ln}\mathcal{N}_{s}(\tau_{2};\alpha)]}
		{\delta {\rm ln}\mathcal{N}_{s}(\tau_{2};\alpha)}
		=\frac{\upgamma}{\digamma_{s}(\alpha,\beta)},
	\end{aligned}
\end{equation}
where $\flat$ is an integer,
and $\upgamma$ is the Euler-Mascheroni constant.
Performing the functional derivative with respect to $\mathcal{N}_{s'}(\tau_{1};\alpha)\ 
(\mathcal{N}_{s}(\tau_{1};\alpha))$ on both sides,
we obtain
\begin{equation} 
	\begin{aligned}
		&
		\frac{\delta \mathcal{N}_{s'}(\tau_{2}';\alpha)}
		{\delta \mathcal{N}_{s'}(\tau_{1};\alpha)}
				\frac{\mathcal{N}_{s'}(\tau_{1};\alpha)}
		{ \mathcal{N}_{s'}(\tau_{2}';\alpha)}=1-
	\mathcal{N}_{s'}(\tau_{2}';\alpha)
	\left[\frac{1}{	\mathcal{N}_{s'}(\tau_{1};\alpha)}
	-
\frac{1}{	\mathcal{N}_{s'}^{2}(\tau_{1};\alpha)}\right],\\
&
		\frac{\delta \mathcal{N}_{s}(\tau_{2};\alpha)}
{\delta \mathcal{N}_{s}(\tau_{1};\alpha)}
\frac{\mathcal{N}_{s}(\tau_{1};\alpha)}
{ \mathcal{N}_{s}(\tau_{2};\alpha)}=1-
\mathcal{N}_{s}(\tau_{2};\alpha)
\left[\frac{1}{	\mathcal{N}_{s}(\tau_{1};\alpha)}
-
\frac{1}{	\mathcal{N}_{s}^{2}(\tau_{1};\alpha)}\right],
	\end{aligned}
\end{equation}
which further leads to
\begin{equation} 
	\begin{aligned}
		&
		\frac{\delta \mathcal{N}_{s'}(\tau_{2}';\alpha)}
		{\delta \mathcal{N}_{s'}(\tau_{1};\alpha)}
	=	\frac{\mathcal{N}_{s'}(\tau_{2}';\alpha)}
	{ \mathcal{N}_{s'}(\tau_{1};\alpha)}-
		\mathcal{N}^{2}_{s'}(\tau_{2}';\alpha)
		\left[\frac{1}{	\mathcal{N}^{2}_{s'}(\tau_{1};\alpha)}
		-
		\frac{1}{	\mathcal{N}_{s'}^{3}(\tau_{1};\alpha)}\right],\\
		&
		\frac{\delta \mathcal{N}_{s}(\tau_{2};\alpha)}
		{\delta \mathcal{N}_{s}(\tau_{1};\alpha)}
	=	\frac{\mathcal{N}_{s}(\tau_{2};\alpha)}
	{ \mathcal{N}_{s}(\tau_{1};\alpha)}
	-
		\mathcal{N}_{s}^{2}(\tau_{2};\alpha)
		\left[\frac{1}{	\mathcal{N}^{2}_{s}(\tau_{1};\alpha)}
		-
		\frac{1}{	\mathcal{N}_{s}^{3}(\tau_{1};\alpha)}\right],
	\end{aligned}
\end{equation}

Inserting the result $\digamma[{\rm ln}\mathcal{N}_{s'}(\tau_{2}';\alpha)]=\frac{{\rm ln} \digamma_{s'}(\alpha,\beta)}
{\digamma_{s'}(\alpha,\beta)}$,
we obtain
\begin{equation} 
	\begin{aligned}
		&
		\left[	\frac{\delta \digamma_{s'}(\alpha,\beta)}{\delta {\rm ln}
			\mathcal{N}_{s'}(\tau_{2}';\alpha)}
		\right]^{-1}
		\digamma_{s'}(\alpha,\beta)
		\frac{\delta \digamma[{\rm ln}\mathcal{N}_{s'}(\tau_{2}';\alpha)]}
		{\delta {\rm ln}\mathcal{N}_{s'}(\tau_{2}';\alpha)}
		\\
		&
		=		
		\left[	\frac{\delta \digamma_{s'}(\alpha,\beta)}{\delta 
			\mathcal{N}_{s'}(\tau_{2}';\alpha)}
		\right]^{-1}
	\frac{	\digamma_{s'}(\alpha,\beta)}{	\mathcal{N}_{s'}(\tau_{2}';\alpha)}
	\left(
	\frac{\delta {\rm ln} \digamma_{s'}(\alpha,\beta)}
		{\delta {\rm ln}\mathcal{N}_{s'}(\tau_{2}';\alpha)}
		\frac{1}{\digamma_{s'}(\alpha,\beta)}
		+ 
		 {\rm ln}\digamma_{s'}(\tau_{2}';\alpha)
	 	\frac{\delta 	\frac{1}{\digamma_{s'}(\alpha,\beta)}}
 	{\delta {\rm ln}\mathcal{N}_{s'}(\tau_{2}';\alpha)}
		\right)\\
	&
		=\frac{\upgamma}{\digamma_{s'}(\alpha,\beta)},\\
		&
		\left[
		\frac{\delta \digamma_{s}(\alpha,\beta)}{\delta {\rm ln}
			\mathcal{N}_{s}(\tau_{2};\alpha)}
		\right]^{-1}
		\digamma_{s}(\alpha,\beta)
		\frac{\delta \digamma[{\rm ln}\mathcal{N}_{s}(\tau_{2};\alpha)]}
		{\delta {\rm ln}\mathcal{N}_{s}(\tau_{2};\alpha)}\\
		&
		=		\left[
		\frac{\delta \digamma_{s}(\alpha,\beta)}{\delta
			\mathcal{N}_{s}(\tau_{2};\alpha)}
		\right]^{-1}
		\frac{\digamma_{s}(\alpha,\beta)}{	\mathcal{N}_{s}(\tau_{2};\alpha)}
\left(\frac{\delta {\rm ln}\digamma_{s}(\alpha,\beta)}	
{\delta {\rm ln}\mathcal{N}_{s}(\tau_{2};\alpha)}
\frac{1}{\digamma_{s}(\alpha,\beta)}
+{\rm ln}	\digamma_{s}(\tau_{2};\alpha)
\frac{\delta \frac{1}{\digamma_{s}(\alpha,\beta)}}
		{\delta {\rm ln}\mathcal{N}_{s}(\tau_{2};\alpha)}\right)\\
		&
		=\frac{\upgamma}{\digamma_{s}(\alpha,\beta)}.
	\end{aligned}
\end{equation}
Under the commuting assumption, the above equations can be reduced to
\begin{equation} 
	\begin{aligned}
	&
		\frac{1}{\mathcal{N}_{s'}(\tau_{2}';\alpha)}
		\left(1-
		{\rm ln}\digamma_{s'}(\tau_{2}';\alpha)\right)
		=\frac{\upgamma}{\digamma_{s'}(\alpha,\beta)},\\
	&
\frac{1}{\mathcal{N}_{s}(\tau_{2};\alpha)}
\left(1-
{\rm ln}\digamma_{s}(\tau_{2};\alpha)\right)
		=\frac{\upgamma}{\digamma_{s}(\alpha,\beta)},
	\end{aligned}
\end{equation}
where we have
\begin{equation} 
	\begin{aligned}
		&
		\upgamma=
		\frac{1}{\mathcal{N}_{s'}(\tau_{2}';\alpha)}
		\left(\digamma_{s'}(\alpha,\beta)-
\digamma_{s'}(\alpha,\beta){\rm ln}\digamma_{s'}(\tau_{2}';\alpha)\right)
=
		\frac{1}{\mathcal{N}_{s}(\tau_{2};\alpha)}
		\left(\digamma_{s}(\alpha,\beta)-
\digamma_{s}(\alpha,\beta){\rm ln}\digamma_{s}(\tau_{2};\alpha)\right).
	\end{aligned}
\end{equation}

\section{Appendix.D: Cutoff condition and the approximately integrable case:
$[\psi^{\dag}_{s}(\tau_{1})]^{-1}{\bf I}[\psi^{\dag}_{s}(\tau_{1})]=1$}

Substituting Eqs.(\ref{6211},\ref{6213}) back to the Eq.(\ref{6212}),
\begin{equation} 
	\begin{aligned}
		\frac{\partial ^{(2)}\psi_{s}^{\dag}(\tau_{1})}{\partial \tau_{1}^{(2)}}
		&=
		\hat{H}^{*}_{1}\frac{\partial \psi_{s}^{\dag}(\tau_{1})}{\partial \tau_{1}}
		+
\psi^{\dag}_{s}(\tau_{1})
		\psi^{\dag}_{s'}(\tau_{1}';\alpha)\psi_{s'}(\tau_{1}';\beta)
		+
\psi^{\dag}_{s'}(\tau_{1}) 
		\psi^{\dag}_{s}(\tau_{1}';\alpha) \psi_{s}(\tau_{1}';\beta)=0,
	\end{aligned}
\end{equation}
and we note that the first order derivative reads
\begin{equation} 
	\begin{aligned}
				\frac{\partial \psi_{s}^{\dag}(\tau_{1})}{\partial \tau_{1}}&=
				\hat{H}^{*}_{1}\psi_{s}^{\dag}(\tau_{1})
				+
				\int dr_{2}' U_{r_{1}-r_{2}'}    
				\psi^{\dag}_{s}(\tau_{1}) \psi^{\dag}_{s'}(\tau_{2}')\psi_{s'}(\tau_{2}')
				+
				\int dr_{2} U_{r_{2}-r_{1}}   
				\psi^{\dag}_{s'}(\tau_{1}) 	\psi^{\dag}_{s}(\tau_{2}) \psi_{s}(\tau_{2})\\
				&=
				\hat{H}^{*}_{1}\psi_{s}^{\dag}(\tau_{1})
				+
				\psi^{\dag}_{s}(\tau_{1}) \mathcal{N}_{s'}(\tau_{1},\alpha)
				+
				\psi^{\dag}_{s'}(\tau_{1})  \mathcal{N}_{s}(\tau_{1},\alpha).
	\end{aligned}
\end{equation}
For the special point $\tau_{1}'$
which represents a collection of all time other than $\tau_{1}$
(but restricted by the correlation between the operators of $\alpha$ and that of $\beta$, see Eq.(\ref{6214})),
we have
$	\psi^{\dag}_{s'}(\tau_{1}';\alpha)\psi_{s'}(\tau_{1}';\beta)=\mathcal{N}_{s'}(\tau_{1},\alpha)$,
$	\psi^{\dag}_{s}(\tau_{1}';\alpha) \psi_{s}(\tau_{1}';\beta)=\mathcal{N}_{s}(\tau_{1},\alpha)$,
and such special point can be identified by the following equation,
\begin{equation} 
	\begin{aligned}
		\hat{H}^{*}_{1}\frac{\partial \psi_{s}^{\dag}(\tau_{1})}{\partial \tau_{1}}
=
		\hat{H}^{*}_{1}\psi_{s}^{\dag}(\tau_{1})-	\frac{\partial \psi_{s}^{\dag}(\tau_{1})}{\partial \tau_{1}}.
		\end{aligned}
\end{equation}
Then we can obtain another expression related to an infinite sum
\begin{equation} 
	\begin{aligned}
		\label{6227}
	\frac{\partial {\rm ln}\psi_{s}^{\dag}(\tau_{1})}{\partial \tau_{1}}
		=
		\frac{\hat{H}^{*}_{1}}{\hat{H}^{*}_{1}+1},
	\end{aligned}
\end{equation}
which is valid only in the case of $
[H^{*}_{1},	\psi_{s}^{\dag}(\tau_{1})]
=[H^{*}_{1},	\frac{\partial \psi_{s}^{\dag}(\tau_{1})}{\partial \tau_{1}}]=0$.
This statement corresponds to the case where the identity is still valid in Eq.(\ref{6225}),
and correspondingly, Eq.(\ref{6222}) becomes
\begin{equation} 
	\begin{aligned}
		&\int d\tau_{1}
		\mathcal{G}_{L;s'}
		\frac{\delta 	\psi^{\dag}_{s'}(\tau_{2}';\alpha)\psi_{s'}(\tau_{2}';\beta)}{\delta 	\tau_{1}}
		\bigg|_{u(\alpha-\beta)=0}\mathcal{G}_{R;s'}
		=\int d\tau_{1}
		\left[\frac{\delta
			\digamma_{s'}(\alpha,\beta)
		}{\delta  \mathcal{N}_{s'}(\tau_{2}';\alpha)}
		\bigg|_{u(\alpha-\beta)=0}
		\frac{\delta 	\psi^{\dag}_{s'}(\tau_{2}';\alpha)\psi_{s'}(\tau_{2}';\beta)}{\delta 	\tau_{1}}
		\bigg|_{u(\alpha-\beta)=0}
		+1\right]\\
		&
		=\int d\tau_{1} [\psi_{s}^{\dag}(\tau_{1})]^{-1}\int d\tau_{1}' \psi^{\dag}_{s}(\tau_{1}')
		\frac{\partial \psi^{\dag}_{s'}(\tau_{1}';\alpha)\psi_{s'}(\tau_{1}';\beta)}{\partial \tau'_{1}}	
		\bigg|_{u(\alpha-\beta)=0}\\
		&
		=
		\int d\tau_{1}\frac{\partial \psi^{\dag}_{s'}(\tau_{1};\alpha)\psi_{s'}(\tau_{1};\beta)}{\partial \tau_{1}}	
		\bigg|_{u(\alpha-\beta)=0}
		+\int d\tau_{1}
		\frac{\psi^{\dag}_{s}(\tau_{1}^{+})}{\psi_{s}^{\dag}(\tau_{1})}	\frac{\partial \psi^{\dag}_{s'}(\tau_{1}^{+};\alpha)\psi_{s'}(\tau_{1}^{+};\beta)}{\partial \tau_{1}^{+}}	
		\bigg|_{u(\alpha-\beta)=0}\\
		&=\digamma_{s'}(\alpha,\beta)+\int d\tau_{1}
		,\\
		&
	\int d\tau_{1}	\mathcal{G}_{L;s}
		\frac{\delta 	\psi^{\dag}_{s}(\tau_{2};\alpha)\psi_{s}(\tau_{2};\beta)}{\delta 	\tau_{1}}
		\bigg|_{u(\alpha-\beta)=0}\mathcal{G}_{R;s}
		=\int d\tau_{1}\left[\frac{\delta \digamma_{s}(\alpha,\beta)
		}{\delta \mathcal{N}_{s}(\tau_{2};\alpha)}
		\bigg|_{u(\alpha-\beta)=0}
		\frac{\delta 	\psi^{\dag}_{s}(\tau_{2};\alpha)\psi_{s}(\tau_{2};\beta)}{\delta 	\tau_{1}}
		\bigg|_{u(\alpha-\beta)=0}
		+1\right]\\
		&=\int d\tau_{1}
		[	\psi^{\dag}_{s'}(\tau_{1})]^{-1}
		\int d\tau_{1}'\psi^{\dag}_{s'}(\tau_{1}') 
		\frac{\partial \psi^{\dag}_{s}(\tau_{1}';\alpha) \psi_{s}(\tau_{1}';\beta)}
		{\partial \tau'_{1}}		
		\bigg|_{u(\alpha-\beta)=0}\\
		&=\int d\tau_{1}
		\frac{\partial \psi^{\dag}_{s}(\tau_{1};\alpha) \psi_{s}(\tau_{1};\beta)}
		{\partial \tau_{1}}		
		\bigg|_{u(\alpha-\beta)=0}
		+	\int d\tau_{1}\frac{\psi^{\dag}_{s'}(\tau_{1}^{+})}{\psi_{s'}^{\dag}(\tau_{1})}
		\frac{\partial \psi^{\dag}_{s}(\tau_{1}^{+};\alpha) \psi_{s}(\tau_{1}^{+};\beta)}
		{\partial \tau_{1}^{+}}		
		\bigg|_{u(\alpha-\beta)=0}\\
				&=\digamma_{s}(\alpha,\beta)+\int d\tau_{1}
		,
	\end{aligned}
\end{equation}
where the integral for the trivial constant 1 in the last line turns to be $\tau_{1}$-independent, thus by performing the $\partial_{\tau_{1}}$ on both sides, we obtain
\begin{equation} 
	\begin{aligned}
		&
		\frac{\partial \psi^{\dag}_{s'}(\tau_{1};\alpha)\psi_{s'}(\tau_{1};\beta)}{\partial \tau_{1}}	
		\bigg|_{u(\alpha-\beta)=0}+
		\frac{\psi^{\dag}_{s}(\tau_{1}^{+})}{\psi_{s}^{\dag}(\tau_{1})}	\frac{\partial \psi^{\dag}_{s'}(\tau_{1}^{+};\alpha)\psi_{s'}(\tau_{1}^{+};\beta)}{\partial \tau_{1}^{+}}	
		\bigg|_{u(\alpha-\beta)=0}\\
		&
	=		\frac{\partial \psi^{\dag}_{s'}(\tau_{1};\alpha)\psi_{s'}(\tau_{1};\beta)}{\partial \tau_{1}}	
	+
		\frac{\partial \psi^{\dag}_{s'}(\tau_{1}^{+};\alpha)\psi_{s'}(\tau_{1}^{+};\beta)}{\partial \tau_{1}}	
		=	\frac{\partial \psi^{\dag}_{s'}(\tau_{1};\alpha)\psi_{s'}(\tau_{1};\beta)}{\partial \tau_{1}}	
		+
		\frac{\partial}{\partial\tau_{1}}
	\left(	\left[	\frac{\partial \psi^{\dag}_{s}(\tau_{1})}{\partial \tau_{1}}
		\right]^{-1}
		{\bf I}[\psi^{\dag}_{s}(\tau_{1})]\right)
,\\
		&
				\frac{\partial \psi^{\dag}_{s}(\tau_{1};\alpha) \psi_{s}(\tau_{1};\beta)}
		{\partial \tau_{1}}		
		\bigg|_{u(\alpha-\beta)=0}+
	\frac{\psi^{\dag}_{s'}(\tau_{1}'^{+})}{\psi_{s'}^{\dag}(\tau_{1})}
		\frac{\partial \psi^{\dag}_{s}(\tau_{1}^{+};\alpha) \psi_{s}(\tau_{1}^{+};\beta)}
		{\partial \tau_{1}^{+}}		
		\bigg|_{u(\alpha-\beta)=0}\\
		&=
				\frac{\partial \psi^{\dag}_{s}(\tau_{1};\alpha) \psi_{s}(\tau_{1};\beta)}
		{\partial \tau_{1}}		
		+
		\frac{\partial \psi^{\dag}_{s}(\tau_{1}^{+};\alpha) \psi_{s}(\tau_{1}^{+};\beta)}
		{\partial \tau_{1}}		
=
				\frac{\partial \psi^{\dag}_{s}(\tau_{1};\alpha) \psi_{s}(\tau_{1};\beta)}
{\partial \tau_{1}}		
+
\frac{\partial}{\partial\tau_{1}}\left(
\left[	\frac{\partial	\psi^{\dag}_{s'}(\tau_{1}) }{\partial \tau_{1}}
\right]^{-1}{\bf I}[\psi^{\dag}_{s'}(\tau_{1})]\right),
	\end{aligned}
\end{equation}
in which case it requires
\begin{equation} 
	\begin{aligned}
&	[\psi^{\dag}_{s}(\tau_{1})]^{-1}{\bf I}[\psi^{\dag}_{s}(\tau_{1})]=1,\\
&	[\psi^{\dag}_{s'}(\tau_{1})]^{-1}{\bf I}[\psi^{\dag}_{s'}(\tau_{1})]=1,
\end{aligned}
\end{equation}
or equivalently,
\begin{equation} 
	\begin{aligned}
		&
 \left[\psi^{\dag}_{s}(\tau_{1}),
		\mathcal{G}_{L;s'}\mathcal{G}^{-1}_{L;s'}\mathcal{G}^{-1}_{R;s'}\mathcal{G}_{R;s'}\right]=0,\\
		&
\left[\psi^{\dag}_{s'}(\tau_{1}),
		\mathcal{G}_{L;s}\mathcal{G}^{-1}_{L;s}\mathcal{G}^{-1}_{R;s}\mathcal{G}_{R;s}\right]=0,
	\end{aligned}
\end{equation}

Further, using Eqs.(\ref{6182},\ref{6226}),
\begin{equation} 
	\begin{aligned}
		\label{62611}
		& 
		\frac{\partial 	\digamma_{s'}(\alpha,\beta)}{\partial \tau_{1}}=
\frac{\delta
			\digamma_{s'}(\alpha,\beta)
		}{\delta  \mathcal{N}_{s'}(\tau_{2}';\alpha)}
		\bigg|_{u(\alpha-\beta)=0}
		\frac{\delta 	\psi^{\dag}_{s'}(\tau_{2}';\alpha)\psi_{s'}(\tau_{2}';\beta)}{\delta 	\tau_{1}}
		\bigg|_{u(\alpha-\beta)=0}
=\frac{\delta
{\rm ln}	\digamma_{s'}(\alpha,\beta)
}{\delta {\rm ln} \mathcal{N}_{s'}(\tau_{2}';\alpha)}	\bigg|_{u(\alpha-\beta)=0}		
\frac{\delta 	\psi^{\dag}_{s'}(\tau_{2}';\alpha)\psi_{s'}(\tau_{2}';\beta)}{\delta 	\tau_{1}}
\bigg|_{u(\alpha-\beta)=0}\\
&		=
\frac{\partial \psi^{\dag}_{s'}(\tau_{1};\alpha)\psi_{s'}(\tau_{1};\beta)}{\partial \tau_{1}}	
\bigg|_{u(\alpha-\beta)=0}
-[\psi^{\dag}_{s}(\tau_{1})]^{-1}{\bf I}[\psi^{\dag}_{s}(\tau_{1})],\\
		& 
\frac{\partial 	\digamma_{s}(\alpha,\beta)}{\partial \tau_{1}}=
\frac{\delta \digamma_{s}(\alpha,\beta)
		}{\delta \mathcal{N}_{s}(\tau_{2};\alpha)}
		\bigg|_{u(\alpha-\beta)=0}
		\frac{\delta 	\psi^{\dag}_{s}(\tau_{2};\alpha)\psi_{s}(\tau_{2};\beta)}{\delta 	\tau_{1}}
		\bigg|_{u(\alpha-\beta)=0}
		=\frac{\delta {\rm ln}\digamma_{s}(\alpha,\beta)
		}{\delta {\rm ln}\mathcal{N}_{s}(\tau_{2};\alpha)}
		\bigg|_{u(\alpha-\beta)=0}
				\frac{\delta 	\psi^{\dag}_{s}(\tau_{2};\alpha)\psi_{s}(\tau_{2};\beta)}{\delta 	\tau_{1}}
		\bigg|_{u(\alpha-\beta)=0}\\
		&=\frac{\partial \psi^{\dag}_{s}(\tau_{1};\alpha) \psi_{s}(\tau_{1};\beta)}
		{\partial \tau_{1}}		
		\bigg|_{u(\alpha-\beta)=0}	-
	[\psi^{\dag}_{s'}(\tau_{1})]^{-1}{\bf I}[\psi^{\dag}_{s'}(\tau_{1})].
	\end{aligned}
\end{equation}
Then under above assumption where $\int d\tau_{1}'f(\tau_{1}')=f(\tau_{1})+f(\tau_{1}^{+})$,
\begin{equation} 
	\begin{aligned}
		&		\frac{\delta
			\digamma_{s'}(\alpha,\beta)
		}{\delta  \mathcal{N}_{s'}(\tau_{2}';\alpha)}
		\bigg|_{u(\alpha-\beta)=0}=\frac{
		\frac{\partial \psi^{\dag}_{s'}(\tau_{1};\alpha)\psi_{s'}(\tau_{1};\beta)}{\partial \tau_{1}}	
		\bigg|_{u(\alpha-\beta)=0}
		-1}{	\frac{\partial \psi^{\dag}_{s'}(\tau_{1};\alpha)\psi_{s'}(\tau_{1};\beta)}{\partial \tau_{1}}	
		\bigg|_{u(\alpha-\beta)=0}}
	=\frac{	\frac{\partial 	\digamma_{s'}(\alpha,\beta)}{\partial \tau_{1}}}{	\frac{\partial 	\digamma_{s'}(\alpha,\beta)}{\partial \tau_{1}}+1},\\
		&
		\frac{\delta \digamma_{s}(\alpha,\beta)
		}{\delta \mathcal{N}_{s}(\tau_{2};\alpha)}
		\bigg|_{u(\alpha-\beta)=0}=\frac{
		\frac{\partial \psi^{\dag}_{s}(\tau_{1};\alpha) \psi_{s}(\tau_{1};\beta)}
		{\partial \tau_{1}}		
		\bigg|_{u(\alpha-\beta)=0}	-
	1}{
		\frac{\partial \psi^{\dag}_{s}(\tau_{1};\alpha) \psi_{s}(\tau_{1};\beta)}
		{\partial \tau_{1}}		
		\bigg|_{u(\alpha-\beta)=0}}
	=\frac{	\frac{\partial 	\digamma_{s}(\alpha,\beta)}{\partial \tau_{1}}}
	{	\frac{\partial 	\digamma_{s}(\alpha,\beta)}{\partial \tau_{1}}+1}.
	\end{aligned}
\end{equation}
Similar to Eq.(\ref{6227}),
\begin{equation} 
	\begin{aligned}
		\label{62612}
		&		\frac{\delta
			\digamma_{s'}(\alpha,\beta)
		}{\delta  \mathcal{N}_{s'}(\tau_{2}';\alpha)}
		\bigg|_{u(\alpha-\beta)=0}
		=\sum_{\flat=0}^{\infty}\left[-\left(\frac{\partial 	\digamma_{s'}(\alpha,\beta)}{\partial \tau_{1}}\right)^{-1}\right]^{\flat}
	,\\
		&
		\frac{\delta \digamma_{s}(\alpha,\beta)
		}{\delta \mathcal{N}_{s}(\tau_{2};\alpha)}
		\bigg|_{u(\alpha-\beta)=0}
	=\sum_{\flat=0}^{\infty}\left[-\left(\frac{\partial 	\digamma_{s}(\alpha,\beta)}{\partial \tau_{1}}\right)^{-1}\right]^{\flat}.
	\end{aligned}
\end{equation}

\section{Appendix.E: Susceptibility as the projector to structureless observable:
From propagator to self-energy}

Consider the diagonal contribution in self-energy matrix $\pmb{\Sigma}$,
\begin{equation} 
	\begin{aligned}
		\label{673}
	&	\sum_{\tau'}
	p_{21}(\tau^{+},\tau',\theta^{*}) G^{-1}_{11}(\tau',\tau_{f};\theta)
		=-\sum_{\tau'}
		G_{22}(\tau^{+},\tau')
		\frac{\delta G_{21}^{-1}(\tau',\tau_{f};\theta)}
		{\delta \theta^{*}(\tau,\tau)} \\
		&
			=\sum_{\tau',\tau'',\tau_{3},\tau_{4}}
			G_{22}(\tau^{+},\tau')
		\left[\delta_{\tau',\tau}\delta_{\tau_{f},\tau}+\frac{\delta \Sigma_{21}(\tau',\tau_{f};\theta)}{\delta G_{21}(\tau_{3},\tau_{4};\theta)}
		\frac{\delta G_{21}(\tau_{3},\tau_{4};\theta)}{\delta \theta^{*}(\tau,\tau)} \right].
	\end{aligned}
\end{equation}
Let the above equation multiplied by $G^{-1}_{22}(\tau^{+},\tau')$ in the left,
\begin{equation} 
	\begin{aligned}
		\label{674}
		&	\sum_{\tau',\tau''}
		G^{-1}_{22}(\tau^{+},\tau')	
		\frac{\delta G_{21}(\tau',\tau'';\theta)}{\delta \theta^{*}(\tau,\tau)} G^{-1}_{11}(\tau'',\tau_{f};\theta)
		=-\sum_{\tau',\tau''}
		G^{-1}_{22}(\tau^{+},\tau')	G_{22}(\tau',\tau'')
		\frac{\delta G_{21}^{-1}(\tau'',\tau_{f};\theta)}
		{\delta \theta^{*}(\tau,\tau)} \\
		&
		=\sum_{\tau',\tau'',\tau_{3},\tau_{4}}
		G^{-1}_{22}(\tau^{+},\tau')G_{22}(\tau',\tau'')
		\left[\delta_{\tau'',\tau}\delta_{\tau_{f},\tau}+\frac{\delta \Sigma_{21}(\tau'',\tau_{f};\theta)}{\delta G_{21}(\tau_{3},\tau_{4};\theta)}
		\frac{\delta G_{21}(\tau_{3},\tau_{4};\theta)}{\delta \theta^{*}(\tau,\tau)} \right],
	\end{aligned}
\end{equation}
filtered by the choose at $\tau''=\tau^{+}$,
above expression reduces to
\begin{equation} 
	\begin{aligned}
		\label{671}
		&	\sum_{\tau'}
		[G^{-1}_{22}(\tau^{+},\tau')	
		\frac{\delta G_{21}(\tau',\tau^{+};\theta)}{\delta \theta^{*}(\tau,\tau)}] G^{-1}_{11}(\tau^{+},\tau_{f};\theta)
		=-
		\frac{\delta G_{21}^{-1}(\tau^{+},\tau_{f};\theta)}
		{\delta \theta^{*}(\tau,\tau)} \\
		&
		=\sum_{\tau_{3},\tau_{4}}
		\left[\frac{\delta \Sigma_{21}(\tau^{+},\tau_{f};\theta)}{\delta G_{21}(\tau_{3},\tau_{4};\theta)}
		\frac{\delta G_{21}(\tau_{3},\tau_{4};\theta)}{\delta \theta^{*}(\tau,\tau)} \right],
	\end{aligned}
\end{equation}
In this way,
we obtain a point of $\tau'$ in Eq.(\ref{673})
or $\tau''$ in Eq.(\ref{674}) with the compressed information density
(in terms of the quantum information scrambling dynamics)
which is about the conservation sector formed between 
$G_{21}$ and $G_{21}^{-1}$.

Similarly,
\begin{equation} 
	\begin{aligned}
		&	\sum_{\tau'}
		p_{21}(\tau^{+},\tau',\theta^{*}) G^{-1}_{11}(\tau',\tau'';\theta)G_{11}(\tau'',\tau_{f};\theta)
		=-\sum_{\tau'}
		G_{22}(\tau^{+},\tau')
		\frac{\delta G_{21}^{-1}(\tau',\tau'';\theta)}
		{\delta \theta^{*}(\tau,\tau)} G_{11}(\tau'',\tau_{f};\theta)\\
		&
		=\sum_{\tau',\tau'',\tau_{3},\tau_{4}}
		G_{22}(\tau^{+},\tau')
		\left[\delta_{\tau',\tau}\delta_{\tau_{f},\tau}+\frac{\delta \Sigma_{21}(\tau',\tau'';\theta)}{\delta G_{21}(\tau_{3},\tau_{4};\theta)}
		\frac{\delta G_{21}(\tau_{3},\tau_{4};\theta)}{\delta \theta^{*}(\tau,\tau)} \right]G_{11}(\tau'',\tau_{f};\theta),
	\end{aligned}
\end{equation}
which, at $\tau'=\tau_{f}$, reduces to
\begin{equation} 
	\begin{aligned}
		&
		q_{21}(\tau^{+},\tau_{f},\theta^{*})
		=-\sum_{\tau''}
		G_{22}(\tau^{+},\tau_{f})
		[\frac{\delta G_{21}^{-1}(\tau_{f},\tau'';\theta)}
		{\delta \theta^{*}(\tau,\tau)} G_{11}(\tau'',\tau_{f};\theta)]\\
		&
		=\sum_{\tau'',\tau_{3},\tau_{4}}
		G_{22}(\tau^{+},\tau_{f})
		\left[\frac{\delta \Sigma_{21}(\tau_{f},\tau'';\theta)}{\delta G_{21}(\tau_{3},\tau_{4};\theta)}
		\frac{\delta G_{21}(\tau_{3},\tau_{4};\theta)}{\delta \theta^{*}(\tau,\tau)} G_{11}(\tau'',\tau_{f};\theta)\right],
	\end{aligned}
\end{equation}

\begin{equation} 
	\begin{aligned}
		&
			q_{21}(\tau^{+},\tau_{f},\theta^{*})
			-	q_{21}(\tau^{+},\tau_{f},\theta^{*})\bigg|_{\theta=0}\\
			&	=\sum_{\tau'',\tau_{3},\tau_{4}}
		G_{22}(\tau^{+},\tau_{f})
		\left[\frac{\delta \Sigma_{21}(\tau_{f},\tau'';\theta)}{\delta G_{21}(\tau_{3},\tau_{4};\theta)}
		\frac{\delta G_{21}(\tau_{3},\tau_{4};\theta)}{\delta \theta^{*}(\tau,\tau)} G_{11}(\tau'',\tau_{f};\theta)\right]\\
&		
		-\sum_{\tau',\tau'',\tau_{3},\tau_{4}}
		G_{22}(\tau^{+},\tau')
	\left[\delta_{\tau',\tau}\delta_{\tau'',\tau}+	\frac{\delta \Sigma_{21}(\tau',\tau'';\theta)}{\delta G_{21}(\tau_{3},\tau_{4})}\bigg|_{\theta=0}
		\frac{\delta G_{21}(\tau_{3},\tau_{4};\theta)}{\delta \theta^{*}(\tau,\tau)} \bigg|_{\theta=0}\right]G_{11}(\tau'',\tau_{f}).
	\end{aligned}
\end{equation}

Combining with the expression of $\frac{-1}{U}\Sigma_{21}(\tau^{+},\tau_{f};\theta)$
in the limit $\Sigma_{21}(\tau^{+},\tau_{f};\theta)=0$,
the correlation between $G_{21}$ and $G_{11}^{-1}$ follows the pattern of linear-kinetic term,
which can be regarded as the integration by parts
\begin{equation} 
	\begin{aligned}
		&	
		G_{21}(\tau',\tau'')
		\frac{\delta G_{11}^{-1}(\tau'',\tau_{f};\theta)}
		{\delta \theta^{*}(\tau,\tau)}
		=\frac{\delta G_{21}(\tau',\tau^{+};\theta)}{\delta \theta^{*}(\tau,\tau)} G^{-1}_{11}(\tau^{+},\tau_{f};\theta).
	\end{aligned}
\end{equation}

\begin{equation} 
	\begin{aligned}
		&	q_{21}(\tau^{+},\tau_{f},\theta^{*})\bigg|_{\theta=0}
		=G_{22}(\tau^{+},\tau)G_{11}(\tau,\tau_{f})
+\sum_{\tau',\tau'',\tau_{3},\tau_{4}}
G_{22}(\tau^{+},\tau')
	\frac{\delta \Sigma_{21}(\tau',\tau'';\theta)}{\delta G_{21}(\tau_{3},\tau_{4})}\bigg|_{\theta=0}
		\frac{\delta G_{21}(\tau_{3},\tau_{4};\theta)}{\delta \theta^{*}(\tau,\tau)} \bigg|_{\theta=0}G_{11}(\tau'',\tau_{f}).
	\end{aligned}
\end{equation}

\begin{equation} 
	\begin{aligned}
		&\sum_{\tau',\tau'',\tau_{3},\tau_{4}}
		G_{22}(\tau^{+},\tau')
	[\sum_{\tau'''}
	G^{-1}_{22}(\tau',\tau''')	
	[\frac{\delta G_{21}(\tau''',\tau';\theta)}{\delta \theta^{*}(\tau,\tau)}\bigg|_{\theta=0}
	 G^{-1}_{11}(\tau',\tau'')]
	]
	G_{11}(\tau'',\tau_{f})\\
	=		&\sum_{\tau',\tau'',\tau_{3},\tau_{4}}
	G_{22}(\tau^{+},\tau')
	[\sum_{\tau'''}
	[	G^{-1}_{22}(\tau',\tau''')	
	G_{22}(\tau''',\tau)
	G_{11}(\tau,\tau')
	G^{-1}_{11}(\tau',\tau'')
	]
	G_{11}(\tau'',\tau_{f})\\
	+	&\sum_{\tau',\tau'',\tau_{3},\tau_{4},\tau'''}
	[G_{22}(\tau^{+},\tau')
	G^{-1}_{22}(\tau',\tau''')]
	\frac{\delta G_{21}(\tau''',\tau';\theta)}{\delta \theta^{*}(\tau,\tau)}\bigg|_{\theta=0}
[	G^{-1}_{11}(\tau',\tau'')
	G_{11}(\tau'',\tau_{f})]
	\end{aligned}
\end{equation}

The product between self-energy and propagator matrices describeing the two particle correlation in opposite spin components
\begin{equation} 
	\begin{aligned}
\frac{-1}{U}\pmb{\Sigma}\pmb{G}=
&\sum_{\tau'}\begin{pmatrix}
	p_{21}(\tau^{+},\tau',\theta^{*}) 
	& p_{22}(\tau^{+},\tau',\theta^{*}) \\
	p_{11}(\tau^{+},\tau',\theta) 
	& p_{12}(\tau^{+},\tau',\theta) 
\end{pmatrix}\\
=&\sum_{\tau',\tau'',\tau_{3},\tau_{4}}
\begin{pmatrix}
	G_{21}(\tau^{+},\tau')&G_{22}(\tau^{+},\tau')\\
	0  &  0
\end{pmatrix}\\
&
\left[
\begin{pmatrix}
	\frac{\delta \Sigma_{11}(\tau',\tau'';\theta)}{\delta G_{21}(\tau_{3},\tau_{4};\theta)}
	\frac{\delta G_{21}(\tau_{3},\tau_{4};\theta)}{\delta \theta^{*}(\tau,\tau)} 
	& \frac{\delta \Sigma_{12}(\tau',\tau'';\theta)}{\delta G_{21}(\tau_{3},\tau_{4};\theta)}
	\frac{\delta G_{21}(\tau_{3},\tau_{4};\theta)}{\delta \theta^{*}(\tau,\tau)}  \\
\delta_{\tau',\tau}\delta_{\tau_{f},\tau}+\frac{\delta \Sigma_{21}(\tau',\tau'';\theta)}{\delta G_{21}(\tau_{3},\tau_{4};\theta)}
	\frac{\delta G_{21}(\tau_{3},\tau_{4};\theta)}{\delta \theta^{*}(\tau,\tau)} 
	&\frac{\delta \Sigma_{22}(\tau',\tau'';\theta)}{\delta G_{21}(\tau_{3},\tau_{4};\theta)}
	\frac{\delta G_{21}(\tau_{3},\tau_{4};\theta)}{\delta \theta^{*}(\tau,\tau)} 
\end{pmatrix}
\begin{pmatrix}
	G_{11}(\tau'',\tau_{f})&G_{12}(\tau'',\tau_{f})\\
	G_{21}(\tau'',\tau_{f}) & G_{22}(\tau'',\tau_{f})
\end{pmatrix}\right]\\
&+
\sum_{\tau',\tau'',\tau_{3},\tau_{4}}
\begin{pmatrix}
	0  &  0\\
	G_{11}(\tau^{+},\tau')&G_{12}(\tau^{+},\tau')
\end{pmatrix}\\
&\left[
\begin{pmatrix}
	\frac{\delta \Sigma_{11}(\tau',\tau'';\theta)}{\delta G_{12}(\tau_{3},\tau_{4};\theta)}
	\frac{\delta G_{12}(\tau_{3},\tau_{4};\theta)}{\delta \theta(\tau,\tau)} 
	& \delta_{\tau',\tau}\delta_{\tau_{f},\tau}+\frac{\delta \Sigma_{12}(\tau',\tau'';\theta)}{\delta G_{12}(\tau_{3},\tau_{4};\theta)}
	\frac{\delta G_{12}(\tau_{3},\tau_{4};\theta)}{\delta \theta(\tau,\tau)}  \\
	\frac{\delta \Sigma_{21}(\tau',\tau'';\theta)}{\delta G_{12}(\tau_{3},\tau_{4};\theta)}
	\frac{\delta G_{12}(\tau_{3},\tau_{4};\theta)}{\delta \theta(\tau,\tau)} 
	&\frac{\delta \Sigma_{22}(\tau',\tau'';\theta)}{\delta G_{12}(\tau_{3},\tau_{4};\theta)}
	\frac{\delta G_{12}(\tau_{3},\tau_{4};\theta)}{\delta \theta(\tau,\tau)} 
\end{pmatrix}
\begin{pmatrix}
	G_{11}(\tau'',\tau_{f})&G_{12}(\tau'',\tau_{f})\\
	G_{21}(\tau'',\tau_{f}) & G_{22}(\tau'',\tau_{f})
\end{pmatrix}\right]\\
=&
\sum_{\tau'}
\begin{pmatrix}
	G_{22}(\tau^{+},\tau';\theta)G_{11}(\tau',\tau_{f};\theta)&
	G_{22}(\tau^{+},\tau';\theta)G_{12}(\tau',\tau_{f};\theta)\\
G_{11}(\tau^{+},\tau';\theta)G_{21}(\tau',\tau_{f};\theta)  &
G_{11}(\tau^{+},\tau';\theta)G_{22}(\tau',\tau_{f};\theta)
\end{pmatrix}\\
&+
\sum_{\tau',\tau'',\tau_{3},\tau_{4}}
\begin{pmatrix}
	G_{21}(\tau^{+},\tau';\theta)&G_{22}(\tau^{+},\tau';\theta)\\
	0  &  0
\end{pmatrix}\\
&
\begin{pmatrix}
	\frac{\delta \Sigma_{11}(\tau',\tau'';\theta)}{\delta G_{21}(\tau_{3},\tau_{4};\theta)}
	\frac{\delta G_{21}(\tau_{3},\tau_{4};\theta)}{\delta \theta^{*}(\tau,\tau)} 
	& \frac{\delta \Sigma_{12}(\tau',\tau'';\theta)}{\delta G_{21}(\tau_{3},\tau_{4};\theta)}
	\frac{\delta G_{21}(\tau_{3},\tau_{4};\theta)}{\delta \theta^{*}(\tau,\tau)}  \\
	\frac{\delta \Sigma_{21}(\tau',\tau'';\theta)}{\delta G_{21}(\tau_{3},\tau_{4};\theta)}
	\frac{\delta G_{21}(\tau_{3},\tau_{4};\theta)}{\delta \theta^{*}(\tau,\tau)} 
	&\frac{\delta \Sigma_{22}(\tau',\tau'';\theta)}{\delta G_{21}(\tau_{3},\tau_{4};\theta)}
	\frac{\delta G_{21}(\tau_{3},\tau_{4};\theta)}{\delta \theta^{*}(\tau,\tau)} 
\end{pmatrix}
\begin{pmatrix}
	G_{11}(\tau'',\tau_{f};\theta)&G_{12}(\tau'',\tau_{f};\theta)\\
	G_{21}(\tau'',\tau_{f};\theta) & G_{22}(\tau'',\tau_{f};\theta)
\end{pmatrix}\\
&+\sum_{\tau',\tau'',\tau_{3},\tau_{4}}
\begin{pmatrix}
	0  &  0\\
	G_{11}(\tau^{+},\tau';\theta)&G_{12}(\tau^{+},\tau';\theta)
\end{pmatrix}
\\
&
\begin{pmatrix}
	\frac{\delta \Sigma_{11}(\tau',\tau'';\theta)}{\delta G_{12}(\tau_{3},\tau_{4};\theta)}
	\frac{\delta G_{12}(\tau_{3},\tau_{4};\theta)}{\delta \theta(\tau,\tau)} 
	& \frac{\delta \Sigma_{12}(\tau',\tau'';\theta)}{\delta G_{12}(\tau_{3},\tau_{4};\theta)}
	\frac{\delta G_{12}(\tau_{3},\tau_{4};\theta)}{\delta \theta(\tau,\tau)}  \\
	\frac{\delta \Sigma_{21}(\tau',\tau'';\theta)}{\delta G_{12}(\tau_{3},\tau_{4};\theta)}
	\frac{\delta G_{12}(\tau_{3},\tau_{4};\theta)}{\delta \theta(\tau,\tau)} 
	&\frac{\delta \Sigma_{22}(\tau',\tau'';\theta)}{\delta G_{12}(\tau_{3},\tau_{4};\theta)}
	\frac{\delta G_{12}(\tau_{3},\tau_{4};\theta)}{\delta \theta(\tau,\tau)} 
\end{pmatrix}
\begin{pmatrix}
	G_{11}(\tau'',\tau_{f};\theta)&G_{12}(\tau'',\tau_{f};\theta)\\
	G_{21}(\tau'',\tau_{f};\theta) & G_{22}(\tau'',\tau_{f};\theta)
\end{pmatrix}\\
\end{aligned}
\end{equation}

Choosing the left-top elements of Eq.(\ref{331}) as an example,
we have the following relations base on the particle number conservation
and the Luttinger-Ward identity,
\begin{equation} 
\begin{aligned}
&\sum_{\tau}\sum_{\tau'}
p_{21}(\tau^{+},\tau',\theta^{*}) 
G^{-1}_{11}(\tau',\tau_{f};\theta)\\
&=G_{22}(\tau^{+},\tau;\theta)\delta_{\tau_{f},\tau}+\sum_{\tau}\sum_{\tau',\tau_{3},\tau_{4}}G_{22}(\tau^{+},\tau')
\frac{\delta \Sigma_{11}(\tau',\tau_{f};\theta)}{\delta G_{21}(\tau_{3},\tau_{4};\theta)}
\frac{\delta G_{21}(\tau_{3},\tau_{4};\theta)}{\delta \theta^{*}(\tau,\tau)},
\\
&=-\sum_{\tau}\sum_{\tau'}G_{22}(\tau^{+},\tau';\theta)
\frac{\delta G^{-1}_{21}(\tau',\tau_{f};\theta)}{\delta \theta^{*}(\tau,\tau)},
\end{aligned}
\end{equation}
where we define the $x$-independent (thus not commutates with the functional derivative of ${\bf G}$) test function as $G^{-1}_{11}(\tau',\tau^{+};\theta)=\frac{\delta\theta^{*}(\tau,\tau)}
{x}=\eta_{\tau}$ through a $x$-dependent $\delta\theta^{*}(\tau,\tau)=\eta_{\tau}x$
as appears in the denorminator,

\renewcommand\refname{References}

\end{document}